\documentclass[a4paper,10pt]{article}
\pdfoutput=1

\usepackage[usenames,dvipsnames,table]{xcolor}
\usepackage{float}
\usepackage[utf8]{inputenc}
\usepackage{jheppub}
\usepackage{comment}
\usepackage{ytableau}
\usepackage{cancel} 
\usepackage{xcolor}
\usepackage{amsmath,bm}
\usepackage{mathtools}
\usepackage{float}
\usepackage{graphicx}
\usepackage{amsmath, amssymb}
\usepackage{youngtab}
\usepackage{changepage}
\usepackage{etoolbox}
\usepackage{mathdots}

\usepackage{mathtools}

\usepackage{pdflscape}

\pretocmd{\eqref}{Eq.\ }{}{}

\def\cI{{\cal I}}

\def\cM{{\cal M}}

\def\cN{{\cal N}}

\def\cW{{\cal W}}

\def\cZ{{\cal Z}}

\def\D{\Delta}

\newcommand{\nn}{\nonumber}
\newcommand{\PFS}[1]{\mathcal{Z}_{#1}}

\newcommand{\dd}[1]{\bm{\delta}_{#1}}
\newcommand{\DCB}[1]{\Delta^{\rm CB}_{#1}}
\newcommand{\mf}[1]{\mathfrak{#1}}

\newcommand{\Kah}{K\"ahler }
\newcommand{\Kahs}{K\"ahlers }
\newcommand{\MM}[1]{\mathcal{M}\left(#1\right)}

\title{From pq-webs to magnetic quivers and back II: \\ distributional indices and 5d theories breaking bad}

\author[a,b]{Sebastiano Garavaglia}
\author[a,b]{Sara Pasquetti}
\author[a,b]{Palash Singh}
\affiliation[a]{Dipartimento di Fisica, Università di Milano--Bicocca, Piazza della Scienza 3, I-20126 Milano, Italy}
\affiliation[b]{INFN, Sezione di Milano--Bicocca, Piazza della Scienza 3, I-20126 Milano, Italy}
\emailAdd{s.garavaglia18@campus.unimib.it}
\emailAdd{sara.pasquetti@gmail.com}
\emailAdd{palash.singh@unimib.it}

\abstract{This is the second of two companion papers in which we study five-dimensional $\mathcal{N}=1$ theories engineered by $(p,q)$ five-brane webs and their associated three-dimensional magnetic quivers. Leveraging the exact parameter dictionary between the $N+\mathrm{SU}(N)_0^{k-1}+N$ theories and their magnetic quivers at infinite coupling, proposed in the first paper, we investigate nilpotent moment-map VEV deformations and their three-dimensional counterparts. We show that, when the Higgsing is excessive, the five-dimensional superconformal index becomes distributional, mirroring a characteristic feature of bad and broken theories in three dimensions. We also develop a five-brane web interpretation of these results, identifying the Hanany--Witten moves associated with the Higgsing cascade and relating inequivalent frames to distinct tessellations of the generalised toric diagram.}

\begin{document} 

\maketitle

\flushbottom

\section{Introduction}
\label{sec:intro}

It is by now well established that supersymmetric partition functions and indices need not be ordinary functions, but may instead acquire a distributional character. Rather than being merely a technical subtlety of localisation, this phenomenon can encode the vacuum structure, spontaneous symmetry breaking, and non-trivial infrared dynamics of the theory.

An early example arises in four-dimensional $\mathcal{N}=1$ theories exhibiting chiral symmetry breaking, whose superconformal indices become distributions \cite{Spiridonov:2014cxa}. This distributional structure reflects a quantum-deformed moduli space on which gauge-invariant operators necessarily acquire expectation values. Upon gauging the corresponding flavour symmetries, these VEVs Higgs the newly introduced gauge group. The delta functions in the index therefore encode the quantum vacuum structure rather than a pathology of the localisation computation.

A closely related phenomenon occurs in three-dimensional $\mathcal{N}=4$ bad theories. Their $S_b^3$ partition functions admit a distributional completion that takes the form of a sum over different frames \cite{Giacomelli:2023zkk,Giacomelli:2024laq}. In bad SQCD, for instance, each frame contains a good theory of lower rank, together with a decoupled free sector and, generically, delta functions imposing the conditions under which monopole operators acquire vacuum expectation values. A particularly interesting subclass is provided by the broken theories introduced in \cite{Comi:2025zwu}. Their partition functions also decompose into sums of distributional frames, but all the frames are related by global-symmetry transformations. This structure reflects spontaneous symmetry breaking: upon gauging the corresponding flavour symmetry, the delta functions constrain the corresponding integration variables and reduce the gauged symmetry to a proper subgroup. This mechanism has proved useful, for example, in understanding channel decompositions of theories of class $\mathcal{S}$ with non-maximal punctures \cite{Comi:2025zwu}.

The purpose of this paper is to show that an analogous phenomenon arises in five-dimensional $\mathcal{N}=1$ theories. We demonstrate that sufficiently large nilpotent Higgsings can render the five-dimensional superconformal index distributional. The resulting periodic delta functions impose relations among external flavour fugacities and organise the index into a sum of frames. We call the corresponding five-dimensional configurations {\it broken} when all the distributional frames are rotated by the flavour symmetry, and {\it bad} when they form multiple inequivalent classes, in direct analogy with the three-dimensional terminology.

We focus on the five-dimensional quiver gauge theories $N+SU(N)_0^{k-1}+N$, engineered by box-shaped $(p,q)$ five-brane webs with $N$ D5 branes and $k$ NS5 branes \cite{Aharony:1997ju,Aharony:1997bh}. We refer to this family as the \emph{box} theories. Equivalently, they can be engineered by M-theory on non-compact toric Calabi--Yau threefolds whose toric diagrams are dual to the corresponding five-brane webs \cite{Katz:1996fh,Leung:1997tw}.

Our main five-dimensional observable is the superconformal index, or equivalently the $S^4\times S^1$ partition function \cite{Kim:2012gu,Terashima:2012ra}. This is a protected observable of the ultraviolet SCFT. When the fixed point admits an infrared Lagrangian gauge theory description, the index can be computed in that frame by supersymmetric localisation, which expresses it as an integral over the Coulomb branch parameters of the modulus squared of the Nekrasov partition function \cite{Nekrasov:2002qd,Nekrasov:2003rj}. For the box theories considered here, the latter can be computed as a refined topological string partition function using the refined topological vertex \cite{Aganagic:2003db,Awata:2005fa,Iqbal:2007ii,Taki:2007dh,Iqbal:2012xm}.

We study deformations generated by nilpotent VEVs for the flavour moment maps. Such a VEV is labelled by a partition of $N$,
\begin{equation}
    \rho=[r_1,r_2,\ldots,r_s],
    \qquad r_1\geq r_2\geq\cdots\geq r_s.
\end{equation}
In the five-brane web, it is realised by terminating groups of parallel external D5 branes on common D7 branes.  In the corresponding generalised toric diagram, this operation introduces
white dots encoding the partition of the external D5 branes \cite{Benini:2009gi,vanBeest:2020kou,Bourget:2023wlb}. At the level of the refined topological string partition function, the VEV is implemented by specialising the external mass parameters \cite{Hayashi:2013qwa,Hayashi:2014hfa,Isachenkov:2014eya}.

\begin{figure}[b!]
    \centering
    \includegraphics[width=1\linewidth]{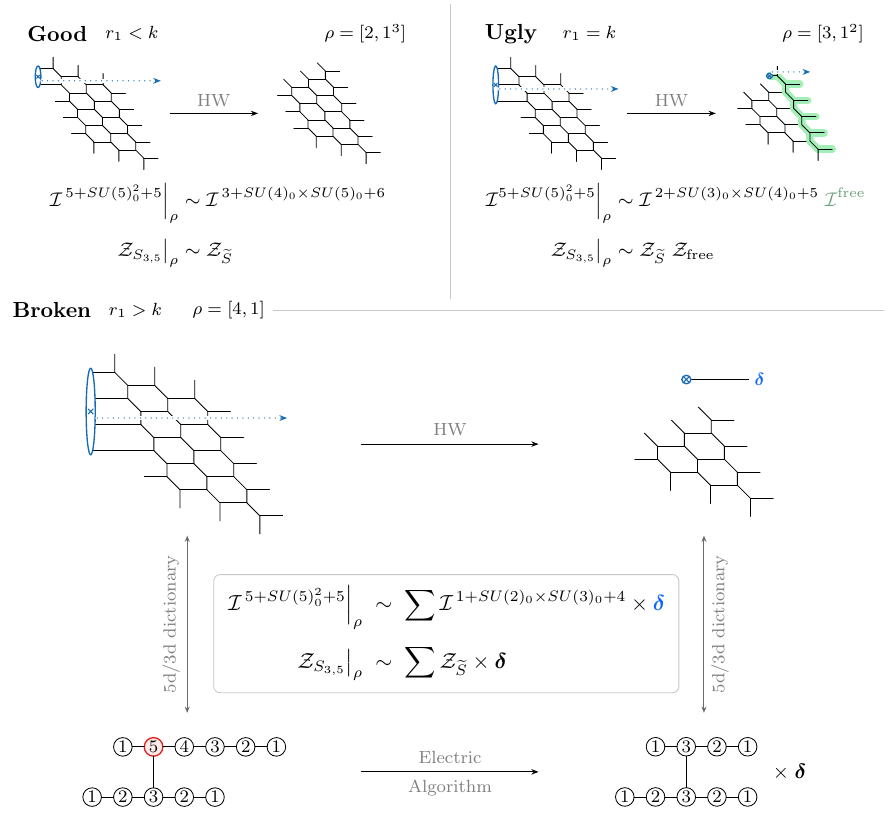}
    \caption{Three nilpotent deformations of the $5+SU(5)_0^2+5$ box theory and its magnetic quiver $S_{3,5}$. The partitions $[2,1^3]$, $[3,1^2]$, and $[4,1]$ produce the good, ugly, and broken cases, respectively. In each case, the 5d/3d dictionary identifies the structure of the superconformal index with that of the $S_b^3$ partition function. Hanany--Witten moves reveal a free sector in the ugly case and a semi-infinite D5 brane associated with an external periodic delta function (in blue) in the broken case. We denote the residual magnetic quiver by $\widetilde S$. The sum in the broken case runs over symmetry-related frames. The equalities are schematic: Goldstone contributions and other prefactors are suppressed.}
    \label{fig:OneSided}
\end{figure}
We first consider one-sided VEVs, in which one of the two $SU(N)$ flavour moment maps of the box theory acquires a nilpotent VEV, as illustrated in Figure~\ref{fig:OneSided}. The mass specialisation moves the poles of the index integrand and can pinch its Coulomb integration contours. Evaluating the corresponding residues produces the index of the infrared theory, together with the Goldstone multiplets generated by the VEV \cite{Gaiotto:2012xa}. For sufficiently large Higgsings, the resulting constraints propagate through the entire box and reach the opposite external boundary. They then survive as periodic delta functions imposing relations among external flavour fugacities, so the index becomes a distribution.

The propagation through the web admits a local description in terms of the strip decomposition of the refined topological string partition function. We derive a rule for how each Jordan block crosses an individual strip and how the resulting constraints pass from one set of gluing variables to the next. Geometrically, the same process is realised by a sequence of Hanany--Witten moves \cite{Hanany:1996ie}. Iterating this rule determines whether the Higgsing terminates inside the box or reaches its opposite boundary.

For a one-sided VEV, the outcome depends on the size $r_1$ of the largest Jordan block relative to the width $k$ of the box:
\begin{equation}\label{eq:intro_single_threshold}
    \begin{array}{ccl}
        r_1<k &:& \text{good case; an ordinary residual theory}, \\[2mm]
        r_1=k &:& \text{ugly case; a residual theory with a decoupled free sector}, \\[2mm]
        r_1>k &:& \text{broken case; a sum of equivalent distributional frames}.
    \end{array}
\end{equation}
In the broken case, the delta functions in the different frames are related by the Weyl symmetry of the original flavour group. If this flavour symmetry is gauged, the delta functions reduce the integration measure to that of the unbroken subgroup. The other factors in each frame describe a residual theory, any decoupled free sectors, and the Goldstone multiplets associated with the initial VEV.

Magnetic quivers play a central role in analysing these results. For the box theories studied here, the strong-coupling magnetic quiver $S_{k,N}$ is a three-dimensional $\mathcal N=4$ unitary gauge theory whose Coulomb branch reproduces the Higgs branch of the five-dimensional theory at its ultraviolet fixed point
\cite{Benini:2009gi,Cremonesi:2015lsa,Ferlito:2016grh,Ferlito:2017xdq,Cabrera:2018jxt,vanBeest:2020kou}.\footnote{More generally, magnetic quivers include non-simply-laced examples that need not admit a conventional Lagrangian description
\cite{BourgetHananyMiketa2020,BourgetGrimmingerMartoneZafrir2021}.}
The $S_{k,N}$ theory has two full $T[U(N)]$ tails and two full $T[U(k)]$ tails attached, respectively, to central $U(N)$ and $U(k)$ gauge nodes.

In the companion paper \cite{one}, we established a dictionary between the five-dimensional flavour masses, instanton counting parameters, and bifundamental abelian masses of the box theory and the FI parameters of $S_{k,N}$. Under this dictionary, a nilpotent VEV for a five-dimensional flavour moment map maps to a monopole VEV in the magnetic quiver. We can therefore compare the effects of the same deformation on the five-dimensional index and the magnetic quiver's $S_b^3$ partition function, including the relations among their parameters.

The criterion in \eqref{eq:intro_single_threshold} has a precise magnetic-quiver counterpart. A nilpotent VEV with largest block $r_1$ reduces the effective number of flavours of the corresponding central $U(N)$ node of $S_{k,N}$. This node is good for $r_1<k$, ugly for $r_1=k$, and bad for $r_1>k$. When only one full tail is deformed, the bad quiver belongs to the broken class. Thus the magnetic quiver becomes broken precisely at the threshold at which the five-dimensional index becomes distributional.

The agreement extends beyond the threshold itself.
The 5d/3d dictionary identifies the supports of the periodic delta functions in the index with those of the Dirac delta functions in the $S_b^3$ partition function. It also matches the number and symmetry properties of the frames, the residual theories and their magnetic quivers, the parameters inherited by these theories, any decoupled free sectors, and the Goldstone contributions. These are independent features of two different supersymmetric observables, computed using different realisations of the Higgsing.

\begin{figure}[t!]
    \centering
    \includegraphics[width=1\linewidth]{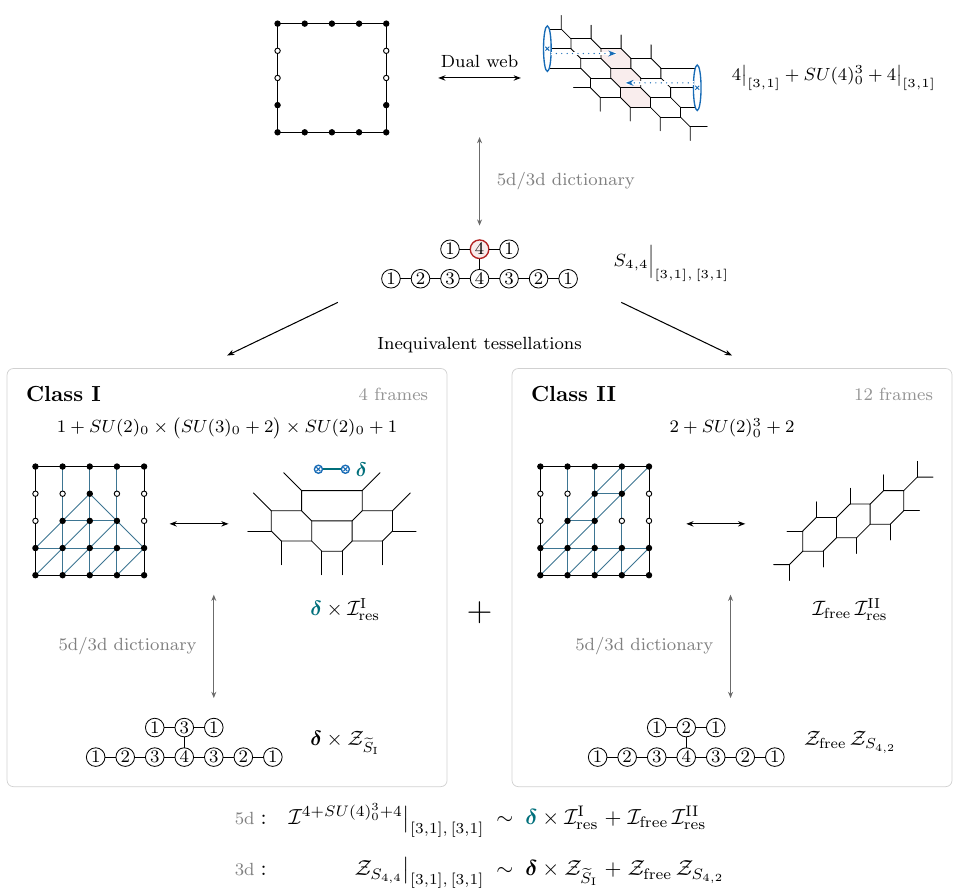}
    \caption{Two $[3,1]$ nilpotent VEVs of the $4+SU(4)_0^3+4$ box theory produce two inequivalent classes of contributions associated to distinct tessellations of the same generalised toric polygon. The 5d/3d dictionary matches the residual theories, delta-function constraints, and free sectors in each class. Class I carries an external delta function in both dimensions, geometrically encoded in five dimensions by the alignment required to detach the finite D5 brane between the D7 branes. Class II has no external delta function and contains an additional free sector. The symbol $\sim$ indicates that the Goldstone contributions associated with the initial nilpotent VEVs have been suppressed.}
    \label{fig:TwoSided}
\end{figure}

We next consider two-sided deformations, in which both $SU(N)$ flavour moment maps acquire nilpotent VEVs. Let $r_L$ and $r_R$ be the sizes of the largest Jordan blocks in their respective partitions. On the magnetic-quiver side, the effective number of flavours of the central $U(N)$ node is
$N_f^{\rm eff}=k+2N-r_L-r_R$. This gives the classification
\begin{equation}\label{eq:intro_double_threshold}
    \begin{array}{ccl}
        r_L+r_R\leq k
        &:& \text{good case; an ordinary residual theory}, \\[2mm]
        r_L+r_R=k+1
        &:& \text{ugly case; a residual theory with a decoupled free sector}, \\[2mm]
        r_L+r_R>k+1
        &:& \text{bad case; a sum containing inequivalent frames}.
    \end{array}
\end{equation}
In the bad regime, the magnetic-quiver partition function can decompose into inequivalent frames. We find a corresponding structure in the five-dimensional index.

Our main example is the $4+SU(4)_0^3+4$ box theory with a $[3,1]$ nilpotent VEV on each side, illustrated in Figure~\ref{fig:TwoSided}. The two Higgsings propagate through the box and eventually constrain the same set of internal Coulomb parameters. Resolving the resulting collision yields sixteen contributions. Four arise when the two constraints act on the same Coulomb parameter (Class I), while twelve arise when they act on distinct parameters (Class II). The classes are inequivalent: Class I retains an external periodic delta function, whereas Class II has no external delta function and contains an additional free sector. Electric dualisation of the deformed magnetic quiver $S_{4,4}$ produces the same two classes, with matching delta functions, residual theories, free sectors, Goldstone contributions, and parameter maps.

The example also suggests a geometric interpretation of five-dimensional bad theories. The generalised toric polygon of the doubly Higgsed configuration admits two classes of tessellations corresponding to the two classes of frames. In this example, the tessellation classes are not related by flop transitions or Hanany--Witten moves. This motivates the proposal that inequivalent tessellation classes of a generalised toric polygon can provide a geometric diagnostic of bad five-dimensional configurations.

Taken together, our results show that magnetic quivers retain detailed information about five-dimensional deformations even when they become bad or broken. They identify the threshold for a distributional index and, in the examples we study, reproduce its frame decomposition, symmetry-breaking pattern, residual theories, free sectors, and parameter dependence. Conversely, the five-brane web and topological string descriptions provide a geometric realisation of distributional structures found in three-dimensional bad and broken theories.

These results suggests several directions for further investigation. A natural next step is the analysis of more general simultaneous nilpotent VEVs, in particular those involving both the $SU(N)$ and $SU(k)$ moment maps. The geometric interpretation that we propose also motivates a systematic prescription for extracting frames, multiplicities, and distributional support directly from the five-brane webs or generalised toric diagrams. Finally, combining the nilpotent deformations studied here with the mass deformations analysed in \cite{one} raises the question of when these operations commute. 

The paper is organised as follows. In Section~\ref{sec:review_3d}, we review three-dimensional bad and broken theories and study nilpotent monopole VEVs in the magnetic quivers $S_{k,N}$. In Section~\ref{sec:review_5d}, we introduce the five-dimensional strip and box partition functions and review the 5d/3d dictionary established in \cite{one}. In Section~\ref{sec:Higgsing}, we study one-sided nilpotent Higgsings. We begin with good and ugly examples, derive the local propagation rule, apply it to general partitions, and establish the criterion \eqref{eq:intro_single_threshold} for broken theories. In Section~\ref{sec:multiplenil_bad}, we consider simultaneous Higgsings of the two flavour symmetries, analyse the resulting bad five-dimensional theory, and discuss its geometric interpretation in terms of inequivalent tessellations. The appendices collect conventions for special functions and three-dimensional partition functions, together with the Goldstone-mode contributions.

\section{3d bad theories and broken MQ }
\label{sec:review_3d}

A central result of this paper is that turning on excessive moment map VEVs in a five-dimensional $\mathcal{N}=1$ theory can render its superconformal index distributional. Under the 5d/3d map proposed in \cite{one}, these deformations correspond to monopole VEVs in the associated magnetic quiver. Such VEVs Higgs selected gauge nodes and can render the resulting magnetic quiver bad or broken, with a correspondingly distributional $S^3_b$ partition function.

In this section, we focus on the three-dimensional side and study how these deformations drive magnetic quivers into bad or broken phases. We first review the relevant results for three-dimensional bad and broken theories and then discuss some examples of the $S_{k,N}$ magnetic quivers that exhibit these behaviours. In the following sections, we will instead analyse the five- and three-dimensional theories in parallel.

The notion of badness in three dimensions is controlled by the scaling dimensions of monopole operators computed using the ultraviolet $R$ charge assignment \cite{Gaiotto:2008ak}. In particular, a unitary gauge node $U(N_c)$ with $N_f$ fundamental hypermultiplets is called good if $N_f\geq 2N_c$, ugly if $N_f=2N_c-1$, and bad if $N_f<2N_c-1$. In the last case, the dimensions of some monopole operators, computed using the na\"ive ultraviolet $R$ charge assignment, violate the unitarity bound. The ultraviolet $R$ symmetry therefore cannot coincide with the superconformal $R$ symmetry in the infrared, and some monopole operators decouple from the interacting sector as free fields.

The appearance of these decoupled operators makes the infrared behaviour of bad theories considerably more intricate \cite{Assel:2017jgo,Assel:2018exy}. This is reflected in the $S^3_b$ partition function, whose defining matrix integral is divergent when evaluated using the na\"ive $R$ charge assignment. As shown in \cite{Giacomelli:2023zkk}, a careful analysis reveals that the apparent divergences of bad-theory partition functions reflect their distributional nature. These partition functions can generally be expressed as sums of inequivalent distributional contributions, or {\it frames}.

\begin{figure}[h!]
    \centering
    \includegraphics[width=\linewidth]{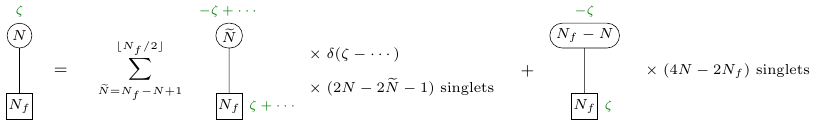}
    \caption{Schematic depiction of the partition function of bad $U(N)$ SQCD with $N\leq N_F<2N-1$ flavours, with FI parameter $\zeta$.}
    \label{fig:badSQCD}
\end{figure}

The simplest example is bad $U(N)$ SQCD with $N\leq N_f<2N-1$ flavours. Its partition function was analysed in \cite{Giacomelli:2023zkk} and takes the form summarized in Figure \ref{fig:badSQCD}. Each frame contains a good SQCD theory of lower rank, together with a decoupled free sector and background contact terms. Except for the last frame, each contribution also contains a Dirac delta function imposing a constraint on the FI parameter. This constraint sets to zero the total charge, defined in \eqref{eq:monopolecharge}, of a suitably dressed monopole operator that acquires a VEV and Higgses $U(N)$ to $U(\widetilde N)$.\footnote{In the so-called \textit{evil} case, $N_f<N$, the residual rank $\widetilde N$ ranges from zero to $\left\lfloor N_f/2\right\rfloor$, and every frame is accompanied by a Dirac delta function.} The last frame, of rank $N_f-N$, was first identified in \cite{Kim:2012uz,Yaakov:2013fza} and is known as the symmetric frame. It carries no Dirac delta function imposing a monopole VEV condition, so the topological symmetry remains unbroken.

An algorithm for computing the partition functions of bad linear unitary quivers was developed in \cite{Giacomelli:2023zkk,Giacomelli:2024laq}. It applies the SQCD dualisation identity locally to underbalanced nodes and iterates these moves until the residual quivers have only good nodes. For a generic bad quiver theory, the result yields a distributional sum of inequivalent frames. More recently, the algorithm has been extended to star-shaped and circular quivers \cite{Comi:2025zwu,Comi:2026lcj}.

A distinguished subclass, referred to as {\it broken theories}, was identified in \cite{Comi:2025zwu}. A broken theory is a bad theory whose partition function decomposes into a sum of frames that are equivalent up to global symmetry transformations. The terminology reflects the spontaneous symmetry breaking exhibited by these theories. Indeed, if one gauges the global symmetry relating the equivalent frames, the delta functions appearing in the partition function constrain the corresponding gauge integral. As a result, the gauged symmetry is effectively Higgsed to a proper subgroup.

The first examples of broken theories identified in \cite{Comi:2025zwu} are star-shaped quivers with an underbalanced central $U(N)$ node, at least one full $T[U(N)]$ tail, and additional quiver tails labelled by partitions $\rho_1,\ldots,\rho_L$ of $N$ encoding the ranks of the gauge nodes in the corresponding tail. Such star-shaped quivers arise as three-dimensional mirrors of theories of class $\mathcal{S}$ \cite{Benini:2010uu}. In this setting, broken theories capture the symmetry-breaking dynamics associated with channel decompositions of theories of class $\mathcal{S}$ on surfaces with non-maximal punctures \cite{Comi:2025zwu}.

Broken theories also arise beyond the class $\mathcal{S}$ setting. In particular, examples have been found in which a broken theory block effectively converts a unitary gauge group into a symplectic one and generates matter in the antisymmetric representation \cite{badusp}. More generally, they arise naturally when gauging Coulomb branch symmetries of unitary quivers \cite{cbgauge}.

We now describe another family of broken theories, obtained by imposing sufficiently large moment map VEVs on magnetic quivers associated with five-dimensional theories at strong coupling.

\subsection{Excessive nilpotent VEVs and $S_{k,N}$ MQ breaking bad}
\label{subsec:sknbreakingbad}

\begin{figure}[h!]
    \centering
    \includegraphics[width=0.5\linewidth]{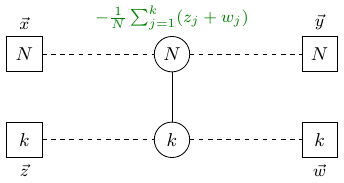}
    \caption{The $S_{k,N}$ magnetic quiver. We use the dashed-line compact notation for the $T[U(N)]$ and $T[U(k)]$ tails.}
    \label{fig:SkN}
\end{figure}

The central three-dimensional theory in this work is $S_{k,N}$, shown in Figure \ref{fig:SkN}. Here and throughout this work, the dashed line notation denotes $T[U(N)]$ and $T[U(k)]$ theories; refer to Appendix \ref{app:3dstuff} for further details. The $S_{k,N}$ theory is the magnetic quiver associated with the five-dimensional $N+SU(N)_0^{k-1}+N$ box theory.

The $S_{k,N}$ theory contains central $U(N)$ and $U(k)$ gauge nodes connected by a bifundamental hypermultiplet. Two full $T[U(N)]$ tails attach to the former, and two $T[U(k)]$ tails attach to the latter. Its global symmetry therefore contains at least $SU(N)^2\times SU(k)^2\times U(1)$. We parameterise the two $T[U(N)]$ tails by $\vec x$ and $\vec y$ subject to $\sum_{i=1}^Nx_i=\sum_{i=1}^Ny_i=0$. The two $T[U(k)]$ tails are instead parameterised by unconstrained $U(k)$ parameters $\vec w$ and $\vec z$.

Since the quiver in Figure \ref{fig:SkN} contains only unitary gauge nodes and bifundamental matter, its overall diagonal $U(1)$ gauge factor decouples. We choose the FI parametrisation so that the FI parameter of the decoupled $U(1)$ factor vanishes, while the central $U(N)$ node carries the displayed FI parameter $$-\frac{1}{N}\sum_{j=1}^k(w_j+z_j) ~.$$ Upon expanding the dashed line notation for the tails, the associated contact terms add $x_N+y_N$ to the FI parameter of the central $U(N)$ node; see Appendix \ref{app:3dstuff}.

Throughout this paper, the interacting magnetic quiver $S_{k,N}$, and its deformations, are understood with the decoupled diagonal $U(1)$ removed. In the conventions of Appendix \ref{app:3dstuff}, its universal contribution is
\begin{equation}
    s_b\left(2m-\frac{iQ}{2}\right) \;\delta(0) ~.
\end{equation}
The double sine factor comes from the neutral adjoint chiral multiplet of the decoupled vector multiplet, while $\delta(0)$ arises from integrating over the diagonal $U(1)$ with vanishing FI parameter. The sphere partition function of the $S_{k,N}$ theory can be expressed as
\begin{equation}
    \begin{split}
         \PFS{S_{k,N}}&(\vec{x},\vec{y},\vec{z},\vec{w},m) = \int {\rm d}\vec{t}_k \, \D^{\rm HB}_k(\vec{t};m) \int {\rm d}\vec{u}_N \, \D^{\rm HB}_N(\vec{u};m) \, e^{-2\pi i\frac{1}{N} \sum_{j=1}^k(w_j+z_j)\sum_{i=1}^N u_i} \\
         &\qquad \times \; \mathcal{Z}^{T[U(N)]}(\vec{u},\vec{x};m) \, \mathcal{Z}^{T[U(N)]}(\vec{u},\vec{y};m) \, \mathcal{Z}^{T[U(k)]} (\vec{t},\vec{w};m) \, \mathcal{Z}^{T[U(k)]}(\vec{t},\vec{ z};m) \\
         &\qquad\quad \times \; \prod_{i=1}^N\prod_{j=1}^k s_b\left( \frac{i Q}{2}-m\pm (u_i-t_j) \right) ~,
    \end{split}
\end{equation}

As we will next discuss, nilpotent VEVs for the Coulomb branch moment maps associated with these tails can drive $S_{k,N}$ into a broken or bad regime. For a $T[U(N)]$ tail, turning on a nilpotent VEV for the Coulomb branch moment map corresponds to
\begin{equation}
    \langle \mathcal C \rangle = J_\rho ~,
\end{equation}
where $J_\rho$ is the Jordan matrix associated with a partition $\rho$ of $N$. This deformation triggers a flow to $T_\rho[U(N)]$, together with a decoupled sector containing the Goldstone multiplets generated by the symmetry breaking \cite{Gaiotto:2008ak}. At the level of the partition function this gives
\begin{equation}
    \left. \mathcal Z^{T[U(N)]}(\vec r,\vec p,m) \right|_{\rho} = \mathcal Z^{T_\rho[U(N)]}(\vec r,\vec p_\rho,m)\; {\rm GM}_{\rm 3d}^{\rho}(\vec p\,{}',\vec h).
\end{equation}
Here $|_\rho$ denotes the nilpotent specialisation: the Coulomb branch parameters belonging to each Jordan block are arranged in an arithmetic progression. The parameters $\vec h$ specify these progressions, while $\vec p\,{}'$  collects the components of $\vec p$ that are not fixed by the nilpotent VEV condition. As reviewed in Appendix \ref{app:GM}, the arithmetic progression condition follows by requiring each moment-map component that acquires a VEV to have vanishing total charge. The same analysis determines the contribution ${\rm GM}_{\rm 3d}^{\rho}(\vec p\,{}',\vec h)$ of the associated Goldstone multiplets.

\subsubsection{Broken regime}
\label{subsec:MQbreakbad}

\begin{figure}[h!]
    \centering
    \includegraphics[width=\linewidth]{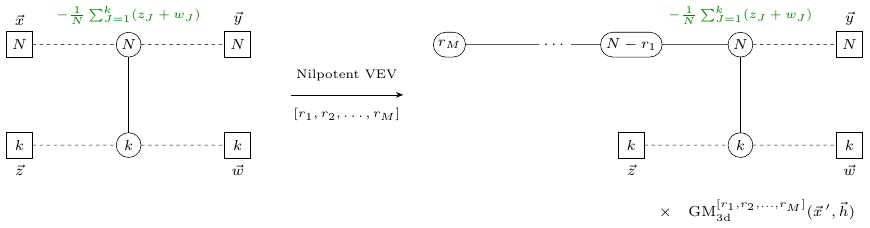}
    \caption{A nilpotent VEV labelled by $\rho$ partially closes one $T[U(N)]$ tail of $S_{k,N}$ to $T_\rho[U(N)]$. This triggers a flow to the deformed quiver on the right along with the associated Goldstone contribution ${\rm GM}^\rho_{\rm 3d}$. FI parameters affected by the deformation are suppressed.}
    \label{fig:SCMQ_VEV}
\end{figure}

We begin by considering the effect of turning on a nilpotent VEV for one of the four tails of the $S_{k,N}$ theory. Consider a nilpotent VEV for the moment map associated with $SU(N)_{\vec x}$, given by the partition
\begin{equation}
    \rho = [r_1,r_2,\dots,r_M] \;\;,\quad r_1\geq r_2 \geq \cdots \geq r_M > 0 \;\;,\quad \sum_{i=1}^M r_i=N ~.
\end{equation}
The resulting quiver is shown on the right of Figure \ref{fig:SCMQ_VEV}.\footnote{We leave the FI parametrisation of the deformed tail and the induced shift of the central FI parameter implicit for a general partition $\rho$. Explicit parametrisations can be obtained case by case by adapting the Higgsing calculation in \cite[Appendix C]{Comi:2025zwu} to the chosen rank $N$ and partition $\rho$.} After this deformation, the effective number of flavours charged under the central $U(N)$ node is
\begin{equation}
    F_{\rm eff} = (N-r_1) + k + (N-1) = 2N-1+k-r_1 ~.
\end{equation}
Therefore the quiver is good for $k>r_1$, ugly for $k=r_1$, and bad for $k<r_1$.
 
In the case of $k<r_1$, applying the electric dualisation algorithm yields a partition function of the characteristic form
\begin{equation}\label{eq:sknbrokengen}
    \begin{split}
        &\PFS{S_{k,N}}(\vec{x}_\rho,\vec{y},\vec{z},\vec{w},m)|_{\rho} \\
        &= {\text{GM}^{\rho}_\text{3d}}(\vec{x}\,{}',\vec h) \sum_{\sigma \in W_{SU(N)}/W_{H \subset SU(N)}} \PFS{\widetilde{S}}(\ldots;m)\times \frac{\D_{H \subset SU(N)} (\ldots;m)}{\D_{SU(N)}(\vec{y};m) } \times  \frac{\dd{\text{Nil}}}{\text{GM}^{\rho'}_{\text {3d}}(\text{$\vec y\,{}'$},\text{$\vec h_{\rho'}$}) } \times \dd{\text{Ex}} ~,
    \end{split}
\end{equation} 
where the prefactor ${\rm GM}^\rho_{\rm 3d}$ contains the contributions from the Goldstone multiplets generated by the original nilpotent VEV $\rho$. The partition function of a broken theory thus contains a sum over equivalent frames, each of which contains the partition function of the residual, possibly interacting, theory $\widetilde S$. Its parameters are combinations of the original parameters and the VEV parameters; we suppress this dependence in the general expression. The sum runs over Weyl cosets, with $\sigma$ acting on the $\vec y$ dependent data of a representative frame. Thus, the frames are related by global symmetry transformations.

The Dirac delta functions encode the monopole VEVs responsible for breaking $SU(N)_{\vec y}$ to a subgroup $H$. We separate them into two sets: $\bm\delta_{\rm Nil}$ and $\bm\delta_{\rm Ex}$. The constraints $\bm\delta_{\rm Nil}$ impose a nilpotent specialisation corresponding to a partition $\rho'$ of $N$.\footnote{The parameters $\vec h_{\rho'}$ of the induced nilpotent VEV are determined by $\bm\delta_{\rm Nil}$ in terms of the original theory's parameters. They need not coincide with the parameters $\vec h$ of the incoming nilpotent VEV.} The other set $\bm\delta_{\rm Ex}$ arises from additional monopole operators that acquire a VEV and generally involves parameters of other global symmetries as well. Combined, these break the $SU(N)$ symmetry down to $H\subset SU(N)$ which, for the cases considered here, is simply an $SU(L)$ subgroup of $SU(N)$. Indeed, the total number of independent constraints on the $SU(N)_{\vec y}$ parameters imposed by the full set of Dirac delta functions is given by
\begin{equation}
    \# {\rm delta} = N-1-{\rm rank}(H) ~.
\end{equation}
The frames are then rotated by the permutation symmetry given by the Weyl group of $SU(N)$ modulo that of $H\subset SU(N)$.

The interpretation of \eqref{eq:sknbrokengen} becomes particularly transparent when $SU(N)_{\vec y}$ is gauged together with a test $T[U(N)]$ tail.\footnote{A $T[U(N)]$ tail is a natural test theory because the five-dimensional construction motivates gluing different $S_{k,N}$ blocks by gauging diagonal combinations of their $SU(N)$ symmetries.} The Dirac delta functions restrict the gauge integral to the locus appropriate to $H$, while the measure ratio converts the $SU(N)$ contribution into the $H$ contribution. We collect the relevant factors into the following combination
\begin{equation}\label{eq:deltacorr3d}
    \Delta^{\rm corr}_{\rm 3d} \;\coloneq\; \left. \frac{1}{{\rm GM}^{\rho'}_{\rm 3d}(\vec y\,{}',\vec h_{\rho'})}\frac{\D_{H \subset SU(N)} (\ldots;m)}{\D_{SU(N)}(\ldots;m) } \right|_{\bm\delta_{\rm Nil},\,\bm\delta_{\rm Ex}} ~.
\end{equation}
which corrects the $SU(N)$ integration measure to the $H\subset SU(N)$ integration measure. The inverse Goldstone factor has a separate role. The constraints $\bm\delta_{\rm Nil}$ induce the corresponding nilpotent deformation of the test tail
\begin{equation}
    \mathcal{Z}^{T[U(N)]}(\vec{r},\vec{y},m)\vert_{\dd{\rm{Nil}}} = \mathcal{Z}^{T_{\rho'}[U(N)]}(\vec{r},\vec{y}_{\rho'},m) \times {\rm GM}^{\rho'}_{\rm 3d}(\vec y\,{}',\vec h_{\rho'}) ~.
\end{equation}
These Goldstone contributions are exactly cancelled by the inverse factor in \eqref{eq:sknbrokengen} (which also appears in our definition in \eqref{eq:deltacorr3d}). Thus the distribution both reduces the gauging to $H$ (along with the appropriate correction of the integration measure) and generically implements a nilpotent deformation of the theory to which it is coupled.

For concreteness, we discuss a concrete example of a broken $S_{k,N}$ theory. Consider $S_{2,4}$ shown at the top left of Figure \ref{fig:VEV_broken}. We impose the $[3,1]$ nilpotent VEV on $SU(4)_{\vec x}$ by setting (as in Appendix \ref{app:GM}):
\begin{equation}
    x_1=h'+\frac{iQ}{2}-m \;\;,\quad x_2=h'-\frac{iQ}{2}+m \;\;,\quad x_3=h'-\frac{3iQ}{2}+3m ~.
\end{equation}
This deformation triggers the flow to the theory depicted on the top right of Figure \ref{fig:VEV_broken} along with some Goldstone multiplets that contribute ${\rm GM}_{\rm 3d}^{[3,1]}(\vec x',h')$. Note that here $\vec x'$ is simply $x_4$ since it is the only parameter in $\vec x$ that does not appear in the VEV condition.

\begin{figure}[h!]
    \centering
    \includegraphics[width=\linewidth]{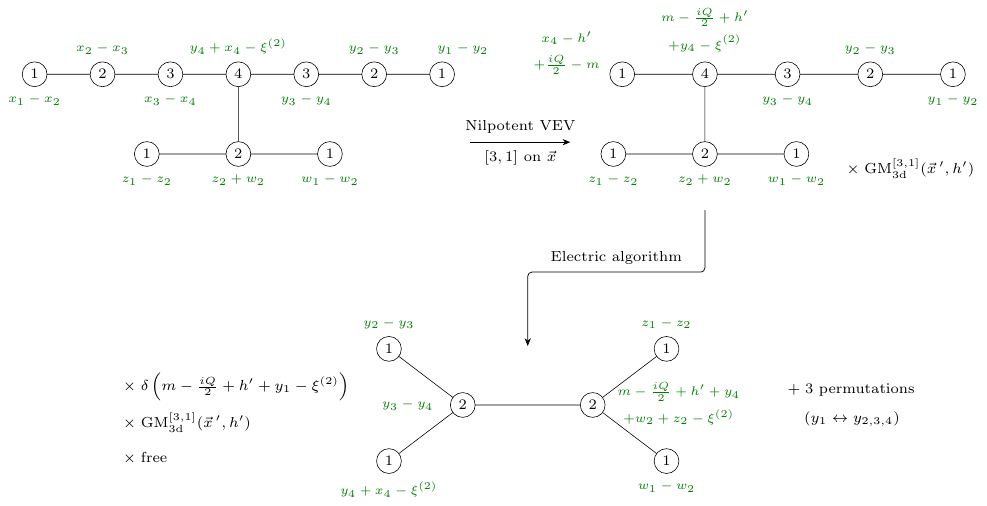}
    \caption{The effect of a $[3,1]$ nilpotent VEV for one of the two $SU(4)$ in $S_{2,4}$. The resulting theory is broken and the electric dualisation algorithm produces equivalent frames related by permutations of $\vec y$. The Dirac delta constraint implements the partial Higgsing of $SU(4)_{\vec y}$ down to $SU(3)$. For brevity, we use the following definition $\xi^{(2)}=\frac14\sum_{i=1}^2 (w_i+z_i)$.}
    \label{fig:VEV_broken}
\end{figure}

The resulting theory is indeed bad since the central $U(4)$ only sees $6<7=2(4)-1$ effective flavours. In fact, it can be seen to be broken by applying the electric dualisation algorithm \cite{Giacomelli:2024laq}. The resulting partition function is given by a sum of four frames equivalent up to a global symmetry transformation and can be expressed in the general form in \eqref{eq:sknbrokengen} by identifying:
\begin{equation}\label{eq: ex3d31}
\left.
\begin{aligned}
    \PFS{\widetilde S} &=  \text{(Quiver in the bottom line of Figure \ref{fig:VEV_broken})} ~, \\[0.69mm]
    \dd{\rm{Nil}} &= 1 ~, \\
    \dd{\rm{Ex}} &= \delta\left(h'-\frac{i Q}{2}+m+y_1-\frac{1}{4}\sum_{j=1}^2(w_j+z_j)\right) ~, \\  
    \Delta^{\rm corr}_{\rm 3d} &= \left.\frac{\DCB{SU(3)}\left(\vec{y'};m\right)}{\DCB{SU(4)}(\vec{y};m)} \right\vert_{\dd{\text{Ex}}} ~, \\
    &= s_b\left(2m-\frac{iQ}{2}\right) \prod_{j=2}^{4}\left[ s_b\left( \frac{iQ}{2}\pm\left(m-\frac{i Q}{2}+h'+y_j-\frac{1}{4}\sum_{i=1}^2(w_i+z_i)\right) \right) \right. \\
    &\hspace{2cm} \times \; \left. s_b\left( 2m-\frac{iQ}{2}\pm\left(m-\frac{i Q}{2}+h'+y_j-\frac{1}{4}\sum_{i=1}^2(w_i+z_i)\right) \right)\right] ~, \\[1.1mm]
    {\rm GM}_{\rm 3d}^{[3,1]}(\vec x',h') &= s_b\left(2m-\frac{iQ}{2}\right)^2 s_b\left(\frac{iQ}{2}\right)^2 s_b\left(-2m+\frac{3iQ}{2}\right) s_b\left(4m-\frac{3iQ}{2}\right) \\
    &\quad \times \; s_b\left(\frac{iQ}{2}\pm\left(m-\frac{i Q}{2}+h'-x_4\right)\right) s_b\left(2m-\frac{iQ}{2}\pm\left(m-\frac{i Q}{2}+h'-x_4\right)\right) ~.
\end{aligned}
\right\vert\;\text{\small Frame 1}
\end{equation}
Here $\vec{y}\,{}'=\{y_2,y_3,y_4\}$ and the other three frames follow by exchanging $y_1$ with $y_2,y_3$, or $y_4$, and the full expression is their sum. Therefore this is an example of a broken theory since the frames are identical up to this permutation. Upon gauging the $SU(4)_{\vec y}$ symmetry, the delta function freezes one integration variable and we can identify all four frames since they are rotated by the Weyl group of the $SU(4)$ being gauged. The $SU(4)$ measure introduced for the gauging is appropriately converted to that of $SU(3)$ by $\Delta_{\rm 3d}^{\rm corr}$ and their multiplicity accounts for the conversion of the Weyl normalisation $1/4!$ to $1/3!$. This operation therefore realises the Higgsing $SU(4)\to SU(3)$ at the level of the sphere partition function. In this case, $\bm\delta_{\rm Nil}=1$ and no geometric progression constraint is imposed on the $\vec y$ parameters. Therefore $\rho'$ is trivial and no inverse Goldstone contribution appears in the expression for $\Delta_{\rm 3d}^{\rm corr}$ in \eqref{eq: ex3d31}.

\subsubsection{Bad regime}

\begin{figure}[h!]
    \centering
    \includegraphics[width=\linewidth]{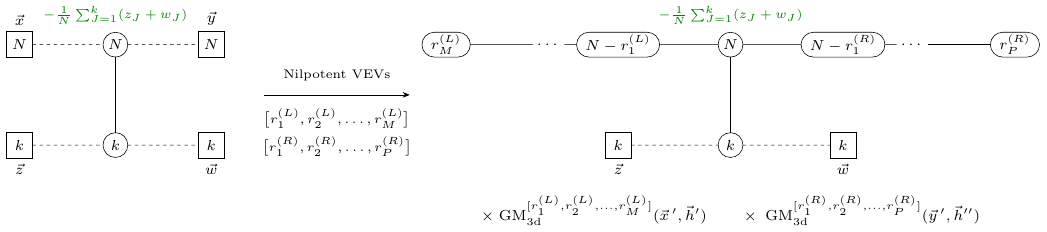}
    \caption{Simultaneous nilpotent deformations of the two $T[U(N)]$ tails, labelled by $\rho_L$ and $\rho_R$. This triggers a flow to the deformed quiver on the right along with the associated Goldstone multiplet contributions ${\rm GM}^{\rho_L}_{\rm 3d}$ and ${\rm GM}^{\rho_R}_{\rm 3d}$. FI parameters affected by the deformations are suppressed.}
    \label{fig:BadVEV_gen}
\end{figure}

We next turn on simultaneous nilpotent VEVs for both $T[U(N)]$ tails in $S_{k,N}$, given by the partitions
\begin{equation}
    \rho_L = [r_1^{(L)},r_2^{(L)},\dots,r_M^{(L)}] \quad,\quad \rho_R = [r_1^{(R)},r_2^{(R)},\dots, r_P^{(R)}] ~.
\end{equation}
The corresponding deformation is shown in Figure \ref{fig:BadVEV_gen}.

The central node now sees $N-r_1^{(L)}+N-r_1^{(R)}+k$ effective flavours and thus turns bad when $k-r_1^{(L)}-r_1^{(R)}+1<0$. For this class of simultaneous deformations, we conjecture that entering the bad regime produces a sum of inequivalent frames, rather than symmetry-related frames characteristic of a broken theory.\footnote{We have also tested this conjecture extensively in examples beyond those displayed here.}

\begin{figure}[b!]
    \centering
    \includegraphics[width=\linewidth]{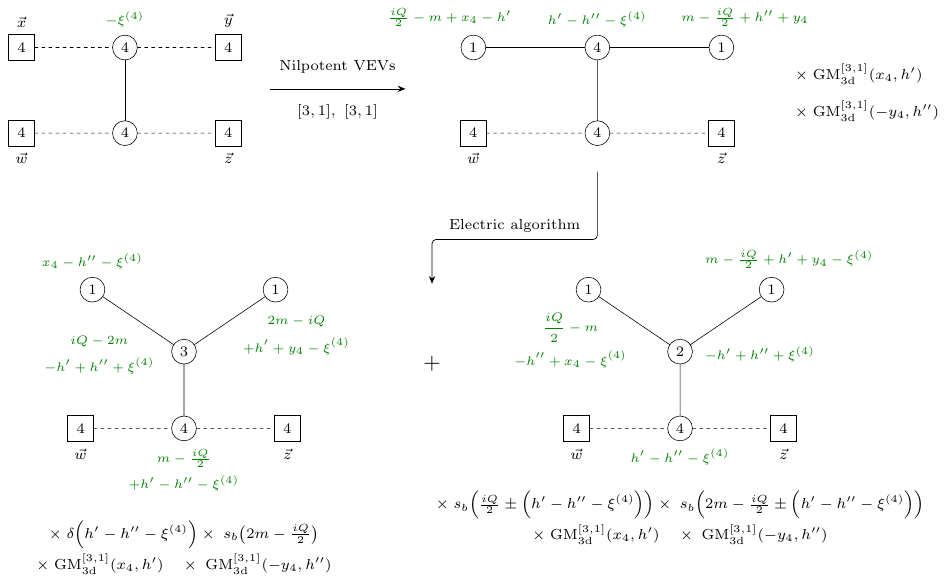}
    \caption{Simultaneous $[3,1]$ nilpotent deformations of $S_{4,4}$ produce a bad central $U(4)$ node. The electric dualisation algorithm gives a sum of two inequivalent frames, with chirals produced by the dualisation on top of the Goldstone modes associated with the nilpotent VEV. For brevity, we use the following definition $\xi^{(4)}=\frac14\sum_{i=1}^4 (w_i+z_i)$.}
    \label{fig:VEV_bad}
\end{figure}

For concreteness, consider the $S_{4,4}$ theory shown in the first line of Figure \ref{fig:VEV_bad}. We impose simultaneous $[3,1]$ nilpotent VEVs on both $SU(4)_{\vec x}$ and $SU(4)_{\vec y}$ via the following parameter tuning
\begin{eqnarray}
    x_1 = h'+\frac{iQ}{2}-m \;\;&,\quad x_2 = h'-\frac{iQ}{2}+m \;\;&,\quad x_3 = h'-\frac{3iQ}{2}+3m ~, \\
    y_1 = -h''-\frac{iQ}{2}+m \;\;&,\quad y_2 = -h''+\frac{iQ}{2}-m \;\;&,\quad y_3 = -h''+\frac{3iQ}{2}-3m ~.
\end{eqnarray}
The flow triggered by these deformations yields the theory on the top right of Figure \ref{fig:VEV_bad} along with the Goldstone modes associated with both nilpotent VEVs.

The resulting theory is bad since the central $U(4)$ node sees only $6<7=2(4)-1$ effective flavours. Moreover, unlike the example considered previously, this theory is not broken since the electric algorithm produces a sum of two inequivalent frames. We present the result in the bottom line of Figure \ref{fig:VEV_bad}. The two residual quivers have different ranks and are not related by a global symmetry transformation, demonstrating the characteristic decomposition of a bad theory into inequivalent frames. The frames also contain contributions from free fields originating from the dualisation, alongside the Goldstone modes associated with the initial nilpotent VEVs.

These examples motivate the following conjecture for nilpotent deformations of $S_{k,N}$:
\begin{itemize}
    \item A nilpotent VEV for the Coulomb-branch moment map associated with any one of the four full tails of $S_{k,N}$ produces a \emph{broken} theory whenever the corresponding central gauge node becomes bad.
    \item Simultaneous non-trivial nilpotent VEVs for both $T[U(N)]$ tails, or for both $T[U(k)]$ tails, produce a \emph{bad} theory with inequivalent frames whenever the corresponding central gauge node becomes bad.
\end{itemize}
The distinction is visible directly in the $S_b^3$ partition functions. In the first case, the partition function decomposes into a sum of equivalent frames rotated by the global symmetry, as expected for a broken theory. In the second, the different frames describe inequivalent residual theories, reproducing the characteristic structure of a bad theory. This conjecture is supported by the examples presented here and by further checks.

We leave for future work the analysis of cases in which nilpotent VEVs are turned on simultaneously for tails of both types, thereby deforming the $SU(N)$ and $SU(k)$ symmetries at the same time. Such deformations can render both central nodes bad and may lead to a richer distributional structure, with several independent delta function constraints and multiple residual frames.

In Sections \ref{sec:Higgsing} and \ref{sec:multiplenil_bad}, we study the five-dimensional counterparts of these phenomena by imposing excessive nilpotent VEVs on the superconformal indices of the corresponding 5d theories. Under the 5d/3d dictionary, the associated magnetic quivers become the bad or broken theories described in this section. We show that the five-dimensional indices mirror the structure of their associated three-dimensional magnetic quiver partition functions: they become distributions in the external flavour fugacities and decompose into sums of  frames, generically supported on delta-function constraints. Inequivalent frames reproduce the structure of a three-dimensional bad theory, whereas equivalent frames, related by global-symmetry transformations, reproduce that of a broken theory. 

\section{5d partition functions and 5d/3d parameter map to $S_{k,N}$}
\label{sec:review_5d}

In this paper, we focus on the family of five-dimensional $\cN=1$ $ N+SU(N)_0^{k-1}+N $ gauge theories. These theories are engineered in Type IIB string theory by box-shaped $(p,q)$ five-brane webs containing $N$ D5 branes and $k$ NS5 branes \cite{Aharony:1997bh,Aharony:1997ju}. Equivalently, they arise from M-theory compactifications on non-compact toric Calabi--Yau threefolds whose toric diagrams are dual to the corresponding five-brane webs \cite{Katz:1996fh,Leung:1997tw}. Owing to the characteristic shape of their brane-web realization, we refer to this family as the box theories. 

Our five-dimensional analysis is carried out directly at the level of the superconformal index, equivalently the $S^4\times S^1$ partition function \cite{Kim:2012gu,Terashima:2012ra}. This observable probes the five-dimensional ultraviolet fixed point for which the box gauge theory provides an infrared Lagrangian description. To compute the superconformal indices of box theories, we use the associated refined topological string partition functions as building blocks \cite{Iqbal:2012xm}. These in turn can be evaluated using the methods of the refined topological vertex \cite{Aganagic:2003db,Iqbal:2007ii}.

For these geometries, the calculation simplifies upon decomposing the box five-brane web into a series of vertical strips glued along their shared horizontal legs. The strip provides a particularly useful building block since its partition function admits a closed-form expression \cite{Iqbal:2004ne,Taki:2007dh}. Physically, gluing these strips then yields the refined topological string partition function of the box theory as a sum over gauge-instanton sectors of the corresponding infrared description.

This section reviews the partition functions and the 5d/3d parameter map needed to study the Higgsings that render these 5d indices distributional. We begin by reviewing the strip and box refined topological string partition functions, and then explain the construction of their 5d superconformal indices, removing the decoupled contributions associated with parallel external branes. Finally, we review the 5d/3d dictionary from \cite{one} relating the five-dimensional parameters to the FI parameters of the associated three-dimensional strong-coupling magnetic quivers $S_{k,N}$. 

\subsection{The strip}
\label{subsec:strip}

\begin{figure}[h!]
    \centering
    \includegraphics[width=0.9\linewidth]{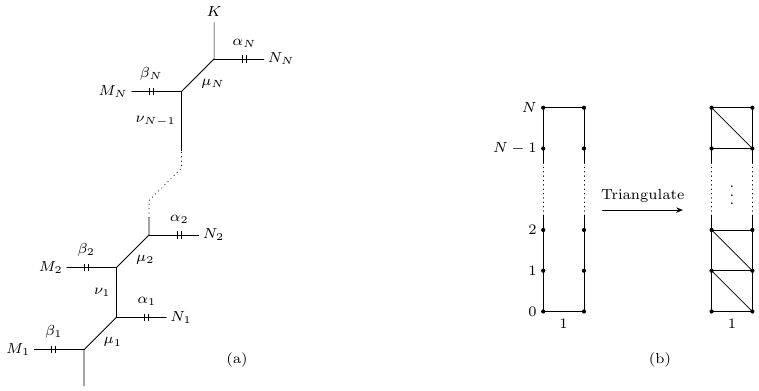}
    \caption{(a) Symmetrical strip of length $N$ that engineers free bifundamental hypermultiplets featuring manifest $SU(N)_{\vec M}\times SU(N)_{\vec N}\times U(1)_K$ global flavour symmetry. These double ticks upon the D5 branes indicate the horizontal preferred direction leading to the partitions $\beta_a,\alpha_a$ turned on along the external D5 branes. The finite internal \Kah parameters are represented by $\mu_a,\nu_a$. (b) The dual toric diagram and the triangulation associated with the resolved five-brane web of panel (a).}
    \label{fig:strip}
\end{figure}

The {\it strip} geometry engineers free bifundamental hypermultiplets and acts as the primary building block of the $(p,q)$ five-brane webs examined here. A general strip can have different number of D5 branes extending to the left and right. We focus on the symmetrical strip featuring $N$ D5 branes stretching on both sides of an NS5 brane, as shown in Figure \ref{fig:strip}(a). The dual toric diagram is simply a rectangle with height $N$ and unit length, whose tessellation is displayed in Figure \ref{fig:strip}(b).

We parameterise the two $SU(N)$ flavour symmetries by fugacities $\vec M, \vec N$, subject to $\prod_{a=1}^N M_a = \prod_{a=1}^N N_a = 1$, together with the fugacity $K$ for the $U(1)_K$ flavour symmetry. They are connected to the internal \Kah parameters by
\begin{equation}\label{eq:strip_par_std}
    \mu_a = \frac{M_aN_a}{K}\sqrt{\frac{\mf q}{\mf t}} \;,\;\; \nu_b = \frac{K}{M_{b+1}N_b}\sqrt{\frac{\mf t}{\mf q}} \qquad
    \begin{gathered}
        ,\;a = 1,\dots,N ~, \\
        \quad\;\;,\; b = 1,\dots,N-1 ~,
    \end{gathered}
\end{equation}
where $\mf q,\mf t$ are exponentiated $\Omega$-deformation parameters defined in \eqref{eq:Omegadef_Region}. These internal \Kahs satisfy $\mu_a\nu_a = \frac{M_{a}}{M_{a+1}}$ and $\nu_a\mu_{a+1} = \frac{N_{a+1}}{N_{a}}$. Thus these products are fixed by ratios of the non-abelian flavour fugacities, while $K$ can be viewed as an intrinsically internal parameter associated with the strip. This parametrisation is naturally adapted to the interpretation of the strip as a theory of free 5d bifundamental hypermultiplets with manifest $SU(N)_{\vec M}\times SU(N)_{\vec N}\times U(1)_K$ global symmetry.

In this paper, we choose the preferred direction for the refined topological vertex to be horizontal. We assign external partitions $\bm\beta=\{\beta_1,\dots,\beta_N\}$ and $\bm\alpha=\{\alpha_1,\dots,\alpha_N\}$ to the left and right external D5 legs, respectively. Notably, the sums over internal partitions in the refined topological vertex computation can be computed explicitly, obtaining a closed form expression of the strip partition function \cite{Iqbal:2004ne,Taki:2007dh}. We separate it into a ``perturbative" and an ``instantonic" part
\begin{equation}
    Z^{\rm strip} (\vec M,\vec N,K;\bm{\beta},\bm{\alpha}) = \cZ_{\rm pert}^{\rm strip} (\vec M,\vec N,K) \; \cZ^{\rm strip}_{\bm{\beta},\bm{\alpha}} (\vec M,\vec N,K) ~.
\end{equation}
The ``perturbative" factor takes the form
\begin{equation}\label{eq:Zstrip_pert}
    \cZ_{\rm pert}^{\rm strip} (\vec M,\vec N,K) = \frac{\prod_{1\leq a<b\leq N} \cM\left( \frac{M_a}{M_b} \right) \cM\left( \frac{N_b}{N_a}\frac{\mf t}{\mf q} \right)}{\prod_{1\leq a<b\leq N }\cM\left( \frac{K}{M_bN_a}\frac{\mf t}{\mf q} \right) \prod_{1\leq a\leq b\leq N }\cM\left( \frac{M_aN_b}{K} \right)} ~,
\end{equation}
while the ``instantonic" part is
\begin{equation}\label{eq:Zstrip_inst}
    \begin{split}
        \cZ^{\rm strip}_{\bm{\beta},\bm{\alpha}} (\vec M,\vec N,K) &= \prod_{a=1}^N \Big[ \mf q^{\frac{||\alpha_a||^2}{2}} \mf t^{\frac{||\,{}^t\beta_a||^2}{2}} \widetilde Z_{\alpha_a}(\mf t,\mf q) \widetilde Z_{\,{}^t\beta_a}(\mf q,\mf t) \Big] \\
        &\quad \times \; \frac{\prod_{1\leq a<b\leq N }\cN_{\alpha_a\beta_b}\left( \frac{K}{M_bN_a}\frac{\mf t}{\mf q} \right) \prod_{1\leq a\leq b\leq N }\cN_{\beta_a\alpha_b}\left( \frac{M_aN_b}{K} \right)}{\prod_{1\leq a<b\leq N} \cN_{\alpha_a\alpha_b}\left( \frac{N_b}{N_a}\frac{\mf t}{\mf q} \right) \cN_{\beta_a\beta_b}\left( \frac{M_a}{M_b} \right)} ~.
    \end{split}
\end{equation}
The special functions in these expressions are defined in Appendix \ref{app:special_func}. Their dependence on $\mf t$, $\mf q$ is suppressed whenever unambiguous. We use the terms ``perturbative" and ``instantonic" provisionally: the isolated strip characterizes a free theory and features no gauge-instanton expansion. The terminology acquires its usual meaning when strips are glued into a configuration engineering a gauge theory, as we discuss next.

\subsection{The $N+SU(N)_0^{k-1}+N$ box}
\label{subsec:box}

\begin{figure}[h!]
  \centering
  \includegraphics[width=\textwidth]{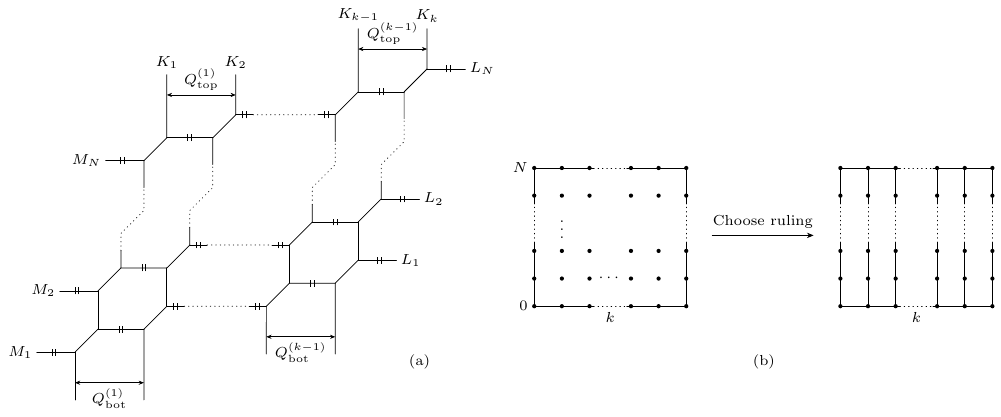}
  \caption{(a) A rectangular five-brane web that engineers the $N+SU(N)_0^{k-1}+N$ box theory. These double ticks denote the horizontal preferred direction, and $K_1,\dots,K_k$ specify the parameters for the component strips. (b) Shows the dual rectangular toric diagram alongside the vertical ruling determining the strip decomposition.}
  \label{fig:braneweb}
\end{figure}

Connecting $k$ symmetrical strips of length $N$ produces the rectangular box five-brane web shown in Figure \ref{fig:braneweb}(a). This brane web engineers the linear quiver gauge theory $N+SU(N)_0^{k-1}+N$, featuring a bifundamental hypermultiplet between every pair of adjacent gauge nodes alongside $N$ fundamental hypermultiplets at both ends. The Chern--Simons level for every gauge group vanishes in this configuration. Its dual toric diagram is a rectangle of height $N$ and length $k$, and the vertical ruling shown in Figure \ref{fig:braneweb}(b) decomposes it into its $k$ constituent strips.

We obtain the corresponding refined topological string partition function by gluing its constituent strips, following the prescription reviewed in \cite[Appendix B]{one}; here we only present the details relevant to the discussion in this paper. Each connected horizontal leg holds a partition $\alpha_a^{(m)}$ alongside a base \Kah parameter $Q_{m,a}$. The gluing then requires a sum over these partitions weighted by $Q_{m,a}^{|\alpha_a^{(m)}|}$. The resulting expression is
\begin{equation}\label{eq:Ztop_SUNkm12N}
    \begin{split}
         &Z^{N+SU(N)_0^{k-1}+N}(\vec Q_{\rm inst},\vec M,\vec L,K_1,\dots,K_k; \bm\beta, \bm\gamma) \\
         &\hspace{2cm} = \sum_{\bm{\alpha}^{(1)}}\sum_{\bm{\alpha}^{(2)}}\cdots\sum_{\bm{\alpha}^{(k-1)}} \left[ \prod_{m=1}^{k-1}\prod_{a=1}^N Q_{m,a}^{|\alpha^{(m)}_{a}|} \right] \; Z^{\rm strip} (\vec M,\vec N^{(1)},K_1; \bm{\beta},\bm{\alpha}^{(1)}) \\
         &\hspace{4cm} \times \; \left[ \prod_{m=1}^{k-1} Z^{\rm strip} (\vec N^{(m)},\vec N^{(m+1)},K_{m+1}; \bm{\alpha}^{(m)},\bm{\alpha}^{(m+1)}) \right]' ~,
    \end{split}
\end{equation}
where we define $\bm\alpha^{(k)} \equiv \bm\gamma$ and $\vec N^{(k)} \equiv \vec L$. The prime denotes alternating parametrisations of the strips. The first strip uses the standard parametrisation introduced above in \eqref{eq:strip_par_std}, while every even-numbered strip uses an alternative one, defined by
\begin{equation}\label{eq:alt_strip}
    \begin{split}
     \left[ Z^{\rm strip}(\vec N, \vec L, K_2; \bm\alpha,\bm\gamma) \right]' = Z^{\rm strip}(\vec N^{-1}, \vec L^{-1}, \frac{\mf q}{\mf t} K_2^{-1}; \bm\alpha,\bm\gamma) ~.
    \end{split}
\end{equation}
This alteration ensures that the parallel leg contributions at each interface combine
into the appropriate $SU(N)$ gauge measure; for more details see \cite[Appendix B]{one}.

The sum over partitions in \eqref{eq:Ztop_SUNkm12N} reproduces the instanton expansion of the $N+SU(N)_0^{k-1}+N$ gauge theory. The relation to the conventional Nekrasov partition function, including the required normalisations and instanton fugacity conventions, is reviewed in \cite[Appendix B]{one}. Here we only need to note that the base parameters in each column are proportional to the corresponding instanton fugacity $$Q_{m,a}\propto Q_{\rm inst}^{(m)} ~,$$ with the proportionality factors determined by the remaining strip parameters and $\mf t,\mf q$. We therefore express the box partition function in terms of the $k-1$ independent instanton fugacities $\vec Q_{\rm inst}$, rather than all the base parameters. Geometrically, the instanton fugacities are given by \cite{Mitev:2014jza}
\begin{equation}
    Q_{\rm inst}^{(i)} = \sqrt{Q_{\rm bot}^{(i)} Q_{\rm top}^{(i)}} \quad,\;\; i=1,\dots,k-1 ~, 
\end{equation}
where $Q_{\rm bot}^{(i)}$ and $Q_{\rm top}^{(i)}$ denote the exponentiated separations between adjacent NS5 branes at the bottom and top of the web, respectively; see Figure \ref{fig:braneweb}.

The fiber and base curves associated with this infrared gauge theory description are fixed by the vertical ruling. For all strips, the preferred direction of the refined topological vertex is chosen to be horizontal. The sums over internal partitions inside each strip have already been carried out, while the remaining sums over the linked horizontal legs yield the instanton expansion for the box theory.

This expression is therefore naturally structured around the weakly coupled description of the gauge theory. An all-orders resummation of the remaining partition sums could provide a representation better suited to studying the strong-coupling regime; we do not undertake such a resummation here. However, information regarding the enhanced ultraviolet flavour symmetry may also be obtained by organising the expansion of the superconformal index into symmetry characters order by order \cite{Kim:2012gu, Bergman:2013aca}.

\subsection{Decoupled contributions and the superconformal index}
\label{subsec:decoupled_index}

The $S^4\times S^1$ partition function, or five-dimensional superconformal index, of $\cN=1$ SCFTs admitting an infrared Lagrangian description can be computed by supersymmetric localisation \cite{Kim:2012gu,Terashima:2012ra}. The result is a matrix integral whose integrand contains perturbative contributions and two copies of the Nekrasov instanton partition function on $\mathbb R^4_{\epsilon_1\epsilon_2}\times S^1$, associated with the two fixed points of $S^4$ \cite{Nekrasov:2002qd,Nekrasov:2003rj}.

In terms of the topological string partition function, the index can be written schematically as \cite{Iqbal:2012xm}:
\begin{equation}\label{eq:index_gen}
    \cI \equiv \cZ^{S^4\times S^1} = \oint \prod_i \left[ \frac{{\rm d}A_i}{2\pi iA_i} \left| M(\mf t,\mf q) \right|^2 \right] \frac{\left| \cZ_{\rm top} \right|^2}{\left| \cZ_{||} \right|^2} ~.
\end{equation}
Here $A_i$ are the exponentiated Coulomb branch parameters of the infrared gauge theory, and $M(\mf t,\mf q)$ is the Cartan factor defined in Appendix \ref{app:special_func}. The factor $\cZ_{||}$ collects contributions associated with parallel external legs of the complete five-brane web. These describe decoupled sectors and must be removed from the physical superconformal index. Finally, the notation $|\cdots|^2$ denotes the pairing of the contributions from the two fixed points and is defined in \eqref{eq:mod_square}.

For an isolated five-brane web, the external partitions are set to be empty. When using the web as a building block for a larger configuration, however, we keep the  partitions on the external D5 legs arbitrary. It is therefore useful to keep the external-partition dependence explicit in the D5 parallel leg factors:
\begin{equation}\label{eq:parallel_D5}
    \begin{aligned}
        \cZ_{||}^{\rm left} (\vec M;\bm\beta) &= \prod_{1\leq a<b\leq N} \frac{ \MM{\frac{M_a}{M_b}} }{ \cN_{\beta_a\beta_b}\left(\frac{M_a}{M_b}\right) } ~, \\
        \cZ_{||}^{\rm alt \; left} (\vec M;\bm\beta) &= \prod_{1\leq a<b\leq N} \frac{ \MM{\frac{M_b}{M_a}} }{ \cN_{\beta_a\beta_b}\left(\frac{M_b}{M_a}\right) } ~,
    \end{aligned}
    \qquad
    \begin{aligned}
        \cZ_{||}^{\rm right} (\vec L;\bm\gamma) &= \prod_{1\leq a<b\leq N} \frac{ \MM{\frac{\mf t}{\mf q}\frac{L_b}{L_a}} }{ \cN_{\gamma_a\gamma_b}\left(\frac{\mf t}{\mf q}\frac{L_b}{L_a}\right) } ~, \\
        \cZ_{||}^{\rm alt \; right} (\vec L;\bm\gamma) &= \prod_{1\leq a<b\leq N} \frac{ \MM{\frac{\mf t}{\mf q}\frac{L_a}{L_b}} }{ \cN_{\gamma_a\gamma_b}\left(\frac{\mf t}{\mf q}\frac{L_a}{L_b}\right) } ~.
    \end{aligned}
\end{equation}
The contributions $\cZ_{||}^{\rm left}$ and $\cZ_{||}^{\rm right}$ correspond to the standard strip parametrisation, while $\cZ_{||}^{\rm alt \; right}$ and $\cZ_{||}^{\rm alt \; left} $ correspond to parallel leg contributions for a strip in the alternate parametrisation. The box also contains contributions from parallel external NS5 branes which we split into those coming from the top and the bottom of the box:
\begin{equation}\label{eq:parallel_NS5}
    \cZ_{||}^{\rm bot}( \vec Q_{\rm bot} ) = \prod_{1\leq i<j\leq k} \MM{\frac{\mf t}{\mf q} \prod_{l=i}^{j-1} Q_{\rm bot}^{(l)}} \;\;,\quad \cZ_{||}^{\rm top}( \vec Q_{\rm top} ) = \prod_{1\leq i<j\leq k} \MM{\prod_{l=i}^{j-1} Q_{\rm top}^{(l)}} ~,
\end{equation}

For the isolated strip, there are no Coulomb moduli to integrate over hence the 5d superconformal index can simply be expressed as
\begin{equation}
     \cI^{\rm strip} (\vec M,\vec N,K) = \left| \frac{ Z^{\rm strip}(\vec M,\vec N,K;\bm\emptyset,\bm\emptyset) }{ \cZ_{||}^{\rm left} (\vec M;\bm\emptyset) \; \cZ_{||}^{\rm right} (\vec N;\bm\emptyset) } \right|^2 = \left| \prod_{a,b=1}^N \cM\left( \frac{M_aN_b}{K} \right) \right|^{-2} ~,
\end{equation}
by setting all external partitions to empty. This is the index of $N^2$ free hypermultiplets organised into an $SU(N)\times SU(N)$ bifundamental, with manifest flavour symmetry $SU(N)\times SU(N)\times U(1)$. The parameter $K$ in \eqref{eq:strip_par_std} can now be identified as a fugacity for this $U(1)$ symmetry.\footnote{The full flavour symmetry group for $N^2$ free hypermultiplets is $USp(2N^2)$. This is shown by redefining $X_{ab}=\frac{M_a N_b}{K}$ which gives the $N^2$ Cartan parameters for $USp(2N^2)$.}

After gauging the $SU(N)$ flavour symmetry with parameters $\vec N$, the variables $\vec M$ and $K$ combine into $U(N)$ flavour fugacities $\widetilde{M}_a=M_a/K$ satisfying $\prod_{a=1}^N\widetilde{M}_a=K^{-N}$. The hypermultiplet factors then depend on $\widetilde M_aN_b$, as appropriate for $N$ fundamental hypermultiplets of the $SU(N)$ gauge group.

For the box, the parallel factors associated with the internal legs are retained, and in fact reconstruct the $SU(N)$ measure under the integral in \eqref{eq:index_gen}. The complete decoupled contribution for the isolated box is therefore
\begin{equation}\label{eq:box_dec}
    \cZ_{||}^{k,N} (\vec M, \vec L, \vec Q_{\rm bot}, \vec Q_{\rm top}) = \cZ_{||}^{\rm top}( \vec Q_{\rm top} ) \; \cZ_{||}^{\rm bot}( \vec Q_{\rm bot} ) \times
    \begin{cases}
        \cZ_{||}^{\rm left}(\vec M;\bm\emptyset) \; \cZ_{||}^{\rm right}(\vec L;\bm\emptyset) \quad\,, \, k \text{ odd}, \\
        \cZ_{||}^{\rm left}(\vec M;\bm\emptyset) \; \cZ_{||}^{\rm alt \; right}(\vec L;\bm\emptyset) , \, k \text{ even}.
    \end{cases}
\end{equation}
Combining the strip contributions and integrating over the Coulomb moduli gives the 5d superconformal index
\begin{equation}\label{eq:Index_SUNkm12N}
    \begin{split}
        &\cI^{N+SU(N)_0^{k-1}+N} (\vec Q_{\rm inst},\vec M,\vec L, K_1, \dots, K_k) = \oint \prod_{m=1}^{k-1} \left[ \frac{1}{N!} \prod_{i=1}^N\frac{{\rm d}N_i^{(m)}}{2\pi iN_i^{(m)}} \times \delta\left( \prod_{i=1}^N N_i^{(m)}\right) \times \right] \\
        &\quad \times \left| M(\mf t,\mf q) \right|^{2(k-1)(N-1)} \times \left| \frac{\cZ_{\rm pert}^{\rm strip} (\vec M, \vec N^{(1)}, K_1) \left[ \prod_{m=1}^{k-1} \cZ^{\rm strip}_{\rm pert} (\vec N^{(m)}, \vec N^{(m+1)}, K_{m+1}) \right]' }{ \cZ_{||}^{k,N} (\vec M, \vec L, \vec Q_{\rm bot}, \vec Q_{\rm top}) } \right|^2  \\
        &\quad \times \left| \sum_{\bm{\alpha}^{(1)}}\cdots\sum_{\bm{\alpha}^{(k-1)}} \; \prod_{m=1}^{k-1}\prod_{a=1}^N Q_{m,a}^{|\alpha^{(m)}_{a}|} \cZ^{\rm strip}_{\bm\emptyset,\bm\alpha^{(1)}} (\vec M, \vec N^{(1)}, K_1) \left[ \prod_{m=1}^{k-1} \cZ^{\rm strip}_{\bm\alpha^{(m)},\bm\alpha^{(m+1)}} (\vec N^{(m)}, \vec N^{(m+1)}, K_{m+1}) \right]' \right|^2 ~,
    \end{split}
\end{equation}
where $\vec N^{(k)}=\vec L$, $\bm\alpha^{(k)} \equiv \bm\gamma = \bm\emptyset$, and $\bm\beta=\bm\emptyset$. The primes once again denote alternating strip parametrisations. Each factor $1/N!$ implements the Weyl quotient while the corresponding Dirac delta functions impose the $SU(N)$ constraint.\footnote{We use $\delta(U)$ to denote the periodic delta function appropriate to a multiplicative fugacity $U$, normalised by
\begin{equation}
    \oint_{|L|=1} \frac{{\rm d}L}{2\pi iL} \;\delta\left(\frac{L}{L_*}\right)f(L) = f(L_*) ~. \nn
\end{equation}}
For $k=1$, the gauge integral and parallel NS5 contributions are absent, and the expression reduces to the strip index.

\subsection{The 5d/3d dictionary}
\label{subsec:dict}

We now review the dictionary relating the 5d mass and instanton parameters of the box theory to the FI parameters of its strong-coupling 3d magnetic quiver $S_{k,N}$, as proposed in \cite{one}. We introduce additive parameters for the mass and strip parameters: the two sets of $SU(N)$ masses take the form
\begin{equation}
    M_i = e^{-\beta m_i} \;,\;\; L_i = e^{-\beta l_i} \;\;,\;\;\; i=1,\dots,N ~,
\end{equation}
with $\sum_im_i=\sum_il_i=0$, while the instanton and $U(1)$ bifundamental masses are
\begin{equation}
    Q_j = e^{-\beta v_j} \;,\;\; j=1,\dots,k-1 \;\;,\quad \sqrt{\frac{\mf t}{\mf q}}K_j = \sqrt{\frac{\mf t}{\mf q}} e^{-\beta k_j} = e^{-\beta h_j} \;,\;\; j=1,\dots,k ~.
\end{equation}
For the dictionary below, we identify
\begin{equation}
    \sqrt{\frac{\mf t}{\mf q}} = e^{-\beta\left(\frac{iQ}{4}\right)} \;\Longrightarrow\;\sqrt{\frac{\mf t}{\mf q}}K_j = e^{-\beta\left(\frac{iQ}{4}+k_j\right)} = e^{-\beta h_j} ~.
\end{equation}

\begin{figure}[h!]
  \centering
  \includegraphics[width=\textwidth]{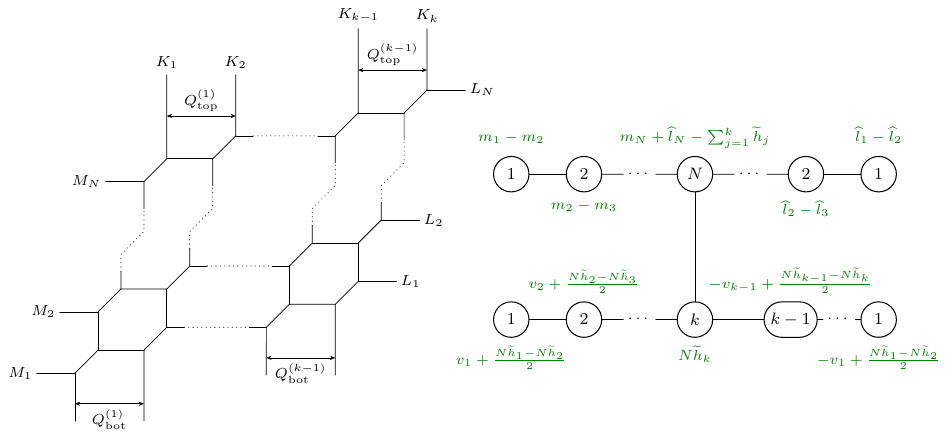}
  \caption{Five-brane web  for the $N+SU(N)_0^{k-1}+N$ box theory and its  associated strong-coupling magnetic quiver $S_{k,N}$. Parameters are connected via the 5d/3d dictionary: specifically, five-dimensional flavour masses correspond to the FI parameters for the two $T[U(N)]$ tails, whereas instanton parameters and  abelian strip masses correspond to the FI parameters associated with the remaining nodes. Instanton parameters  can be expressed as $Q^{(i)}_{\rm inst}=\sqrt{Q_{\rm bot}^{(i)}Q_{\rm top}^{(i)}}$.}
  \label{fig:Dictionary}
\end{figure}

The map between the 5d and 3d parameters is then given by\footnote{This map is defined at the conformal point where the three-dimensional $\cN=4$ theory is endowed with its superconformal $R$ charge assignment. In the three-dimensional partition function computations we work in the $\mathcal N=2^*$  setup. The $\mathcal N=2^*$  refinement of nilpotent specialisations used in this paper is explained in Subsection \ref{subsubsec:compare_good}.}
\begin{equation}\label{eq:dict}
    \boxed{
    \begin{gathered}
        \vec x = \vec m \;\;,\;\;\; \vec y = \vec{\widehat l} ~, \\
        \vec w = \left( \frac{N\widetilde h_1}{2}+\sum_{i=1}^{k-1}v_i \,,\; \frac{N\widetilde h_2}{2}+\sum_{i=2}^{k-1}v_i \,,\; \dots \;, \frac{N\widetilde h_{k-1}}{2}+v_{k-1} \,,\; \frac{N\widetilde h_k}{2} \right) ~, \\
        \vec z = \left( \frac{N\widetilde h_1}{2}-\sum_{i=1}^{k-1}v_i \,,\; \frac{N\widetilde h_2}{2}-\sum_{i=2}^{k-1}v_i \,,\; \dots \;, \frac{N\widetilde h_{k-1}}{2}-v_{k-1} \,,\; \frac{N\widetilde h_k}{2} \right).
    \end{gathered}}
\end{equation}
where we have defined
\begin{equation}\label{eq:hl_redef}
    \widetilde h_i \coloneq
    \begin{cases}
        -h_i \;&,\; i \text{ even} \\
        h_i \;&,\; i \text{ odd}
    \end{cases} \quad,\quad \widehat l_i = (-1)^{k+1} l_i ~.
\end{equation}
The five-brane web and the magnetic quiver, with their parameters related by this dictionary, are shown in Figure \ref{fig:Dictionary}. The signs in this dictionary reflect the alternate strip parametrisation used in the topological string partition function of $N+SU(N)_0^{k-1}+N$ above. For a strip at the $i$th position, the substitutions for even $i$ defined in \eqref{eq:alt_strip} yield:
\begin{equation}
    \sqrt{\frac{\mf t}{\mf q}} K_i \longrightarrow \sqrt{\frac{\mf q}{\mf t}} K_i^{-1} \quad,\quad\; L_i \longrightarrow L_i^{-1} ~,
\end{equation}
implying the definitions in \eqref{eq:hl_redef}.

\section{One-sided nilpotent Higgsings: good, ugly, and broken theories}
\label{sec:Higgsing}

In this and the following section, we study nilpotent VEVs for the moment-map operators associated with the $SU(N)$ flavour symmetries of the  $N+SU(N)_0^{k-1}+N$ box theories, constructed from $k$ strips as in Section \ref{subsec:box}.

We will show that sufficiently large, or \emph{excessive}, nilpotent deformations can alter the analytic structure of the five-dimensional superconformal index, causing it to become distributional.

Two distinct behaviours can arise. When the index decomposes into a sum of equivalent, symmetry-related distributional frames, we refer to the resulting five-dimensional theory as \emph{broken}; this is the regime studied in the present section. More general deformations may instead produce a distributional sum of inequivalent frames. In analogy with the corresponding three-dimensional magnetic quivers, we refer to these theories as \emph{bad}; they will be analysed in the following section. This distinction mirrors the behaviour of  the associated magnetic quiver $S_{k,N}$ becoming  broken and bad as reviewed in Section \ref{sec:review_3d}.

 A nilpotent VEV imposed on one of the two $SU(N)$ moment maps of the box theory  triggers a sequence of Higgsings that may propagate from the corresponding edge into the interior of the web. If the box is sufficiently wide, the sequence terminates in the interior and produces the index of an ordinary Higgsed theory, possibly accompanied by an additional free sector. If, instead, the non-trivial Higgsing constraints reach the opposite edge, the result becomes a distributional sum of equivalent frames, characteristic of a broken theory. Whether this transition occurs depends on both the partition $\rho$ labelling the nilpotent VEV and the width $k$ of the box. 

As we will show in the following sections, a nilpotent VEV for an
$SU(N)$ moment map of the box theory, labelled by a partition of $N$,
\begin{equation}
    \rho=[r_1,r_2,\ldots,r_s],
    \qquad r_1\geq r_2\geq\cdots\geq r_s,
\end{equation}
renders the theory broken when $r_1>k$.

Remarkably, this is precisely the threshold at which the associated three-dimensional magnetic quiver $S_{k,N}$ becomes broken, as discussed in Section~\ref{subsec:MQbreakbad}.

We illustrate the mechanism through a sequence of increasingly non-trivial examples. We begin with a good Higgsing, for which the ordinary residue prescription produces the index of a conventional infrared theory and the associated magnetic quiver remains good. We then consider an ugly example in which the Higgsing generates an additional free sector, again in agreement with the three-dimensional description. We subsequently turn to genuinely broken configurations: the $[2,1^{N-2}]$ deformation of the strip provides the simplest non-trivial distributional example. We then analyse the $[N-1,1]$ VEV followed by general nilpotent deformations of the strip, emphasizing the relation between sequential Higgsing and Hanany--Witten moves. We conclude with excessive Higgsing of the box, extending the analysis to interacting five-dimensional theories by studying the VEV propagation through the interior of the box.

In the explicit examples, we perform the five- and three-dimensional analyses independently and compare their results using the 5d/3d parameter map. These examples will also reveal a simple five-brane interpretation of the distributional answer. We postpone its detailed discussion until after the first broken configurations have been analysed, when the origin and role of the different delta-function constraints will be clear.

The technical origin of the distributional behaviour is the pinching of integration contours in the superconformal index. As the mass parameters are tuned to implement a VEV, poles move in the complex plane and the integration contour may become trapped between colliding sets of poles. The corresponding integrals are then evaluated by taking residues at the pinching poles, yielding the index of the infrared theory reached along the Higgs-branch flow, together with the expected contributions of the decoupled Goldstone multiplets \cite{Gaiotto:2012xa}.

As we will see, the contour-pinching conditions can be expressed as Dirac delta functions constraining gauge variables. These constraints are imposed by the gauge integrations and propagate from one strip to the next until the Higgsing cascade terminates. When $r_1>k$, some constraints reach the opposite boundary of the box and act on external mass parameters instead. The resulting index is therefore distributional in those parameters.
To make this distributional action explicit, we glue the broken theory to a symmetric length $N$ strip, which acts as a test theory.

Following the conventions of \eqref{eq:Ztop_SUNkm12N}, we impose the nilpotent VEV $\rho$ on the $SU(N)_{\vec M}$ moment map of the box, while $SU(N)_{\vec L}$ is the symmetry gauged upon attaching the test strip. For the broken theories obtained by imposing a nilpotent VEV on one edge of the box theory, the ``modulus square" of the refined topological string partition function takes the following general schematic form:

\begin{equation}\label{eq:broken_formula}
    \left| \frac{ Z^{\rm box} }{ \cZ_{||}^{\rm left} } \right|^2_{\rho} = {\rm GM}_{\rm 5d}^\rho \times  \sum_{\cW_{SU(N)}/\cW_{H\subset SU(N)}} \bm\delta_{\rm pert} \; \left| \bm\delta_{\rm inst} \; \frac{ Z^{\rm residual} }{ \cZ_{||}^{\rm residual} } \right|^2 \;\; \Delta^{\rm corr}_{\rm 5d} ~.
\end{equation}
Let us now break down this expression and precisely define each contribution. On the LHS of \eqref{eq:broken_formula}, $Z^{\rm box}|_{\rho}$ denotes the refined topological string partition function of the box with the nilpotent VEV $\rho$ imposed on the $SU(N)_{\vec M}$ moment map.\footnote{Note that by default, this edge is perpendicular to the chosen preferred direction when gluing the strips. As such, for the $N+SU(N)_0^{k-1}+N$ box theory, this corresponds to a nilpotent VEV for the $SU(N)$ moment map.} We remove the decoupled parallel-leg contribution associated with the Higgsed $\vec M$ edge while we  retain the contribution on the $\vec L$ side, along which the broken configuration is to be glued to the test theory.\footnote{Accordingly, we set the external partitions on the $\vec M$ side to be empty $\bm\beta=\bm\emptyset$ and retain arbitrary external partitions on the $\vec L$ side until gluing.} Crucially, this factor participates in the reconstruction of the residual gauge measure after gluing and is contained in $\Delta^{\rm corr}_{\rm 5d}$.

The original nilpotent VEV $\rho$ produces Goldstone multiplets whose contributions appear on the RHS of \eqref{eq:broken_formula} as ${\rm GM}_{\rm 5d}^\rho$. We explain how to compute these in Appendix \ref{app:GM}. The factors $\bm\delta_{\rm pert}$ and $\bm\delta_{\rm inst}$ denote products of Dirac and discrete delta functions acting on $\vec L$ and the associated external partitions, respectively. Upon gluing, their combined action partially Higgses the $SU(N)_{\vec L}$ gauge group to the subgroup $H_{\vec L'}\subset SU(N)_{\vec L}$. Here $\vec L'$ denote the surviving Cartan variables that remain unfrozen under the delta action and thus parameterise the residual gauge group $H_{\vec L'}$. We split the perturbative deltas into two sets:
\begin{equation}
    \bm\delta_{\rm pert} = \bm\delta_{\rm pert}^{\rm Ex} \times \bm\delta_{\rm pert}^{\rm Nil} ~.
\end{equation}
The set $\bm\delta_{\rm pert}^{\rm Nil}$ denotes the Dirac deltas that impose the geometric progression constraints associated with an induced nilpotent VEV $\rho'$ on $SU(N)_{\vec L}$. The remaining Dirac delta functions are collected in $\bm\delta_{\rm pert}^{\rm Ex}$.

While the delta functions freeze/trivialise the parameters/partitions, the factor $\Delta^{\rm corr}_{\rm 5d}$, evaluated on the support of these constraints, converts the original $SU(N)_{\vec L}$ gluing measure into the measure appropriate to $H_{\vec L'}$. In the presence of an induced nilpotent VEV $\rho'$ on $SU(N)_{\vec L}$, the factor $\Delta^{\rm corr}_{\rm 5d}$ also contains $1/{\rm GM}_{\rm 5d}^{\rho'}$ that cancels the contributions of the Goldstone multiplets generated by this induced VEV.

The factor $Z^{\rm residual}$ is the refined topological string partition function of the residual theory, while $\cZ_{||}^{\rm residual}$ removes the decoupled contributions from the parallel external legs in the five-brane web  associated with the residual theory. 

The sum on the RHS of \eqref{eq:broken_formula} runs over the Weyl group of $SU(N)$ modulo that of the residual subgroup $H\subset SU(N)$, generating equivalent frames related by Weyl transformations. The number of frames is  $|\mathcal W_{SU(N)}|/|\mathcal W_{H}|$; upon gauging, this multiplicity also converts the original Weyl normalisation into that appropriate to $H$. This precisely mirrors the defining property of a broken three-dimensional magnetic quiver \eqref{eq:sknbrokengen}, whose partition function decomposes into symmetry-related frames.

The similarity between the five-dimensional \eqref{eq:broken_formula} and three-dimensional \eqref{eq:sknbrokengen} formulas reflects a physical correspondence, rather than an accidental resemblance between the final expressions. Indeed, through the examples developed in the rest of the section, we will establish a one-to-one map between the various terms in the partition function of a broken five-dimensional theory and those in the partition function of its associated broken magnetic quiver.

In general, the map proceeds as follows:\footnote{The five-dimensional expressions also depend on partitions, which have no counterpart in the three-dimensional partition function. Therefore for this comparison, we set external partitions to empty.}
\begin{itemize}
    \item {\it Delta functions and frames}: The Dirac deltas in \eqref{eq:broken_formula} encode the tuning of the 5d mass parameters corresponding to a moment map operator acquiring a VEV. Via the 5d/3d dictionary (\ref{eq:dict}), these translate into constraints on the FI parameters corresponding to a monopole VEV. Comparing the partition function of a broken 5d theory in \eqref{eq:broken_formula} with that of its associated broken magnetic quiver in \eqref{eq:sknbrokengen}, we find that the 3d delta functions impose precisely the additive counterparts of the 5d multiplicative constraints: $\bm\delta_{\rm pert}^{\rm Ex} \longleftarrow \bm\delta_{\rm Ex}$ and $\bm\delta_{\rm pert}^{\rm Nil} \longleftarrow \bm\delta_{\rm Nil}$. It follows that the symmetry-breaking data entering the two partition functions coincide: the unbroken subgroup $H\subset SU(N)$, the number of frames $|\mathcal W_{SU(N)}|/|\mathcal W_H|$, and the nilpotent VEV $\rho'$ induced by $\bm\delta_{\rm Nil}$ are all the same on both sides.
    \item {\it Residual theory}: The magnetic quiver associated with the residual 5d theory coincides with the residual 3d theory produced by the electric dualisation algorithm in \eqref{eq:sknbrokengen}. This is one of the strongest tests of the proposed 5d/3d map. Conversely, the 3d analysis can be used to identify the residual 5d theory and its five-brane web, even for configurations with white dots on multiple/adjacent sides of the box.
    \item {\it Goldstone modes}: The Goldstone modes are fixed by the pair $(SU(N),\rho)$. Both ${\rm GM}_{\rm 5d}^\rho$ and ${\rm GM}_{\rm 3d}^\rho$ are obtained from the same analysis described in Appendix \ref{app:GM}.
    \item {\it Measure correction}: The factor $\Delta_{\rm 5d}^{\rm corr}$ is the direct counterpart of $\Delta_{\rm 3d}^{\rm corr}$, defined in \eqref{eq:deltacorr3d}. Both convert the appropriate $SU(N)$ gauging measure into the $H$ measure, implementing the same symmetry-breaking pattern $SU(N)\to H$ and parameter constraints determined by the matching delta functions. Both also contain the inverse Goldstone contribution $1/{\rm GM}^{\rho'}$ which cancels the contribution generated due to the induced nilpotent VEV in the test theory.
\end{itemize}

In this paper, we consider nilpotent VEVs for the moment maps of one or both $SU(N)$ flavour symmetries of the five-dimensional theory. We do not consider VEVs for the enhanced $SU(k)$ symmetries, whose moment maps involve instanton operators in the chosen gauge theory description. Implementing these deformations directly in this description would require resumming the relevant instanton contributions, which lies beyond the present analysis. We therefore restrict to deformations for which we can perform both the five-dimensional and three-dimensional magnetic quiver calculations independently, and compare their results.

\subsubsection*{A comment on the $r$-rule}

It is useful to compare the distributional structure described above with the $r$-rule for generalised toric polygons (GTPs). Given a GTP $P$, one can associate to it the combinatorial quantity \cite{vanBeest:2020kou}
\begin{equation}\label{eq:rrule}
    r(P) = {\rm Area}(P)+1-\frac{1}{2}\sum_{\alpha}\sum_{x=1}^{b_{\alpha}}\mu_{\alpha,x}^{\,2} ~.
\end{equation}
Here $\alpha$ labels the external edges of $P$. The $\alpha$-th edge is divided by its black vertices into $b_\alpha$ segments, and $\mu_{\alpha,x}$ denotes the lattice length of the $x$-th segment. In the absence of white dots, all these segments have unit length, $\mu_{\alpha,x}=1$.

In its conventional application, the $r$-rule requires $r(P)\geq0$, and $r(P)$ is interpreted as the rank of the five-dimensional theory associated with the GTP. For the broken configurations studied here, however, $r(P)$ underestimates the rank of the residual theory appearing in each frame. For broken configurations obtained from one-sided nilpotent VEVs on the box theory, this discrepancy is accounted for precisely by the periodic delta functions and the induced nilpotent VEV, giving the relation
\begin{equation}\label{eq:rrule_broken}
    r(P) = r({\rm residual})-n_{\delta}-\sum_{i=1}^{L}\sum_{j=1}^{\rho_i'-1}j ~.
\end{equation}
Here $r({\rm residual})$ is the rank of the residual five-dimensional theory obtained after resolving the degeneration, $n_{\delta}$ is the number of independent periodic delta function constraints appearing in each frame, and $\rho_1',\ldots,\rho_L'$ denote the sizes of the non-trivial Jordan blocks of the induced nilpotent VEV $\rho'$.

The relation in \eqref{eq:rrule_broken} therefore shows that throughout this family of broken configurations, $r(P)$ is strictly smaller than the residual rank. Note that negative values are also possible, but do not imply a negative physical rank, the latter being determined by the residual theory in each frame. The conventional rank interpretation of $r(P)$ must therefore be refined for this family of distributional configurations.\footnote{The situation is more intricate for the bad configurations discussed in Section \ref{sec:multiplenil_bad}. Their indices decompose into inequivalent, rather than symmetry-related, frames, which are realised by inequivalent classes of tessellations of the same GTP. In this case, $r(P)$ does not by itself determine the frame decomposition or the associated infrared theories.}

\subsubsection*{Nilpotent VEVs realisation}

\begin{figure}[h!]
    \centering
    \includegraphics[width=0.85\linewidth]{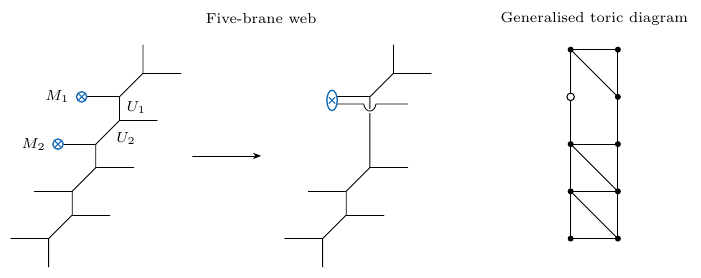}
    \caption{The $[2,1^2]$ nilpotent deformation ends two D5 branes on a common D7 brane (blue), producing one white dot on the boundary of the generalised toric diagram. Here $U_1,U_2$ are the internal \Kah parameters, while $M_1,M_2$ are external mass fugacities.}
    \label{fig:D7Whitedot}
\end{figure}

Consider the $(p,q)$ five-brane web of the box theory. Without loss of generality, we focus on nilpotent VEVs for the $SU(N)$ moment map associated with the $N$ D5 branes extending to the left. In the five-brane web, these VEVs are realised by ending multiple D5 branes on a common D7 brane \cite{DeWolfe:1999hj,Benini:2009gi}. Each part of the partition labelling the VEV specifies the number of D5 branes ending on one D7 brane. The accompanying Figure \ref{fig:D7Whitedot} illustrates the $[2,1^2]$ deformation, together with relevant mass fugacities and internal \Kah parameters. Two of the four D5 branes end on a common D7 brane, while the remaining two are left unchanged. In the dual generalised toric diagram, this replaces the corresponding black dot with a white dot. The geometric interpretation of such nilpotent deformations in terms of generalised toric polygons and Calabi--Yau geometries was studied in \cite{Bourget:2023wlb}, including rectangular box examples. Here we follow the deformation at the level of the refined partition function and the superconformal index.

We can translate this geometric operation into a parameter restriction at the level of the refined topological string partition function or the superconformal index. Bringing the two D5 branes together requires shrinking the internal segments between them. In the unrefined limit, this corresponds to $U_{1,2}=e^{-\beta u_{1,2}}\longrightarrow1$, and hence $\frac{M_1}{M_2}=e^{-\beta(m_1-m_2)}\longrightarrow e^0=1$. For the refined topological string partition function, the corresponding condition becomes \cite{Hayashi:2013qwa,Hayashi:2014hfa}\footnote{Through the five-dimensional AGT correspondence, such Higgsing conditions are related to degenerations of primary fields in $q$-deformed Toda theory. The case of a semi-degenerate primary is analysed in \cite{Isachenkov:2014eya}.}
\begin{equation}
    \frac{M_2}{M_1}=\frac{\mf q}{\mf t} \;\text{ or }\; \frac{M_2}{M_1}=\frac{\mf t}{\mf q} ~,
\end{equation}
where we will generally choose the former.

If three D5 branes, labelled by $M_1,M_2,M_3$, end on the same D7 brane, the conditions are
\begin{equation}
    \frac{M_2}{M_1} = \frac{\mf q}{\mf t} \quad,\quad \frac{M_3}{M_2} = \frac{\mf q}{\mf t} \quad \text{with}\quad\frac{M_3}{M_1} = \left(\frac{\mf q}{\mf t}\right)^2 ~.
\end{equation}
These conditions impose a geometric progression on the mass fugacities, analogous to the nilpotent specialisation in the 3d magnetic quivers discussed in Appendix \ref{app:GM}. Since only the ratios within the block are fixed, we can use the freedom to produce a geometric progression centered around $K'$:
\begin{equation}
    M_1 = \frac{\mf t}{\mf q}K' \quad,\quad M_2 = K' \quad,\quad M_3 = \frac{\mf q}{\mf t}K' ~.
\end{equation}
We will use this prescription to implement nilpotent VEVs for the $SU(N)$ moment maps of the five-dimensional theories studied below.

\subsection{Warm up: a good case}
\label{subsec:good}

We begin with a $[2,1^{N-2}]$ nilpotent VEV for the $SU(N)_{\vec M}$ moment map of the $N+SU(N)_0^2+N$ box theory. This deformation produces the $(N-2)+SU(N-1)_0\times SU(N)_0+(N+1)$ gauge theory, together with Goldstone multiplets associated with the VEV. We study the corresponding deformation of the strong-coupling magnetic quiver and compare the two results using the 5d/3d dictionary. This provides a further consistency check of the map, complementing those performed in the companion paper \cite{one}.

This example introduces the contour-pinching mechanism used throughout the following subsections. The pinching generates delta functions that, in the good case, are saturated by the gauge integrations, leaving no distributional constraints on external parameters. In the broken and bad cases studied later, some delta functions instead survive as constraints on external parameters, making the resulting observables distributional.

\subsubsection{5d calculation}
\label{subsubsec:5dgood}

The superconformal index of the $N+SU(N)_0^2+N$ box theory can be expressed as
\begin{equation}\label{eq:Good_index}
    \begin{split}
        & \cI^{N+SU(N)_0^2+N} (\vec M,\vec L, K_1, K_2, K_3, \vec Q_{\rm inst}) \\
        &\quad = \frac{1}{(N!)^2} \, \oint \prod_{i=1}^N \left[ \frac{{\rm d}N_i^{(1)}}{2\pi iN_i^{(1)}} \frac{{\rm d}N_i^{(2)}}{2\pi iN_i^{(2)}} \right] \times \delta\left( \prod_{i=1}^N N_i^{(1)} \right) \delta\left( \prod_{i=1}^N N_i^{(2)} \right) \left| M(\mf t,\mf q) \right|^{4(N-1)} \\
        &\qquad \times \; \left| \frac{\cZ_{\rm pert}^{\rm strip} (\vec M, \vec N^{(1)}, K_1) \; \cZ_{\rm pert}^{\rm alt \; strip} (\vec N^{(1)},\vec N^{(2)}, K_2) \; \cZ_{\rm pert}^{\rm strip} (\vec N^{(2)}, \vec L, K_3)}{ \cZ_{||}^{\rm left} (\vec M;\bm\emptyset) \; \cZ_{||}^{\rm right} (\vec L;\bm\emptyset) \; \cZ_{||}^{\rm top}( \vec Q_{\rm top} ) \; \cZ_{||}^{\rm bot}( \vec Q_{\rm bot} ) } \right|^2  \\
        &\qquad \times \left| \sum_{\bm\alpha^{(1)},\bm\alpha^{(2)}} \left(\prod_{a=1}^N Q_{a,1}^{|\alpha_a^{(1)}|} Q_{a,2}^{|\alpha_a^{(2)}|}\right) \, \cZ^{\rm strip}_{\bm\emptyset,\bm\alpha^{(1)}} (\vec M, \vec N^{(1)}, K_1) \right. \\
        &\hspace{4.5cm} \times \; \cZ^{\rm alt \; strip}_{\bm\alpha^{(1)},\bm\alpha^{(2)}} (\vec N^{(1)}, \vec N^{(2)}, K_2) \; \cZ^{\rm strip}_{\bm\alpha^{(2)},\bm\emptyset} (\vec N^{(2)}, \vec L, K_3) \, \Big|^2 ~.
    \end{split}
\end{equation}
The contributions from the external parallel legs are defined in Eqs.\ (\ref{eq:parallel_D5}) and (\ref{eq:parallel_NS5}). In this case they take the form
\begin{equation}
    \begin{split}
        \cZ_{||}^{\rm left}(\vec M;\bm\emptyset) = \prod_{1\leq a<b\leq N} \MM{\frac{M_a}{M_b}} \;\;&,\quad \cZ_{||}^{\rm right}(\vec L;\bm\emptyset) = \prod_{1\leq a<b\leq N} \MM{\frac{\mf t}{\mf q}\frac{L_b}{L_a}} ~, \\
        \cZ_{||}^{\rm top}( \vec Q_{\rm top} ) = \prod_{1\leq i<j\leq 3} \MM{\prod_{l=i}^{j-1}Q_{\rm top}^{(l)}} \;\;&,\quad \cZ_{||}^{\rm bot}( \vec Q_{\rm bot} ) = \prod_{1\leq i<j\leq 3} \MM{\frac{\mf t}{\mf q}\prod_{l=i}^{j-1}Q_{\rm bot}^{(l)}} ~.
    \end{split}
\end{equation}
The contributions from the external parallel NS5 branes are independent of the gauge fugacities. We therefore suppress them in the intermediate contour calculations and reinstate them when identifying the full residual index.

Following the prescription introduced above, we implement the nilpotent VEV by tuning the external mass parameters to
\begin{equation}\label{eq:GoodVEV_211}
    M_1 = \sqrt{\frac{\mf t}{\mf q}} K' d \;\;,\quad M_2 = \sqrt{\frac{\mf q}{\mf t}} K' d^{-1} ~,
\end{equation}
where $d$ regulates the contour pinching and is eventually taken to one. The relevant poles responsible for pinching the integration contour originate from the perturbative contribution of the first strip,
\begin{equation}\label{eq:good_inter}
    \left| \cZ^{\rm strip}_{\rm pert} (\vec M,\vec N^{(1)},K_1) \right|^2 = \left| \frac{\prod_{1\leq a<b\leq N} \cM\left( \frac{M_a}{M_b} \right) \cM\left( \frac{\mf t}{\mf q} \frac{N_b^{(1)}}{N_a^{(1)}} \right)}{\prod_{1\leq a<b\leq N} \cM\left( \frac{\mf t}{\mf q}\frac{K_1}{M_bN_a^{(1)}} \right) \prod_{1\leq a\leq b\leq N} \MM{\frac{M_aN_b^{(1)}}{K_1}}} \right|^2 ~.
\end{equation}
There are $N$ Weyl-related choices of the gauge fugacity that will be frozen by the VEV. We first analyse the branch involving $N_1^{(1)}$.

Substituting the tuning for the $[2,1^{N-2}]$ nilpotent VEV \eqref{eq:GoodVEV_211} into \eqref{eq:good_inter}, the factor responsible for the pinching is
\begin{equation}
    \left| \frac{ \MM{\frac{\mf t}{\mf q}d^2} }{ \MM{ \sqrt{\frac{\mf t}{\mf q}} \frac{K_1}{N_1^{(1)}K'd} } \MM{ \sqrt{\frac{\mf t^3}{\mf q^3}} \frac{K_1\,d}{N_1^{(1)}K'} } } \right|^2 ~. \nn
\end{equation}
The numerator vanishes as $d\to1$, while the denominator produces four families of poles, obtained from the zeroes listed in \eqref{eq:Msq_zeroes}:
\begin{equation}
    \begin{split}
        (1) : \; N_1^{(1)} = \frac{K_1}{K'd} \, \mf t^{-n_1-\frac12} \mf q^{n_2+\frac12} \quad,&\quad (\bar{1}) : \; N_1^{(1)} = \frac{K_1}{K'd} \, \mf t^{n_1+\frac12} \mf q^{-n_2-\frac12} ~, \\
        (2) : \; N_1^{(1)} = \frac{K_1\,d}{K'} \, \mf t^{\frac12-n_1} \mf q^{n_2-\frac12} \quad,&\quad (\bar 2) : \; N_1^{(1)} = \frac{K_1\,d}{K'} \, \mf t^{n_1+\frac32} \mf q^{-n_2-\frac32} ~,
    \end{split}
\end{equation}
where $n_1,n_2\in\mathbb Z_{\geq0}$. In our conventions $|\mf t|>1$ and $|\mf q|<1$, therefore family $(\bar 1)$ extends towards infinity, whereas family $(2)$ extends towards the origin. The integration contour over $N_1^{(1)}$ separates these two families. As $d\to1$, their leading poles, corresponding to $n_1=n_2=0$, collide at
\begin{equation}
    N_1^{(1)} = \sqrt{\frac{\mf t}{\mf q}} \frac{K_1}{K'} ~,
\end{equation}
pinching the integration contour. Using the local behaviour of the paired MacMahon functions given in \eqref{eq:Mstr_def}, the pinching factor has the following distributional limit under the $N_1^{(1)}$ integration:
\begin{equation}
    \lim_{d\rightarrow 1} \left| \frac{ \MM{\frac{\mf t}{\mf q}d^2} }{ \MM{ \sqrt{\frac{\mf t}{\mf q}} \frac{K_1}{N_1^{(1)}K'd} } \MM{ \sqrt{\frac{\mf t^3}{\mf q^3}} \frac{K_1\,d}{N_1^{(1)}K'} } } \right|^2 = \frac{1}{\left|M(\mf t,\mf q)\right|^2} \times \delta\left( \sqrt{\frac{\mf t}{\mf q}} \frac{K_1}{K'N_1^{(1)}} \right) ~.
\end{equation}

Similar pinchings can also be considered for the remaining $N_a$ variables, leading to a total of $N$ pinchings:
\begin{equation}\label{eq:Strip211_polepinch_Good}
    N_a^{(1)} = \sqrt{\frac{\mf t}{\mf q}}\frac{K_1}{K'} \quad,\text{ for } a=1,\dots,N ~.
\end{equation}
These pinchings lead to a sum over the respective residues. The perturbative part of the integrand of the superconformal index in \eqref{eq:Good_index} then simplifies to the following:
\begin{equation}\label{eq:Good_index_N11frozen}
    \boxed{\begin{aligned}
        &\left| \frac{ \cZ_{\rm pert}^{\rm strip} (\vec M, \vec N^{(1)}, K_1) \; \cZ_{\rm pert}^{\rm alt \; strip} (\vec N^{(1)},\vec N^{(2)}, K_2) \; \cZ_{\rm pert}^{\rm strip} (\vec N^{(2)}, \vec L, K_3) }{ \cZ_{||}^{\rm left}(\vec M;\bm\emptyset) \; \cZ_{||}^{\rm right} (\vec L;\bm\emptyset) } \right|^2_{\vec M\to[2,1^{N-2}]_{K'}} \\[1.5mm]
        &\quad = \delta\left( \sqrt{\frac{\mf t}{\mf q}}\frac{K_1}{K'N_1^{(1)}} \right) \times |M(\mf t,\mf q)|^{-2} \times \left| \MM{\frac{\mf t}{\mf q}}\prod_{a=3}^N\MM{\sqrt{\frac{\mf t}{\mf q}}\frac{K'}{M_a}} \right|^{-2} \\[1.5mm]
        &\qquad \times \left| \frac{ \prod_{2\leq a<b\leq N} \MM{\frac{N_a^{(1)}}{N_b^{(1)}}} \MM{\frac{N_b^{(1)}}{N_a^{(1)}}} \prod_{1\leq a<b\leq N} \MM{\frac{N_a^{(2)}}{N_b^{(2)}}} \MM{\frac{N_b^{(2)}}{N_a^{(2)}}} }{ \prod_{\substack{a=3\\b=2}}^N \MM{\frac{M_aN_b^{(1)}}{K_1}} \prod_{\substack{a=2\\b=1}}^N \MM{\frac{N_a^{(1)}N_b^{(2)}}{K_2}} \prod_{a=1}^N \left[ \MM{\sqrt{\frac{\mf t}{\mf q}}\frac{K_1N_a^{(2)}}{K'K_2}} \prod_{b=1}^N \MM{\frac{N_a^{(2)}L_b}{K_3}} \right] } \right|^2 \\[1.5mm]
        &\qquad\quad + \text{ $N-1$ permutations } \left( N_1^{(1)} \longleftrightarrow N_{a\neq1}^{(1)} \right) ~,
    \end{aligned}}
\end{equation}
The displayed term corresponds to the branch in which $N_1^{(1)}$ is frozen, with the remaining terms obtained by Weyl permutations. We have explicitly decoupled the contributions from parallel D5 branes on both edges of the box. The top and bottom decoupling factors are independent of the gauge fugacities and hence unaffected by the contour pinchings.

\begin{figure}[h!]
    \centering
    \includegraphics[width=\linewidth]{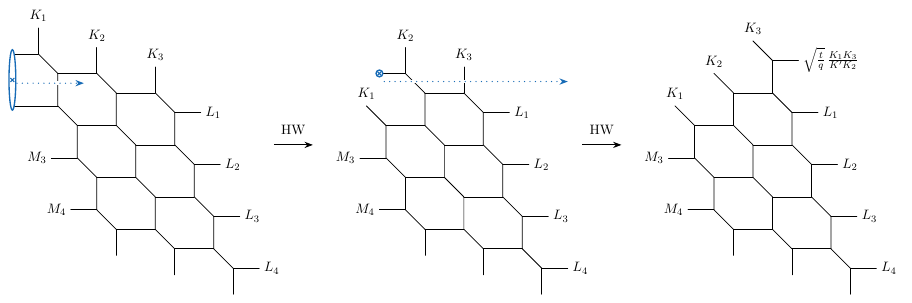}
    \caption{Five-brane web realisation of the $[2,1^{N-2}]$ nilpotent VEV in the $N+SU(N)_0^2+N$ theory. On the left, two external D5 branes end on the same D7 brane, implementing the VEV. Hanany--Witten moves give the web on the right, which engineers the $(N-2)+SU(N-1)_0\times SU(N)_0+(N+1)$ gauge theory.}
    \label{fig:Braneweb_Good}
\end{figure}

We now identify the factors in the expression \eqref{eq:Good_index_N11frozen} and their action inside the superconformal index \eqref{eq:Good_index}. The factor $\left|M(\mf t,\mf q)\right|^{-2}$ removes one Cartan contribution from the original factor $|M(\mf t,\mf q)|^{4(N-1)}$, consistent with the reduction of the total gauge rank by one. The remaining gauge-dependent factors can be identified as the perturbative part of the refined topological string partition function of the $(N-2)+SU(N-1)_0\times SU(N)_0+(N+1)$ gauge theory.\footnote{After freezing $N_1^{(1)}$, the surviving gauge fugacities satisfy $\prod_{a=2}^N N_a^{(1)}=\sqrt{\mf q/\mf t}\,K'/K_1\neq1$. The canonical $SU(N-1)$ unit-product constraint can be restored by a common rescaling of these fugacities, accompanied by a rescaling of $K_1$ and $K_2$ to preserve the matter arguments. Their ratio, and hence the additional flavour fugacity in \eqref{eq:good_newvar}, remains unchanged. We retain the inherited variables here and discuss these redefinitions explicitly in the next subsection.} The fugacities $M_3,\dots,M_N$ parameterise its $N-2$ flavours on the left, while $L_1,\dots,L_N$ parameterise $N$ of the right flavours. The additional factor
\begin{equation}\label{extraflavourgoodvev}
    \left| \prod_{a=1}^N \MM{\sqrt{\frac{\mf t}{\mf q}}\,\frac{K_1N_a^{(2)}}{K'K_2}} \right|^{-2}
\end{equation}
describes an additional fundamental hypermultiplet of the second $SU(N)_{\vec N^{(2)}}$ gauge group. Comparing its argument with those of the other fundamental flavours gives
\begin{equation}
    \sqrt{\frac{\mf t}{\mf q}}\,\frac{K_1N_a^{(2)}}{K'K_2} = \frac{N_a^{(2)}L_{N+1}}{K_3} ~,
\end{equation}
where
\begin{equation}\label{eq:good_newvar}
    L_{N+1} = \sqrt{\frac{\mf t}{\mf q}}\,\frac{K_1K_3}{K'K_2} ~.
\end{equation}
Therefore (\ref{extraflavourgoodvev}) describes the additional fundamental hypermultiplet that enlarges the flavour symmetry at this end of the residual quiver from $SU(N)$ to $SU(N+1)$, with the $N+1$ flavour fugacities parameterised by $L_1,\dots,L_N,L_{N+1}$. This identification is also realised by Hanany--Witten moves in the five-brane web as illustrated in Figure \ref{fig:Braneweb_Good}.

Writing $\vec M'=(M_3,\dots,M_N)$ and $\vec N^{(1)}{}'=(N_2^{(1)},\dots,N_N^{(1)})$, the normalised perturbative contribution of the residual theory in the displayed branch is
\begin{equation}
    \begin{split}
        &\left| \frac{\cZ^{\rm residual}_{\rm pert}(\vec M',\vec N^{(1)}{}',\vec N^{(2)},\vec L,K_1,K_2,K_3,K')}{\cZ_{||,\,{\rm pert}}^{\rm residual}(\vec M',\vec L,L_{N+1};\bm\emptyset,\bm\emptyset)} \right|^2 = \left| \prod_{a=1}^N \MM{\sqrt{\frac{\mf t}{\mf q}}\frac{K_1N_a^{(2)}}{K'K_2}} \right|^{-2} \\
        &\quad \times \; \left| \frac{ \prod_{2\leq a<b\leq N} \MM{\frac{N_a^{(1)}}{N_b^{(1)}}} \MM{\frac{N_b^{(1)}}{N_a^{(1)}}} \prod_{1\leq a<b\leq N} \MM{\frac{N_a^{(2)}}{N_b^{(2)}}} \MM{\frac{N_b^{(2)}}{N_a^{(2)}}} }{ \prod_{a=3}^N\prod_{b=2}^N \MM{\frac{M_aN_b^{(1)}}{K_1}} \prod_{a=2}^N\prod_{b=1}^N \MM{\frac{N_a^{(1)}N_b^{(2)}}{K_2}} \prod_{a=1}^N \prod_{b=1}^N \MM{\frac{N_a^{(2)}L_b}{K_3}} } \right|^2 ~,
    \end{split}
\end{equation}
Here the contribution from the external parallel D5 legs of the residual five-brane web is\footnote{The top and bottom decoupling factors associated with the residual web in Figure \ref{fig:Braneweb_Good} are kept separate and included in the final index.}
\begin{equation}
    \cZ_{||,\,{\rm pert}}^{\rm residual}(\vec M',\vec L,L_{N+1};\bm\emptyset,\bm\emptyset) = \prod_{3\leq a<b\leq N} \MM{\frac{M_a}{M_b}}\prod_{1\leq a<b\leq N+1} \MM{\frac{\mf t}{\mf q}\frac{L_b}{L_a}} ~.
\end{equation}
The second product includes the pairs involving the additional external D5 leg parameterised by $L_{N+1}$. The remaining factor in \eqref{eq:Good_index_N11frozen} gives the Goldstone contribution associated with the $[2,1^{N-2}]$ nilpotent VEV:
\begin{equation}\label{eq:chirals_211}
    {\rm GM}_{\rm 5d}^{[2,1^{N-2}]}(\vec M',K';\bm\emptyset) = \left| \MM{\frac{\mf t}{\mf q}} \prod_{a=3}^N \MM{\sqrt{\frac{\mf t}{\mf q}}\frac{K'}{M_a}} \right|^{-2} ~,
\end{equation}
where, as before, $\vec M'=(M_3,\dots,M_N)$. This agrees with the independent computation of the Goldstone contribution obtained from the prescription reviewed in Appendix \ref{app:GM}.

Although we have displayed only the branch in which $N_1^{(1)}$ is frozen, the deformed index receives contributions from all $N$ branches in \eqref{eq:Strip211_polepinch_Good}. These are related by the original $SU(N)$ Weyl group, with each branch invariant under the Weyl subgroup of the residual $SU(N-1)$. after relabelling the intergation variables, they give equal contributions to \eqref{eq:Good_index}. Their multiplicity $N=N!/(N-1)!$ therefore converts the Weyl factor of the first gauge group from $1/N!$ to $1/(N-1)!$, as required for the residual $SU(N-1)$ gauge group.

The instantonic contributions in \eqref{eq:Good_index} after the contour pinching induced by the nilpotent VEV undergo similar simplifications. On the branch where $N_1^{(1)}$ is frozen, the first strip contains the factor
\begin{equation}
    \left. \cN_{\alpha_1^{(1)}\emptyset}\left( \sqrt{\frac{\mf t^3}{\mf q^3}}\frac{K_1}{N_1^{(1)}K'} \right) \times (\cdots) \; \right|_{N_1^{(1)}\rightarrow \sqrt{\frac{\mf t}{\mf q}}\frac{K_1}{K'}} = \; \cN_{\alpha_1^{(1)}\emptyset}\left( \frac{\mf t}{\mf q} \right) \times (\cdots) \;=\; \delta_{\alpha_1^{(1)},\emptyset} \times (\cdots) ~,
\end{equation}
where we use the identity \eqref{eq:Ntrivial_identity}. This discrete delta function trivialises the sum over the partition $\alpha_1^{(1)}$ as expected from the Higgsing of the $SU(N)$ gauge group to $SU(N-1)$. The remaining strip factors and gluing weights reorganise into the instanton contribution of the residual $(N-2)+SU(N-1)_0\times SU(N)_0+(N+1)$ gauge theory.

Combining the perturbative and instantonic contributions, the final superconformal index obtained by implementing the $[2,1^{N-2}]$ nilpotent VEV on \eqref{eq:Good_index} gives\footnote{In the residual five-brane web of Figure \ref{fig:Braneweb_Good}, the Hanany--Witten moves turn the upper parallel NS5 legs into parallel $(1,-1)$ legs with the same separation parameters under the VEV. Their decoupled contribution is therefore precisely $\cZ_{||}^{\rm top}(\vec Q_{\rm top})$, while the lower contribution remains $\cZ_{||}^{\rm bot}(\vec Q_{\rm bot})$.}
\begin{equation}\label{5dindexgoodcase1}
    \boxed{\begin{split}
        &\cI^{N+SU(N)_0^2+N} (\vec M, \vec L, K_1,K_2,K_3,\vec Q_{\rm inst}) \Big\vert_{\vec M\to[2,1^{N-2}]_{K'}} \\[0.1cm]
        &\quad = \cI^{(N-2)+SU(N-1)_0\times SU(N)_0+(N+1)} (\vec M', \vec L, K_1,K_2,K_3,K',\vec Q_{\rm inst}) \times {\rm GM}_{\rm 5d}^{[2,1^{N-2}]}(\vec M', K';\bm\emptyset) ~,
    \end{split}}
\end{equation}
The residual index is written in the inherited variables, with $L_{N+1}$ fixed by \eqref{eq:good_newvar}. This is in agreement with the residual theory engineered by the five-brane web in Figure \ref{fig:Braneweb_Good}.

\subsubsection{3d counterpart and the 5d/3d map}
\label{subsubsec:compare_good}

\begin{figure}[h!]
    \centering
    \includegraphics[width=\linewidth]{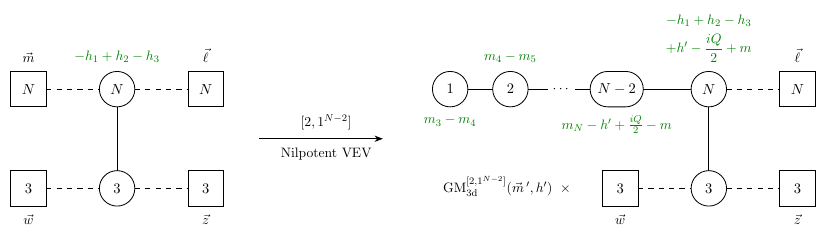}
    \caption{The $[2,1^{N-2}]$ nilpotent VEV partially closes a $T[U(N)]$ tail of the $S_{3,N}$ magnetic quiver, producing the good residual quiver and the associated Goldstone contribution. }
    \label{fig:3dgood_VEV}
\end{figure}

We now study the three-dimensional counterpart of the $[2,1^{N-2}]$ nilpotent VEV. The five-dimensional mass tuning is
\begin{equation}\label{eq:n4map}
    M_1 = \sqrt{\frac{\mf t}{\mf q}} K'  \;\;,\quad M_2 = \sqrt{\frac{\mf q}{\mf t}} K' \quad\Longrightarrow\quad m_1=h' \;\;,\quad m_2=h' - \frac{iQ}{2} ~.
\end{equation}
Via the 5d/3d dictionary in \ref{eq:dict}, this condition identifies a monopole VEV in the associated $S_{3,N}$ magnetic quiver, shown in the first line of Figure \ref{fig:3dgood_VEV}. The relevant monopole has magnetic flux $-1$ under the first node of the corresponding $T[U(N)]$ tail and has vanishing total charge when $m_1-m_2-iQ/2=0$, using the charge conventions of \eqref{eq:monopolecharge}, thus acquiring a VEV.

This specialisation corresponds to the canonical $R$ charge assignment $m=iQ/4$ at the $\cN=4$ superconformal point. For the exact $S^3_b$ analysis, we retain the full $\cN=2^*$ refinement parameterised by the complexified axial mass $m$. The monopole VEV condition then becomes
\begin{equation}\label{eq:VEV211_N2star}
    m_1=h' \;\;,\quad m_2=h'-iQ+2m ~.
\end{equation}
This provides the $\mathcal N=2^*$ refined version of the map. As can be easily confirmed, it precisely agrees with \eqref{eq:n4map} at the canonical $R$-charge value.

This monopole VEV triggers the flow from the $T[U(N)]$ tail to the $T_{[2,1^{N-2}]}[U(N)]$ tail, producing the good quiver on the right of Figure \ref{fig:3dgood_VEV} together with the associated Goldstone contribution. This is the strong-coupling magnetic quiver of the residual five-dimensional theory. It contains $N-3$ balanced nodes corresponding to an $SU(N-2)$ topological symmetry parameterised by $m_3,\dots,m_N$, together with $N$ balanced nodes corresponding to an $SU(N+1)$ topological symmetry parameterised by $l_1,\dots,l_N$ and $l_{N+1}=h_1-h_2+h_3-h'+iQ/2-m$. This additional Cartan parameter is precisely the image, under the refined 5d/3d map, of the five-dimensional flavour fugacity $L_{N+1}=\sqrt{\mf t/\mf q}\,(K_1K_3)/(K'K_2)$ identified in the 5d computation. Together with the mapping of $M_j$ with $m_j$ and $L_j$ with $l_j$, this precisely matches the parameters of the $SU(N-2)$ and $SU(N+1)$ flavour symmetries of the residual five-dimensional theory.

The contribution of the Goldstone modes associated with the Higgsing induced by the monopole VEV can be computed using the prescription reviewed in Appendix \ref{app:GM}. In the present case, it is given by
\begin{equation}\label{eq:21n2gold}
    {\rm GM}_{\rm 3d}^{[2,1^{N-2}]}(\vec m',h') = s_b\left(\frac{iQ}{2}\right) s_b\left(2m-\frac{iQ}{2}\right) \prod_{j=3}^N s_b\left(m\pm\left(m-\frac{iQ}{2}+h'-m_j\right)\right) ~,
\end{equation}
where $\vec m'=(m_3,\ldots,m_N)$. The 5d/3d dictionary identifies the same nilpotent orbit $\rho=[2,1^{N-2}]$ on both sides, and hence the corresponding symmetry-breaking pattern and Goldstone directions. The contributions ${\rm GM}_{\rm 3d}^{[2,1^{N-2}]}(\vec m',h')$ and ${\rm GM}_{\rm 5d}^{[2,1^{N-2}]}(\vec M',K')$ can then be obtained from these directions using the appropriate partition function building blocks in each dimension. The details of this construction are reviewed in Appendix \ref{app:GM}.

The full 5d/3d comparison is summarised in Table \ref{tab:Comp_Good}.
\begin{table}[htbp]
    \centering
    \setlength{\tabcolsep}{8pt}
    \renewcommand{\arraystretch}{1.4}

    \begin{tabular}{>{\centering\arraybackslash}m{0.46\linewidth} | >{\centering\arraybackslash}m{0.46\linewidth}}
        \textbf{5d} & \textbf{3d} \\ \hline
        \includegraphics[width=\linewidth]{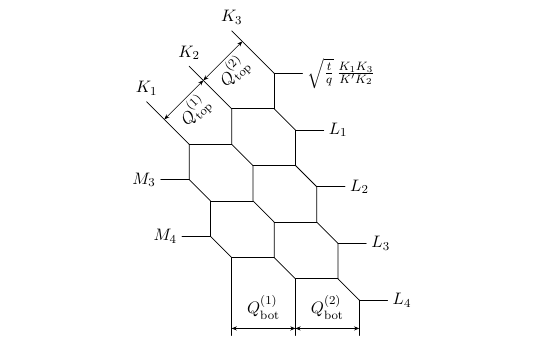} &
        \includegraphics[width=\linewidth]{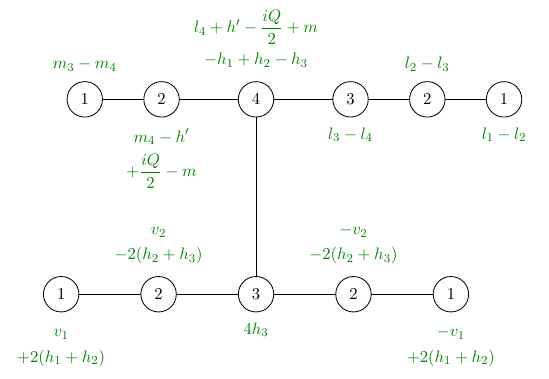} \\
    \end{tabular}
    \caption{Comparison of the residual five-brane web and its three-dimensional strong-coupling magnetic quiver for $N=4$ and $k=3$.}
    \label{tab:Comp_Good}
\end{table}

We conclude by emphasising that the good case produces ordinary observables on both sides of the 5d/3d correspondence. The delta functions generated by the contour pinching freeze the gauge fugacities and are saturated by the corresponding integrations, leaving no constraints on external parameters. After factoring out the Goldstone contributions, the final five-dimensional superconformal index and the three-dimensional $S^3_b$ partition function of the magnetic quiver therefore each consist of a single, non-distributional frame, with no decoupled free sector.

\subsection{Warm up: an ``ugly'' case}
\label{subsec:ugly}

We now consider the $[2,1^{N-2}]$ nilpotent VEV for the $N\times2$ box, corresponding to the five-dimensional $N+SU(N)_0+N$ gauge theory. This configuration lies at the threshold between the good and broken regimes: the largest part of the partition is $r_1=2$, equal to the width $k=2$ of the box. Correspondingly, the central $U(N)$ node of the deformed magnetic quiver has $2N-1$ effective flavours and is therefore ugly.

On the five-dimensional side, the contour pinching leaves no distributional constraint on the external parameters. After separating the Goldstone contribution, the result factorises into the ordinary superconformal index of an interacting residual theory and a decoupled free sector. This provides the intermediate case between the good Higgsing studied in Subsection \ref{subsec:good} and the genuinely distributional examples considered in subsequent subsections.

\subsubsection{5d computation}
\label{subsubsec:5dugly}

The calculation closely parallels that of Section \ref{subsec:good}, so we emphasise the features that differ in the present case. Before turning on the nilpotent VEV, the superconformal index of the $N+SU(N)_0+N$ theory is
\begin{equation}\label{eq:Ugly_index}
    \begin{split}
        & \cI^{N+SU(N)_0+N} (\vec M,\vec L, K_1, K_2, Q_{\rm inst}) \\
        &= \frac{1}{N!} \, \oint \prod_{i=1}^N\frac{{\rm d}N_i}{2\pi iN_i} \times \delta\left( \prod_{i=1}^N N_i \right) \times \left| \frac{\cZ_{\rm pert}^{\rm strip} (\vec M, \vec N, K_1) \; \cZ_{\rm pert}^{\rm alt \; strip} (\vec N,\vec L, K_2)}{ \cZ_{||}^{\rm left} (\vec M;\bm\emptyset) \; \cZ_{||}^{\rm alt \; right} (\vec L;\bm\emptyset) \; \cZ_{||}^{\rm top}(\vec Q_{\rm top}) \; \cZ_{||}^{\rm bot}(\vec Q_{\rm bot})} \right|^2  \\
        &\hspace{2cm} \times \left| M(\mf t,\mf q) \right|^{2(N-1)} \times \left| \sum_{\bm\alpha} \prod_{a=1}^N Q_a^{|\alpha_a|} \, \cZ^{\rm strip}_{\bm\emptyset,\bm\alpha} (\vec M, \vec N, K_1) \; \cZ^{\rm alt \; strip}_{\bm\alpha,\bm\emptyset} (\vec N, \vec L, K_2) \right|^2 ~.
    \end{split}
\end{equation}
The external parallel leg contributions are defined in Eqs.\ (\ref{eq:parallel_D5}) and (\ref{eq:parallel_NS5}). As in the good case, the top and bottom contributions are independent of the gauge fugacities and are suppressed in the intermediate computations. We implement the nilpotent VEV through the same mass tuning as in \eqref{eq:GoodVEV_211}:
\begin{equation}\label{eq:UglyVEV_211}
    M_1 = \sqrt{\frac{\mf t}{\mf q}} K' d \;\;,\quad M_2 = \sqrt{\frac{\mf q}{\mf t}} K' d^{-1} ~.
\end{equation}
As before, the limit $d\to1$ produces $N$ Weyl-related pinching branches,
\begin{equation}
    N_a = \sqrt{\frac{\mf t}{\mf q}}\frac{K_1}{K'} \;\;,\quad a=1,\ldots,N ~.
\end{equation}
The corresponding residues are related by permutations of the $SU(N)$ gauge fugacities. We display the branch in which the $N_1$ contour is pinched; the remaining $N-1$ contributions are obtained by Weyl permutations.

Within the gauge integral, the freezing of $N_1$ is represented by the multiplicative delta function
\begin{equation}
    \delta\left( \sqrt{\frac{\mf t}{\mf q}}\,\frac{K_1}{K'N_1} \right) ~.
\end{equation}
The perturbative contribution then becomes
\begin{equation}\label{eq:Ugly_index_N1frozen}
    \boxed{\begin{split}
        &\left| \frac{ \cZ_{\rm pert}^{\rm strip} (\vec M, \vec N, K_1) \; \cZ_{\rm pert}^{\rm alt \; strip} (\vec N,\vec L, K_2) }{ \cZ_{||}^{\rm left}(\vec M;\bm\emptyset) \; \cZ_{||}^{\rm alt \; right} (\vec L;\bm\emptyset) } \right|^2_{\vec M\to[2,1^{N-2}]_{K'}} \\
        &\quad = \delta\left( \sqrt{\frac{\mf t}{\mf q}}\frac{K_1}{K'N_1} \right) \times | M(\mf t,\mf q) |^{-2} \times \left| \MM{\frac{\mf t}{\mf q}}\prod_{a=3}^N\MM{\sqrt{\frac{\mf t}{\mf q}}\frac{K'}{M_a}} \right|^{-2} \\
        &\qquad \times \; \left| \prod_{a=1}^N \MM{\sqrt{\frac{\mf t}{\mf q}}\frac{K_1L_a}{K'K_2}}\right|^{-2} \times \left| \frac{ \prod_{2\leq a<b\leq N} \MM{\frac{N_a}{N_b}} \MM{\frac{N_b}{N_a}} }{ \prod_{a=3}^N\prod_{b=2}^N \MM{\frac{M_aN_b}{K_1}} \prod_{a=2}^N\prod_{b=1}^N \MM{\frac{N_aL_b}{K_2}} } \right|^2 \\[1.5mm]
        &\qquad\quad + \text{ $N-1$ permutations } \left( N_1 \longleftrightarrow N_{a\neq1} \right) ~.
    \end{split}}
\end{equation}
Apart from the delta function, each branch contains four factors. First, $\left|M(\mf t,\mf q)\right|^{-2}$ removes one refined MacMahon factor from the original gauge measure, as required by the reduction of the gauge group from $SU(N)$ to $SU(N-1)$. Second, the factor
\begin{equation}\label{eq:ugly_free_5d}
    \prod_{a=1}^N \left| \MM{ \sqrt{\frac{\mf t}{\mf q}}\, \frac{K_1L_a}{K'K_2} } \right|^{-2} \;\eqcolon\; \cI^{\rm free}(\vec L,K_1,K_2,K')
\end{equation}
is independent of the remaining gauge fugacities and describes a decoupled sector of $N$ free hypermultiplets. Third, the final contribution on the second line of \eqref{eq:Ugly_index_N1frozen} can be identified as the Goldstone contribution associated with the $[2,1^{N-2}]$ nilpotent VEV: ${\rm GM}_{\rm 5d}^{[2,1^{N-2}]}$. This agrees with the independent prescription reviewed in Appendix \ref{app:GM}, as well as the analogous contribution in Subsection \ref{subsec:good}.

\begin{figure}[t]
    \centering
    \includegraphics[width=\linewidth]{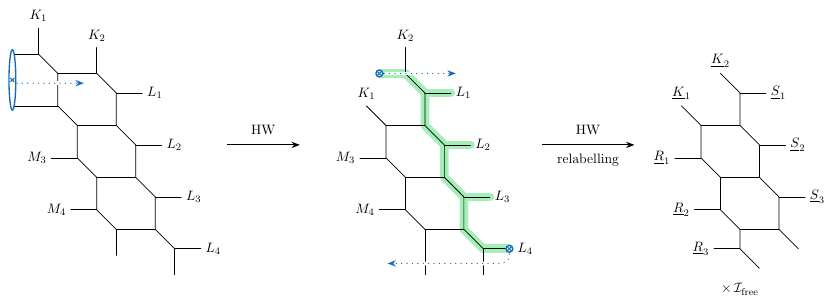}
    \caption{The $[2,1^2]$ nilpotent deformation of the $4+SU(4)_0+4$ box theory and subsequent Hanany--Witten dual presentations. The highlighted part of the second strip in the middle web accounts for the four free hypermultiplets accompanying the residual $2+SU(3)_0+4$ theory. In the final transition, the D7 brane associated with the free sector is moved to the far right, while the D7 brane associated with $L_4$ is moved to the far left. The resulting Hanany--Witten dual $3+SU(3)_0+3$ residual theory is shown in the canonical underlined parameters.}
    \label{fig:Braneweb_Ugly}
\end{figure}

Finally, writing $\vec M'=(M_3,\dots,M_N)$ and $\vec N'=(N_2,\dots,N_N)$, the remaining gauge-dependent contribution is
\begin{equation}
    \left| \frac{ \cZ_{\rm pert}^{\rm residual} (\vec M', \vec N', \vec L, K_1, K_2, K') }{ \cZ_{||,\,{\rm pert}}^{\rm residual} (\vec M', \vec L;\bm\emptyset,\bm\emptyset) } \right|^2
    = \left| \frac{ \prod_{2\leq a<b\leq N} \MM{\frac{N_a}{N_b}} \MM{\frac{N_b}{N_a}} }{ \prod_{a=3}^N\prod_{b=2}^N \MM{\frac{M_aN_b}{K_1}} \prod_{a=2}^N\prod_{b=1}^N \MM{\frac{N_aL_b}{K_2}} } \right|^2 ~.
\end{equation}
This is the perturbative contribution of the $(N-2)+SU(N-1)_0+N$ gauge theory, with the external D5 leg contributions decoupled. Moving the D7 brane associated with $L_N$ from the right to the left of the five-brane web gives an equivalent $(N-1)+SU(N-1)_0+(N-1)$ gauge theory presentation, as illustrated in Figure \ref{fig:Braneweb_Ugly}. The two $SU(N-1)$ flavour symmetries are then parameterised by
\begin{equation}
    \vec R = \left( M_3,\dots,M_N,\frac{L_NK_1}{K_2} \right) \;\;,\quad \vec S = (L_1,\dots,L_{N-1}) ~.
\end{equation}
As in the good case, the transferred flavour fugacity is fixed by preserving its matter argument, {\it i.e.}, by imposing $R_{N-1}N_a/K_1=N_aL_N/K_2$.

Further, as in the previous subsection, the inherited residual gauge fugacities parameterising the $SU(N-1)$ gauge group do not satisfy the unit-product constraint, since freezing $N_1$ leaves
\begin{equation}
    \prod_{a=2}^N N_a = N_1^{-1} =\sqrt{\frac{\mf q}{\mf t}}\,\frac{K'}{K_1} \neq 1 ~.
\end{equation}
Due to the above Hanany--Witten transition, the same is true for the two $SU(N-1)$ flavour fugacities which satisfy $\prod_{a=1}^{N-1} R_a=(L_NK_1)/(K_2K'^2)$ and $\prod_{a=1}^{N-1}L_a=L_N^{-1}$. By introducing the common rescalings
\begin{equation}\label{eq:NRS_rescale}
    \begin{split}
        \underline{\vec N}{}' &\coloneq \frac{\vec N'}{c_N} \;\;,\quad c_N = \left(\sqrt{\frac{\mf q}{\mf t}}\frac{K'}{K_1}\right)^{\frac{1}{N-1}} ~, \\
        \underline{\vec R}{}' &\coloneq \frac{\vec R}{c_R} \;\;,\quad c_R = \left(\frac{L_NK_1}{K'^2K_2}\right)^{\frac{1}{N-1}} \quad,\qquad \underline{\vec S}{}' \coloneq \frac{\vec S}{c_S} \;\;,\quad c_S = L_N^{-\frac{1}{N-1}} ~,
    \end{split}
\end{equation}
we obtain the canonical $SU(N-1)$ parametrisations that now satisfy $\prod_{a=2}^N \underline{N}_a = \prod_{a=1}^{N-1} \underline{R}_a = \prod_{a=1}^{N-1} \underline{S}_a = 1$. These redefinitions preserve the vector-multiplet ratios in \eqref{eq:Ugly_index_N1frozen}, while the matter arguments can be preserved by a simultaneous rescaling of the abelian parameters $K_1$ and $K_2$:
\begin{equation}\label{eq:K_rescale}
    \underline{K}_1 \coloneq \frac{K_1}{c_N\,c_R} = K_1\,\left( \sqrt{\frac{\mf t}{\mf q}}\frac{K'K_2}{L_N} \right)^{\frac{1}{N-1}} \;\;,\quad \underline{K}_2 \coloneq \frac{K_2}{c_N\,c_S} = K_2\,\left( \sqrt{\frac{\mf t}{\mf q}}\frac{K_1L_N}{K'} \right)^{\frac{1}{N-1}} ~.
\end{equation}
In these variables, the perturbative contribution in the Hanany--Witten dual frame takes the canonical form
\begin{equation}\label{eq:uglyres}
    \begin{split}
        \left| \frac{ \cZ_{\rm pert}^{\rm residual} (\vec M', \vec N', \vec L, K_1, K_2, K') }{ \cZ_{||,\,{\rm pert}}^{\rm residual} (\vec M', \vec L;\bm\emptyset,\bm\emptyset) } \right|^2 &= \left| \frac{ \underline{\cZ}_{\rm pert}^{\rm residual} (\underline{\vec R}, \underline{\vec N}{}', \underline{\vec S}, \underline{K}_1, \underline{K}_2) }{ \underline{\cZ}_{||,\,{\rm pert}}^{\rm residual} (\underline{\vec R}, \underline{\vec S};\bm\emptyset,\bm\emptyset) } \right|^2_{\rm HW} \\[3mm]
        &= \left| \frac{ \prod_{2\leq a<b\leq N} \MM{\frac{\underline{N}_a}{\underline{N}_b}} \MM{\frac{\underline{N}_b}{\underline{N}_a}} }{ \prod_{a=1}^{N-1}\prod_{b=2}^N \MM{\frac{\underline{R}_a\underline{N}_b}{\underline{K}_1}} \prod_{a=2}^N\prod_{b=1}^{N-1} \MM{\frac{\underline{N}_a\underline{S}_b}{\underline{K}_2}} } \right|^2 ~.
    \end{split}
\end{equation}
The contributions from external parallel D5 legs in the two presentations are
\begin{equation}
    \begin{split}
        \cZ_{||,\,{\rm pert}}^{\rm residual} (\vec M',\vec L;\bm\emptyset,\bm\emptyset) &= \prod_{3\leq a<b\leq N} \MM{\frac{M_a}{M_b}} \prod_{1\leq a<b\leq N} \MM{\frac{\mf t}{\mf q}\frac{L_a}{L_b}} ~, \\
        \underline{\cZ}_{||,\,{\rm pert}}^{\rm residual} (\underline{\vec R},\underline{\vec S};\bm\emptyset,\bm\emptyset) &= \prod_{1\leq a<b\leq N-1} \MM{\frac{\underline{R}_a}{\underline{R}_b}} \prod_{1\leq a<b\leq N-1} \MM{\frac{\mf t}{\mf q}\frac{\underline{S}_a}{\underline{S}_b}} ~. \\
    \end{split}
\end{equation}
The free and Goldstone contributions are retained in the inherited external parameters.

As in Subsection \ref{subsec:good}, the instanton contribution contains a factor that sets the partition associated with the frozen gauge fugacity to be empty using \eqref{eq:Ntrivial_identity}. The remaining contributions reorganise into the instanton contributions of the residual $(N-1)+SU(N-1)_0+(N-1)$ gauge theory in the Hanany--Witten dual presentation. The $N$ pinching branches give equal contributions after relabelling the integration variables. Their multiplicity converts the original Weyl factor $1/N!$ into $1/(N-1)!$, appropriate to the residual $SU(N-1)$ gauge group. Combining these contributions gives\footnote{The Hanany--Witten moves displayed in Figure \ref{fig:Braneweb_Ugly} preserve the top and bottom decoupling contributions and $Q_{\rm inst}$ under the VEV condition.}
\begin{equation}\label{eq:5dugly}
    \boxed{\begin{split}
        &\cI^{N+SU(N)_0+N} (\vec M,\vec L,K_1,K_2, Q_{\rm inst}) \Big\vert_{\vec M\to[2,1^{N-2}]_{K'}} = \cI^{\rm free} (\vec L,K_1,K_2,K') \\[0.1cm]
        &\hspace{2cm} \times \; \cI^{(N-1)+SU(N-1)_0+(N-1)} (\underline{\vec R},\underline{\vec S},\underline{K}_1,\underline{K}_2, Q_{\rm inst}) \times {\rm GM}_{\rm 5d}^{[2,1^{N-2}]} (\vec M',K';\bm\emptyset) ~.
    \end{split}}
\end{equation}

We stress that the key difference from the good case is the additional decoupled free sector. Its five-brane web interpretation follows from the second strip: the nilpotent VEV freezes the internal leg at $N_1=\sqrt{\mf t/\mf q}K_1/K'$, and the contributions connecting this leg to the $N$ external legs parameterised by $\vec L$ become independent of the remaining gauge fugacities. These contributions therefore survive the Higgsing as $N$ free hypermultiplets, alongside the interacting residual theory. These degrees of freedom are represented by the highlighted part of the second web in Figure \ref{fig:Braneweb_Ugly}.

\subsubsection{3d counterpart and the 5d/3d map}
\label{subsubsec:compare_ugly}

\begin{figure}[h!]
    \centering
    \includegraphics[width=\linewidth]{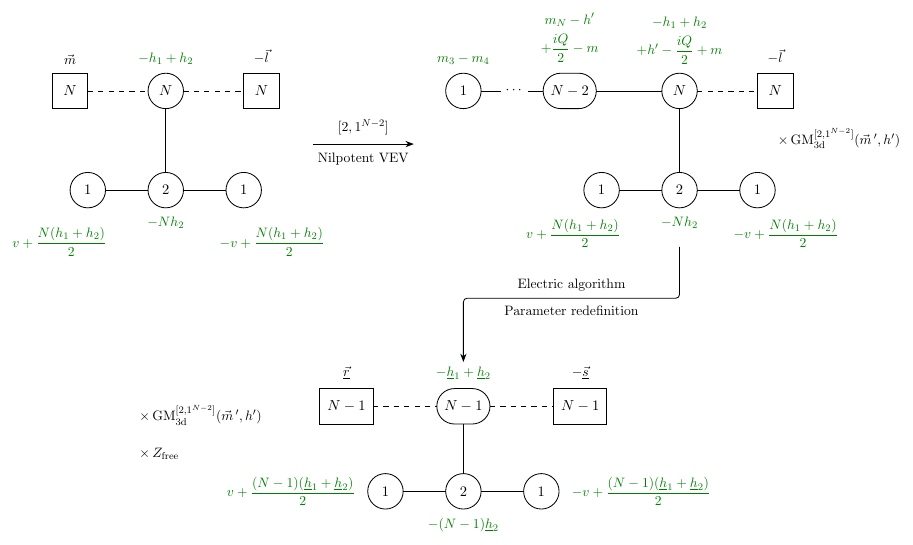}
    \caption{Three-dimensional counterpart of the ugly Higgsing. The $[2,1^{N-2}]$ monopole VEV partially closes the left $T[U(N)]$ tail of $S_{2,N}$, producing the Goldstone contribution and rendering the central $U(N)$ node ugly. Electric dualisation, followed by the parameter redefinitions described in the text, gives the magnetic quiver of the $(N-1)+SU(N-1)_0+(N-1)$ box theory in canonical underlined parameters, together with $N$ free hypermultiplets.}
    \label{fig:3dUgly_VEV}
\end{figure}

We now study the three-dimensional counterpart of the preceding deformation. As explained in Subsection \ref{subsubsec:compare_good}, the $[2,1^{N-2}]$ nilpotent VEV with $\cN=2^*$ refinement requires
\begin{equation}
    m_1 = h' \;\;,\quad m_2 = h'-iQ+2m ~.
\end{equation}
This VEV partially closes the left $T[U(N)]$ tail of the $S_{2,N}$ magnetic quiver to $T_{[2,1^{N-2}]}[U(N)]$ and generates the Goldstone contribution ${\rm GM}_{\rm 3d}^{[2,1^{N-2}]}(\vec m',h')$, where $\vec m'=(m_3,\dots,m_N)$. The central $U(N)$ node of the resulting quiver has $2N-1$ effective flavours and is therefore ugly. Applying the electric dualisation algorithm yields a good interacting magnetic quiver together with $N$ decoupled free hypermultiplets, as shown in Figure \ref{fig:3dUgly_VEV}.

The interacting quiver is the strong-coupling magnetic quiver of the $(N-1)+SU(N-1)_0+(N-1)$ gauge theory. Its two $T[U(N-1)]$ tails are parameterised by
\begin{equation}
    \vec r = (m_3,\dots,m_N,h_1-h_2+l_N) \;\;,\quad \vec s = (l_1,\dots,l_{N-1}) ~.
\end{equation}
In particular, $r_{N-1}=h_1-h_2+l_N$ is the additive counterpart of the transferred flavour fugacity $R_{N-1}=L_NK_1/K_2$ in the five-dimensional Hanany--Witten presentation. To express the residual quiver in canonical parameters, we centre the two flavour vectors:
\begin{equation}
    \underline r_i \coloneq r_i - \frac{h_1-h_2+l_N-2h'+iQ-2m}{N-1} \;\;,\quad \underline s_i \coloneq s_i + \frac{l_N}{N-1} ~,
\end{equation}
for $i=1,\dots,N-1$, leading to the tracelessness constraint $\sum_{i=1}^{N-1} \underline{r}_i = \sum_{i=1}^{N-1} \underline{s}_i = 0$. We simultaneously redefine the parameters $h_1,h_2$ as
\begin{equation}
    \underline{h}_1 \coloneq h_1 + \frac{h'+h_2-l_N+m-iQ/2}{N-1} \;\;,\quad \underline{h}_2 \coloneq h_2 + \frac{h_1-h'+l_N-m+iQ/2}{N-1} ~.
\end{equation}
These shifts put the FI parameters into the canonical form shown in the final quiver in Figure \ref{fig:3dUgly_VEV}. At the canonical $R$ charge assignment $m=iQ/4$, these shifts reproduce the five-dimensional rescalings of the flavour and abelian parameters in Eqs.\ (\ref{eq:NRS_rescale}) and (\ref{eq:K_rescale}) under the 5d/3d dictionary. The interacting magnetic quiver therefore matches the residual five-brane web in the Hanany--Witten dual presentation.

The electric dualisation algorithm also produces a decoupled sector of $N$ free hypermultiplets, whose contribution is
\begin{equation}
    \cZ_{\rm free} (\vec l,h_1,h_2,h';m) = \prod_{j=1}^N s_b\left( m\pm\left(h'-\frac{iQ}{2}+m-h_1+h_2-l_j\right) \right) ~.
\end{equation}
At $m=iQ/4$, this contribution matches the five-dimensional free sector in \eqref{eq:ugly_free_5d} factor by factor. For each $j$, the double-sine pair corresponds to the inverse modulus-squared MacMahon factor. Both descriptions therefore contain $N$ free hypermultiplets with matching charges under the identified global symmetries. The Goldstone contributions likewise correspond to the same $[2,1^{N-2}]$ nilpotent VEV, as in the good case.

The full 5d/3d comparison is summarised in Table \ref{tab:Comp_Ugly}.
\begin{table}[h!]
    \centering
    \setlength{\tabcolsep}{8pt}
    \renewcommand{\arraystretch}{1.4}

    \begin{tabular}{>{\centering\arraybackslash}m{0.46\linewidth} | >{\centering\arraybackslash}m{0.46\linewidth}}
        \textbf{5d} & \textbf{3d} \\ \hline
        \includegraphics[width=\linewidth]{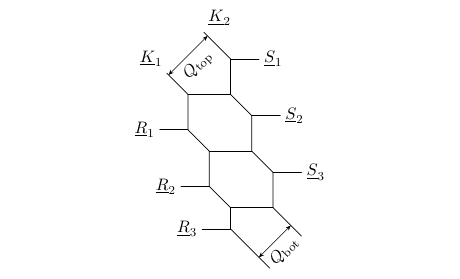} &
        \includegraphics[width=\linewidth]{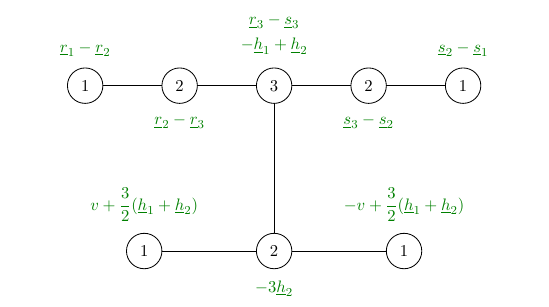} \\[6mm]
        $\times ~\cI_{\rm free}(\vec L,K_1,K_2,K')$ & $\times ~ \cZ_{\rm free}(\vec l,h_1,h_2,h';m)$ \\
    \end{tabular}
    \caption{Comparison of the residual five-brane web and its three-dimensional strong-coupling magnetic quiver for the ugly deformation at $N=4$ and $k=2$.}
    \label{tab:Comp_Ugly}
\end{table}

We conclude by emphasising that the ugly case produces no residual distributional constraint on either side of the 5d/3d correspondence. In five dimensions, the delta function generated by contour pinching fixes a gauge fugacity and is saturated by the corresponding integration. After factoring out the Goldstone contribution, the resulting observable is an ordinary function describing an interacting residual theory along with the contribution of $N$ free hypermultiplets. Likewise, the $S_b^3$ partition function of the ugly magnetic quiver factorises into an interacting good theory and a decoupled free sector, without any residual delta function imposing a constraint on the external parameters. In the broken configurations studied next, delta functions remain supported on external flavour parameters, and the resulting observables are genuinely distributional.

\subsection{Breaking the strip with the $[2,1^{N-2}]$ VEV}
\label{subsec:Strip211}

In this subsection, we present our first example of a broken configuration in five dimensions. We impose a $[2,1^{N-2}]$ nilpotent VEV for one of the $SU(N)$ moment maps of a symmetric length-$N$ strip, the $k=1$ member of the family of box theories. Unlike the examples considered above, the strip has no gauge integral of its own. The delta functions generated by the deformation therefore remain supported on external parameters.

\subsubsection{5d computation}
\label{subsubsec:Strip211_5d}

We implement the $[2,1^{N-2}]$ nilpotent VEV through the same specialisation of the mass parameters used in the previous two subsections:
\begin{equation}\label{eq:StripVEV_211_masses}
    M_1=\sqrt{\frac{\mf t}{\mf q}}K'd \;\;,\quad M_2=\sqrt{\frac{\mf q}{\mf t}} K'd^{-1} ~.
\end{equation}
The parameter $d$ resolves the pole collision; after gluing the strip to a test theory, it regulates the corresponding pinching of the $SU(N)_{\vec N}$ integration contour. We study this action in the limit $d\to1$. Since most of the calculation closely parallels the analysis in the previous two subsections, we only present the essential steps.

We begin with the modulus square of the ``perturbative'' part of the refined topological string partition function of the strip,
\begin{equation}
    \left| \cZ^{\rm strip}_{\rm pert} (\vec M,\vec N,K) \right|^2 = \left| \frac{\prod_{1\leq a<b\leq N} \cM\left( \frac{M_a}{M_b} \right) \cM\left( \frac{\mf t}{\mf q} \frac{N_b}{N_a} \right)}{\prod_{1\leq a<b\leq N} \cM\left( \frac{\mf t}{\mf q}\frac{K}{M_bN_a} \right) \prod_{1\leq a\leq b\leq N} \MM{\frac{M_aN_b}{K}}} \right|^2 ~.
\end{equation}
The specialisation in \eqref{eq:StripVEV_211_masses} produces $N$ Weyl-related pinching branches
\begin{equation}
    N_a=\sqrt{\frac{\mf t}{\mf q}}\frac{K}{K'} \quad,\qquad a=1,\ldots,N.
\end{equation}
The crucial difference from the good and ugly cases studied in the previous two subsections is that the strip has no gauge integral of its own. Here, the variables $N_a$ are external flavour fugacities, and hence the pole collision cannot be saturated by an integration within the strip partition function. It instead produces a delta function supported on an external fugacity, whose action is defined by gluing the Higgsed strip to a test theory and integrating over the resulting $SU(N)_{\vec N}$ gauge variables.

Accordingly, when studying the distributional action of the Higgsed strip, we only decouple the parallel-leg contribution associated with the Higgsed edge (taken to be the left one throughout this section). The contribution on the right edge is retained because the test strip is glued along this edge. We likewise retain non-trivial external partitions on the right: after gluing, these become the partitions summed over in the instanton contributions of the newly gauged $SU(N)_{\vec N}$ symmetry. The following equalities are therefore understood as distributional identities in the external fugacities $\vec N$. Here and below, $\delta$ denotes the periodic delta function appropriate to multiplicative fugacities.

The ``perturbative" contribution of the Higgsed strip can then be expressed as
\begin{equation}\label{eq:Strip211_broken_pert}
    \boxed{\begin{split}
        &\left| \frac{ \cZ_{\rm pert}^{\rm strip} (\vec M, \vec N, K) }{ \cZ_{||}^{\rm left}(\vec M;\bm\emptyset) } \right|^2_{\vec M\to[2,1^{N-2}]_{K'}} = \delta\left( \sqrt{\frac{\mf t}{\mf q}}\frac{K}{K' N_1} \right) \times \left| \MM{\frac{\mf t}{\mf q}}\prod_{a=3}^N\MM{\sqrt{\frac{\mf t}{\mf q}}\frac{K'}{M_a}} \right|^{-2} \\
        &\hspace{3cm} \times \; \left| \frac{ \prod_{2\leq a<b\leq N} \MM{\frac{\mf t}{\mf q}\frac{N_b}{N_a}} }{ M(\mf t,\mf q) \times \prod_{a=2}^N \MM{\sqrt{\frac{\mf q}{\mf t}}\frac{N_aK'}{K}} } \right|^2 \times \left| \prod_{a=3}^N \prod_{b=2}^N \MM{\frac{M_aN_b}{K}} \right|^{-2} \\[1.5mm]
        &\hspace{2.5cm} + \text{ $N-1 \;$ permutations } \left( N_1 \longleftrightarrow N_{a\neq1} \right) ~.
    \end{split}}
\end{equation}
Note that unlike Eqs.\ (\ref{eq:Good_index_N11frozen}) and (\ref{eq:Ugly_index_N1frozen}), we only decouple the parallel leg contributions on the left side of the strip given in \eqref{eq:parallel_D5}, as explained above.

The frame displayed in \eqref{eq:Strip211_broken_pert} is the one in which $N_1$ is frozen. The remaining $N-1$ frames are obtained by Weyl permutations of the variables $N_a$. Upon gluing to a test strip by gauging the $SU(N)_{\vec N}$ symmetry, these $N$ frames contribute equally after relabelling the integration variables. Their multiplicity converts the original Weyl factor into $$\frac{1}{N!} \frac{|\mathcal W_{SU(N)}|}{|\mathcal W_{SU(N-1)}|} = \frac{N}{N!} = \frac{1}{(N-1)!} ~,$$
as required for the residual $SU(N-1)$ gauge group.

We next evaluate the effect of the nilpotent VEV on the ``instantonic'' part of the strip partition function. As in Subsection \ref{subsubsec:5dgood}, the partition associated with the frozen fugacity is forced to be empty. For the branch freezing $N_1$, we find
\begin{equation}\label{eq:Strip211_broken_inst}
    \begin{split}
        &\left.\lim_{d\rightarrow1} \cZ^{\rm strip}_{\bm\emptyset,\bm\alpha} (\vec M,\vec N,K)\right|_{M_i\rightarrow[2,1^{N-2}]_{K'},\,N_1\rightarrow \sqrt{\frac{\mf t}{\mf q}}\frac{K}{K'd}} = \delta_{\alpha_1,\emptyset} \;\times\; \prod_{a=2}^N \Big[ \mf q^{\frac{||\alpha_a||^2}{2}} \widetilde Z_{\alpha_a}(\mf t,\mf q) \Big] \\[0.15cm]
        &\hspace{1cm} \times \; \prod_{b=2}^N \cN_{\emptyset\alpha_b}\left( \sqrt{\frac{\mf q}{\mf t}}\frac{N_bK'}{K} \right) \frac{\prod_{a=3}^N \left[ \prod_{b=2}^{a-1} \cN_{\alpha_b\emptyset}\left( \frac{\mf t}{\mf q}\frac{K}{M_aN_b} \right) \prod_{b=a}^N \cN_{\emptyset\alpha_b}\left( \frac{M_aN_b}{K} \right) \right]}{\prod_{2\leq a<b\leq N} \cN_{\alpha_a\alpha_b}\left( \frac{\mf t}{\mf q} \frac{N_b}{N_a} \right)} ~.
    \end{split}
\end{equation}
Combining the ``perturbative" and ``instantonic" contributions from Eqs.\ (\ref{eq:Strip211_broken_pert}) and (\ref{eq:Strip211_broken_inst}), we obtain the paired refined topological string partition function of the strip with a $[2,1^{N-2}]$ VEV
\begin{equation}\label{eq:Strip211_broken}
    \boxed{\begin{split}
        &\left| \frac{ Z^{\rm strip} (\vec M, \vec N, K;\bm\emptyset,\bm\alpha) }{ \cZ_{||}^{\rm left}(\vec M;\bm\emptyset) } \right|^2_{\vec M\to[2,1^{N-2}]_{K'}} = \delta\left( \sqrt{\frac{\mf t}{\mf q}}\frac{K}{K'N_1} \right) \times \left| \MM{\frac{\mf t}{\mf q}}\prod_{a=3}^N\MM{\sqrt{\frac{\mf t}{\mf q}}\frac{K'}{M_a}} \right|^{-2} \\
        &\qquad \times \; \left| \frac{1}{M(\mf t,\mf q)} \prod_{2\leq a<b\leq N} \frac{\cM\left( \frac{\mf t}{\mf q} \frac{N_b}{N_a} \right)}{\cN_{\alpha_a\alpha_b}\left( \frac{\mf t}{\mf q} \frac{N_b}{N_a} \right)} \times \prod_{b=2}^N \frac{\cN_{\emptyset\alpha_b}\left( \sqrt{\frac{\mf q}{\mf t}}\frac{N_bK'}{K} \right)}{\cM\left( \sqrt{\frac{\mf q}{\mf t}}\frac{N_bK'}{K} \right)} \right|^2 \\[1.5mm]
        &\qquad \times \; \left| \delta_{\alpha_1,\emptyset} \;\times\; \prod_{b=2}^N \Big[ \mf q^{\frac{||\alpha_b||^2}{2}} \widetilde Z_{\alpha_b}(\mf t,\mf q) \Big] \prod_{a=3}^N \left[ \prod_{b=2}^{a-1} \frac{\cN_{\alpha_b\emptyset}\left( \frac{\mf t}{\mf q}\frac{K}{M_aN_b} \right)}{\cM\left( \frac{\mf t}{\mf q}\frac{K}{M_aN_b} \right)} \prod_{b=a}^N \frac{\cN_{\emptyset\alpha_b}\left( \frac{M_aN_b}{K} \right)}{\cM\left( \frac{M_aN_b}{K} \right)} \right] \right|^2 \\[1.5mm]
        &\qquad\quad + \text{ $N-1$ permutations } \left( N_1 \longleftrightarrow N_{a\neq1} \;,\;\; \alpha_1 \longleftrightarrow \alpha_{a\neq1} \right) ~,
    \end{split}}
\end{equation}
Unlike in the good and ugly cases, the periodic delta function in \eqref{eq:Strip211_broken} is not saturated by a gauge integration and remains supported on an external flavour fugacity. Consequently, the Higgsed strip defines a genuinely distributional object. This is our first explicit example of a broken five-dimensional configuration.

To identify the residual theory and the factors responsible for the reduction of the gluing measure, we reorganise \eqref{eq:Strip211_broken} in the general form of \eqref{eq:broken_formula}. For the first frame, the different contributions are
\begin{equation}\label{eq:Strip211_broken_gen}
    \left.
    \begin{split}
        \bm{\delta}_{\rm pert}^{\rm Nil} &= 1 ~, \\
        \bm{\delta}_{\rm pert}^{\rm Ex} &= \delta\left( \sqrt{\frac{\mf t}{\mf q}}\frac{K}{K'N_1} \right) ~, \\
        \bm{\delta}_{\rm inst} &= \delta_{\alpha_1,\emptyset} ~, \\
        \frac{ Z^{\rm residual} }{ \cZ_{||}^{\rm residual} } &= \prod_{b=2}^N \Big[ \mf q^{\frac{||\alpha_b||^2}{2}} \widetilde Z_{\alpha_b}(\mf t,\mf q) \Big] \prod_{a=3}^N \left[ \prod_{b=2}^{a-1} \frac{\cN_{\alpha_b\emptyset}\left( \frac{\mf t}{\mf q}\frac{K}{M_aN_b} \right)}{\cM\left( \frac{\mf t}{\mf q}\frac{K}{M_aN_b} \right)} \prod_{b=a}^N \frac{\cN_{\emptyset\alpha_b}\left( \frac{M_aN_b}{K} \right)}{\cM\left( \frac{M_aN_b}{K} \right)} \right] ~, \\
        {\rm GM}_{\rm 5d}^{[2,1^{N-2}]} &= \left| \MM{\frac{\mf t}{\mf q}}\prod_{a=3}^N\MM{\sqrt{\frac{\mf t}{\mf q}}\frac{K'}{M_a}} \right|^{-2} ~, \\[3mm]
        \Delta^{\rm corr}_{\rm 5d} &= \left| \frac{1}{M(\mf t,\mf q)} \times \frac{ \cZ_{||}^{\rm right} (\vec N';\bm\alpha') \; \cZ_{||}^{\rm alt \; left} (\vec N',\bm\alpha') }{ \cZ_{||}^{\rm alt \; left} (\vec N,\bm\alpha) } \right|^2_{\bm\delta_{\rm pert}^{\rm Ex},\,\bm\delta_{\rm inst}} \\
        &= \left| \frac{1}{M(\mf t,\mf q)} \prod_{2\leq a<b\leq N} \frac{\cM\left( \frac{\mf t}{\mf q} \frac{N_b}{N_a} \right)}{\cN_{\alpha_a\alpha_b}\left( \frac{\mf t}{\mf q} \frac{N_b}{N_a} \right)} \prod_{b=2}^N \frac{\cN_{\emptyset\alpha_b}\left( \sqrt{\frac{\mf q}{\mf t}}\frac{N_bK'}{K} \right)}{\cM\left( \sqrt{\frac{\mf q}{\mf t}}\frac{N_bK'}{K} \right)} \right|^2 ~.
    \end{split}
    \quad\right\vert\;\text{\small Frame 1}
\end{equation}
Here $\vec M'=(M_3,\dots,M_N)$, $\vec N'=(N_2,\dots,N_N)$, and $\bm\alpha'=(\alpha_2,\dots,\alpha_N)$, while the external parallel leg contribution associated with the residual theory is obtained from \eqref{eq:parallel_D5}
\begin{equation}
    \cZ_{||}^{\rm residual} = \prod_{3\leq a<b\leq N} \MM{\frac{M_a}{M_b}} \times \prod_{2\leq a<b\leq N} \frac{ \MM{\frac{\mf t}{\mf q}\frac{N_b}{N_a}} }{ \cN_{\alpha_a\alpha_b}\left(\frac{\mf t}{\mf q}\frac{N_b}{N_a}\right) } ~.
\end{equation}
The remaining $N-1$ frames are obtained by simultaneously permuting the external fugacities $N_a$ and associated partitions $\alpha_a$. They are labelled by elements of $\mathcal W_{SU(N)}/\mathcal W_{SU(N-1)}$ and constitute the symmetry-related frames characteristic of a broken theory.

The distributional action of \eqref{eq:Strip211_broken} partially Higgses the $SU(N)_{\vec N}$ gauge symmetry produced by gluing to a test strip down to $SU(N-1)_{\vec N'}$. The periodic 
Dirac delta $\bm\delta_{\rm pert}^{\rm Ex}$ freezes one Coulomb modulus, while the discrete delta $\bm\delta_{\rm inst}$ trivialises the associated partition. On the support of these constraints, the factor $\Delta^{\rm corr}_{\rm 5d}$ supplies the correction needed to reconstruct the $SU(N-1)_{\vec N'}$ gluing measure. The residual theory is an asymmetric strip engineering a free  $SU(N-2)\times SU(N-1)$ bifundamental hypermultiplet. The non-trivial external partitions are carried by the $SU(N-1)$ side, which is subsequently glued to the test strip. Finally, the Goldstone modes generated by the original $[2,1^{N-2}]$ nilpotent VEV are collected in ${\rm GM}_{\rm 5d}^{[2,1^{N-2}]}(\vec M', K';\bm\emptyset)$.

\begin{figure}[h!]
    \centering
    \includegraphics[width=0.66\linewidth]{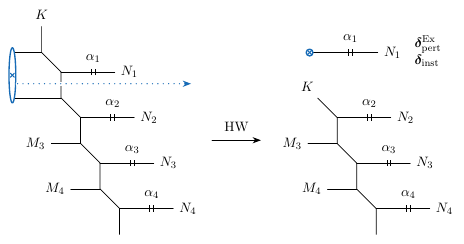}
    \caption{Five-brane-web realisation of the $[2,1^2]$ deformation of the symmetric length-four strip, implemented by ending two D5 branes on the same D7 brane. After the Hanany--Witten move, the configuration decomposes into a residual asymmetric strip and a semi-infinite D5 brane ending on the D7 brane. The asymmetric strip engineers the residual free bifundamental sector, while the semi-infinite D5 brane encodes the distributional part.}
    \label{fig:Braneweb_211}
\end{figure}

The five-brane-web interpretation of the distributional answer is illustrated in Figure \ref{fig:Braneweb_211} for $N=4$. The $[2,1^2]$ nilpotent VEV is implemented by terminating two D5 branes on the same D7 brane. After moving this D7 brane across the NS5 brane through a Hanany--Witten transition, the resulting configuration separates into two distinct components: a residual asymmetric strip and a semi-infinite D5 brane ending on the D7 brane.

The asymmetric strip contains two D5 branes on the left and three on the right and therefore engineers the residual free bifundamental hypermultiplet. The remaining semi-infinite D5 brane, instead, does not belong to the residual strip and geometrically encodes the distributional constraint. In the refined topological-string partition function, this constraint has both a perturbative and an instantonic component. The periodic delta function $\bm{\delta}_{\rm pert}^{\rm Ex}$ fixes the position of one of the D5 branes on the gluing side, or equivalently one of the fugacities $N_a$, while the discrete delta $\bm{\delta}_{\rm inst}$ trivialises the partition attached to the same external leg.

For general $N$, the $N$ possible choices of the external D5 brane whose position is fixed correspond precisely to the $N$ Weyl-related frames. Upon gluing this configuration to a symmetric length-$N$ test strip, the semi-infinite D5 brane identifies the gauge direction removed by the Higgsing. The remaining web couples the residual asymmetric strip to the test theory through the unbroken $SU(N-1)$ gauge symmetry, with the gluing measure reconstructed by $\Delta_{\rm 5d}^{\rm corr}$. This gives the geometric realisation of the distributional action in \eqref{eq:Strip211_broken}.

We next derive the same symmetry-breaking pattern and distributional structure independently from the associated three-dimensional strong-coupling magnetic quiver. The subsequent examples develop the five-brane web interpretation of broken theories further, leading to the general discussion in Subsection \ref{subsec:general_NilHW}.

\subsubsection{3d counterpart and the 5d/3d map}
\label{subsubsec:strip211_3d}

We now study the effect of the $[2,1^{N-2}]$ nilpotent VEV on the $S_{1,N}$ magnetic quiver associated with the symmetric length-$N$ strip. The undeformed magnetic quiver is depicted on left of Figure \ref{fig:3dstrip211}. In the conventions reviewed in Appendix \ref{app:3dstuff}, its $S_b^3$ partition function is:
\begin{equation}
    \begin{split}
        \mathcal{Z}_{S_{1,N}}(\vec{m},\vec{n},Nh,m) &= \int {\rm d}t\,\D^{\rm HB}_1(\vec{t};m) e^{2\pi i t N h} \int {\rm d}\vec{u}_N \; \D^{\rm HB}_N(\vec{u};m) \; e^{2\pi i (-h)\sum_{i=1}^N u_i} \\
        &\times \; \mathcal{Z}^{T[U(N)]}(\vec{u},\vec{m};m) \mathcal{Z}^{T[U(N)]}(\vec{u},\vec{n};m) \times \; \prod_{i=1}^N s_b\left(\frac{i Q}{2}-m\pm (u_i-t)\right).
    \end{split}
\end{equation}
The overall diagonal $U(1)$ gauge factor is removed according to the prescription described in Subsection \ref{subsec:sknbreakingbad}. The central $U(N)$ node sees $2N-1$ effective flavours and is therefore ugly. Applying the electric algorithm to this undeformed theory reduces it to a collection of $N^2$ free hypermultiplets.

To study the nilpotent deformation, we apply the VEV to the ugly magnetic quiver before dualising it. As explained in the preceding subsections, the $\cN=2^*$ map implements the $[2,1^{N-2}]$ VEV on the left $T[U(N)]$ tail through
\begin{equation}
    m_1 = h' \;\;,\quad m_2 = h'-iQ+2m ~.
\end{equation}
The resulting quiver is shown on the right of Figure \ref{fig:3dstrip211}. Closing the left tail reduces the number of effective flavours at the central $U(N)$ node from $2N-1$ to $2N-2$, making this node bad. Since the opposite $T[U(N)]$ tail remains full, the deformed star-shaped quiver lies in the broken regime described in Subsection \ref{subsec:MQbreakbad}. Applying the electric dualisation algorithm produces a sum of $N$ symmetry-related frames.

\begin{figure}[h!]
    \centering
    \includegraphics[width=\linewidth]{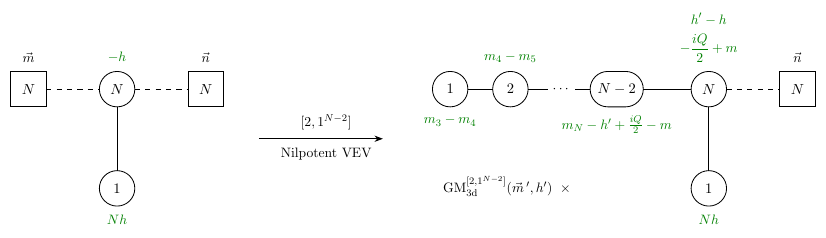}
    \caption{The $[2,1^{N-2}]$ nilpotent deformation of the $S_{1,N}$ magnetic quiver. The VEV partially closes the left $T[U(N)]$ tail to $T_{[2,1^{N-2}]}[U(N)]$ and produces the Goldstone contribution ${\rm GM}_{\rm 3d}^{[2,1^{N-2}]}$. The central node of the resulting quiver sees $2N-2$ effective flavours and is therefore bad. Since the opposite $T[U(N)]$ tail remains full, the resulting star-shaped quiver belongs to the broken class.}
    \label{fig:3dstrip211}
\end{figure}
By convention, the overall diagonal $U(1)$ is removed when we write the parent magnetic quiver, and we apply the same convention to an interacting residual quiver. Here the algorithm leaves an entirely free theory, so there is no residual quiver to which that convention can be applied. We must therefore remove the decoupled $U(1)$ contribution explicitly
from the algorithm's output, implemented by dividing by the factor $s_b(iQ/2)s_b(2m-iQ/2)$.\footnote{Here we are using the distributional identification
$s_b(iQ/2)\sim\delta(0)$ as in \cite{Giacomelli:2023zkk}.}
After this normalization, the contribution of the first frame to the partition function takes the general form in \eqref{eq:sknbrokengen}, with
\begin{equation}\label{eq:3dbroken_strip211}
    \left.
    \begin{aligned}
        \PFS{\text{residual}} &=\; \prod_{j=3}^N\prod_{i=2}^{N} \bigg[ s_b\left(m\pm\left(m_j+n_i-h\right)\right)\bigg] ~, \\
        \dd{\rm{Nil}} &=\; 1 ~, \\
        \dd{\rm{Ex}} &=\; \delta\left(n_1+h'-h-\frac{iQ}{2}+m\right) ~, \\
        \Delta^{\rm corr}_{\rm 3d} &= \left.\frac{\DCB{SU(N-1)}(\vec{n}';m)}{\DCB{SU(N)}(\vec{n};m)} \right\vert_{\dd{\text{Ex}}} \\
        &= s_b\left(2m-\frac{iQ}{2}\right) \prod_{j=2}^{N}\bigg[s_b\left(\frac{iQ}{2}\pm\left(h'-\frac{iQ}{2}+m+n_j-h\right)\right) \\
        &\hspace{2.4cm} \times \; s_b\left(2m-\frac{iQ}{2}\pm\left(h'-\frac{iQ}{2}+m+n_j-h\right)\right)\bigg] ~, \\
        {\rm GM}_{\rm 3d}^{[2,1^{N-2}]} &= s_b\left( m\pm\left(\frac{iQ}{2}-m\right) \right) \prod_{j=3}^N\left[s_b\left(m\pm\left(h'-m_j-\frac{iQ}{2}+m\right)\right)\right] ~,
    \end{aligned}
    \quad\right\vert\;\text{\small Frame 1}
\end{equation}
with $\vec m'=(m_3,\dots,m_N),\vec{n}'=(n_2,\dots,n_N)$. The remaining $N-1$ frames are obtained by exchanging $n_1$ with one of the parameters $n_{a\neq1}$, with the same exchange made in the delta-function support, residual contributions, and measure correction. Each frame realises the same breaking $SU(N)_{\vec n} \rightarrow SU(N-1)_{\vec n'}$.

The independent calculations in 5d \eqref{eq:Strip211_broken} and in 3d \eqref{eq:3dbroken_strip211} can now be compared using the 5d/3d dictionary \eqref{eq:dict}. We compare the first frame explicitly; Weyl symmetry extends the matching to the remaining frames.

\begin{table}[h!]
    \centering
    \setlength{\tabcolsep}{8pt}
    \renewcommand{\arraystretch}{1.4}

    \begin{tabular}{>{\centering\arraybackslash}m{0.46\linewidth} | >{\centering\arraybackslash}m{0.46\linewidth}}
        \textbf{5d} & \textbf{3d} \\ \hline
        $\Delta_{\rm 5d}^{\rm corr}$ & $\Delta_{\rm 3d}^{\rm corr}$ \\
        $\bm\delta_{\rm pert}^{\rm Ex}$ & $\bm\delta_{\rm Ex}$ \\
        \includegraphics[width=\linewidth]{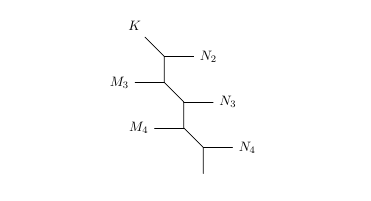} &
        \includegraphics[width=\linewidth]{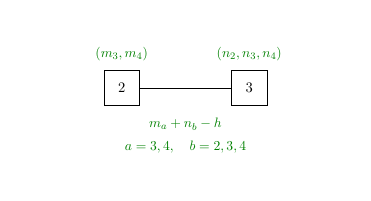}
    \end{tabular}
    \caption{Comparison between representative frames of the five-dimensional broken strip and its three-dimensional magnetic quiver after the $[2,1^{N-2}]$ nilpotent deformation for the case $N=4$.}
    \label{tab:Comp_211}
\end{table}

\begin{itemize}
    \item \textit{Delta functions and measure correction}: Both results have $\bm\delta_{\rm Nil}=\bm\delta_{\rm pert}^{\rm Nil}=1$, so no non-trivial nilpotent VEV is imposed on the test theory. The surviving constraint is instead carried by $\bm\delta_{\rm Ex}$ and $\bm\delta_{\rm pert}^{\rm Ex}$. At $m=iQ/4$, the three-dimensional condition is $$n_1 = h-h'+iQ/4 ~,$$ precisely the additive counterpart of the 5d condition $$N_1 = \sqrt{\frac{\mf t}{\mf q}}\,\frac{K}{K'} ~.$$ The total $N=|\mathcal W_{SU(N)}|/|\mathcal W_{SU(N-1)}|$ frames are related by Weyl permutations. Upon gluing to a test theory, the constraints then break $SU(N)\longrightarrow SU(N-1)$ in each frame. The corresponding measures are corrected by $\Delta_{\rm 3d}^{\rm corr}$ and $\Delta_{\rm 5d}^{\rm corr}$. The 5d results additionally contain $\bm\delta_{\rm inst}$ which trivialises the partition associated with the frozen fugacity. This has no independent three-dimensional counterpart.
    \item \textit{Residual theory}: Both residual theories describe a free $SU(N-2)\times SU(N-1)$ bifundamental hypermultiplet. At $m=iQ/4$, its $(N-1)(N-2)$ components contribute
    $$\prod_{i=3}^N\prod_{j=2}^Ns_b\left(\frac{iQ}{4}\pm(m_i+n_j-h)\right) \;\;\text{ in 3d} \;,\;\;\text{and}\quad \prod_{i=3}^N\prod_{j=2}^N\left|\MM{\frac{M_iN_j}{K}}\right|^{-2} \;\;\text{ in 5d ,}$$ where the latter is the contribution obtained after setting external partitions to be empty. The arguments of each factor can map into the other under the 5d/3d dictionary \eqref{eq:dict}.
    \item \textit{Goldstone modes}: The original nilpotent VEV is labelled by the same pair $(SU(N),\rho)$ in both descriptions. Their common spectrum and charges follow from the broken-generator decomposition reviewed in Appendix \ref{app:GM}. The two expressions ${\rm GM}_{\rm 3d}^{[2,1^{N-2}]}$ and ${\rm GM}_{\rm 5d}^{[2,1^{N-2}]}$ implement this using the respective 5d/3d building blocks, with their parameters related by the dictionary.
\end{itemize}

The comparison therefore reproduces the complete data controlling the broken configuration: the delta-function support, the unbroken subgroup, the number of symmetry-related frames, the residual free sector, the Goldstone spectrum, and the reduction of the gluing measure in both five and three dimensions. The matching in the case $N=4$ is summarised in Table \ref{tab:Comp_211}.

\subsection{Breaking the strip with the $[N-1,1]$ VEV}
\label{subsec:stripminimal}

We now continue the analysis of nilpotent deformations of one side of the symmetric length-$N$ strip by turning on a $[N-1,1]$ VEV for the moment map associated with the left $SU(N)$ symmetry. We will show that this deformation produces a broken theory whose different frames are related by the Weyl group of the original $SU(N)$ symmetry. A new feature in this example is that the incoming $[N-1,1]$ VEV induces a second nilpotent condition of type $[N-2,1^2]$ on the opposite side of the strip. We assume $N\geq4$ since for $N=3$ the incoming partition is $[2,1]$ and the analysis is identical to that of the previous subsection.

\subsubsection{5d computation}
\label{subsubsubsec:stripminimal_5d}

The $[N-1,1]$ nilpotent VEV can be realised by the following tuning of mass parameters
\begin{equation}\label{eq:StripVEV_minimal}
    M_i = \left(\frac{\mf t}{\mf q}\right)^{\frac{N}{2}-i} K' d_i \;\;,\quad \text{ for } i=1,\dots,N-1 \;\;,\quad \text{ where }\; \prod_{i=1}^{N-1}d_i=1
\end{equation}
The parameters $d_i$ regulate the pole collisions, and the nilpotent limit is $d_i\to1$. As in the previous examples, these collisions should be interpreted after gluing the Higgsed strip to a test strip: the variables $\vec N$ are then integrated over, and the colliding poles pinch their integration contours. We therefore remove only the parallel leg contributions along the left edge, retaining the right edge contribution and the external partitions required for this gluing.

Unlike the deformation studied previously, this VEV produces a sequence of dependent pinchings: taking one residue alters the pole structure relevant to the next. We illustrate this mechanism explicitly for $N=4$ before presenting the general result. The case $N=3$ reduces to the $[2,1]$ deformation studied in the previous subsection.

\paragraph{The $N=4$ example:} For $N=4$, the tuning  produces the following three factors in the numerator that vanish in the limit $d_i\to1$:
\begin{equation}
    \left| \MM{\frac{\mf t}{\mf q}\frac{d_1}{d_2}}\MM{\left(\frac{\mf t}{\mf q}\right)^2\frac{d_1}{d_3}}\MM{\frac{\mf t}{\mf q}\frac{d_2}{d_3}} \right|^2 ~. \nonumber
\end{equation}

Unlike the $[2,1^{N-2}]$ case, these do not correspond to independent simple pinchings that can all be listed in advance, but rather must be studied sequentially: after choosing a first pinching, the pole structure of the corresponding residue determines the subsequent pinching.

We first select the branch selected by the colliding zeros of the denominator factors $\MM{M_1N_1/K}$ and
$\MM{\frac{\mf t}{\mf q}\frac{K}{M_2N_1}}$, using  \eqref{eq:Minv_identity} we find:
\[
N_1=\frac{\mf t}{\mf q}\frac{K}{M_1}
    =\frac{K}{K'd_1} \;\;,
\quad
N_1=\frac{K}{M_2}
    =\frac{K}{K'd_2} ~,
\]
These poles approach the gluing contour from opposite
sides and collide as $d_1/d_2\to1$
freezing
\begin{equation}
    N_1 = \frac{K}{K'd_1} ~.
\end{equation}

After taking this first residue, the corresponding contribution to the perturbative part becomes
\begin{equation}\label{eq:StripMin_Inter}
    \begin{split}
        & \lim_{d_i\rightarrow1}\left|\left.\cZ_{\rm pert}^{\rm strip} (\vec M,\vec N,K)\right|_{M_i\rightarrow[3,1]_{K'}\,,\;N_1\rightarrow \frac{K}{K'd_1}} \right|^2 \\
        &\hspace{2cm} =  \; \delta\left(\frac{K}{K'N_1}\right) \; \left| \frac{ \prod_{2\leq a<b\leq 4} \MM{\frac{\mf t}{\mf q}\frac{N_b}{N_a}} }{ M(\mf t,\mf q) \times \prod_{a=2}^4 \MM{\frac{\mf t}{\mf q}\frac{K}{M_4N_a}} } \right|^2 \\
        &\hspace{2.5cm} \times \; \lim_{d_i\to1} \left| \frac{ \MM{\frac{\mf t}{\mf q}\frac{d_2}{d_3}} \MM{\frac{K'\,d_2}{M_4}} \MM{\frac{\mf q}{\mf t}\frac{K'\,d_3}{M_4}} }{ \prod_{2\leq a\leq b\leq 3} \MM{\left(\frac{\mf t}{\mf q}\right)^{b-1} \frac{K}{K'\,N_a\,d_b}} \MM{\left(\frac{\mf t}{\mf q}\right)^{2-a} \frac{K'\,N_{b+1}\,d_a}{K}} } \;\right|^2 ~.
    \end{split}
\end{equation}
The first pinching uses the zero associated with $M_1/M_2\to \frac{\mf t }{\mf q}\frac{d_1}{ d_2}$, while the factor involving the zero $ 
M_1/M_3\to \frac{\mf t^2 }{\mf q^2}\frac{d_1}{ d_3}$
cancels within the residue. 
The remaining zero
$\MM{M_2/M_3}$ accompanies the collision of the denominator poles
that pinch the $N_2$ contour as $d_2/d_3\to1$. Choosing the branch
\begin{equation}
    N_2 = \frac{\mf t}{\mf q}\frac{K}{K'd_2} ~,
\end{equation}
and taking the corresponding residue gives
\begin{equation}\label{eq:Strip_31_broken_pert2}
    \begin{split}
        & \lim_{d_i\rightarrow1}\left|\cZ_{\rm pert}^{\rm strip} (\vec M,\vec N,K) \right|^2_{M_i\rightarrow[3,1]_{K'},\,N_1\rightarrow \frac{K}{K'\,d_1}\;,\,N_2\rightarrow \frac{\mf t}{\mf q}\frac{K}{K'\,d_2}} \\
        &\qquad = \delta\left(\frac{K}{K'N_1}\right) \delta\left(\frac{\mf t}{\mf q}\frac{K}{K'N_2}\right) \left| \frac{ \MM{\frac{\mf q}{\mf t}\frac{K'}{M_4}} \MM{\frac{\mf t}{\mf q}\frac{N_4}{N_3}} }{ M(\mf t,\mf q)^2 \times \prod_{a=3}^4 \MM{\frac{\mf t}{\mf q}\frac{K}{M_4N_a}} \MM{\frac{\mf q}{\mf t}\frac{K'N_a}{K}} } \right|^2 ~.
    \end{split}
\end{equation}
Thus, this branch freezes $N_1$ and $N_2$ through two successive pinchings, with the second determined by the pole structure remaining after the first residue.

This branch is only one representative. More generally, the first pinching can involve any $N_a$, while the second can involve any distinct $N_b$. The frames are therefore labelled by ordered pairs
\begin{equation}\label{eq:polepinching_minimal}
    (a,b) \;:\quad N_a = \frac{K}{K'} \quad,\quad  N_b = \frac{\mf t}{\mf q}\frac{K}{K'} \;,\quad a\neq b.
\end{equation}
For $N=4$, there are $4\times3=12$ such frames. They are related by the Weyl permutations of the original $SU(4)$ flavour fugacities. Removing the left parallel leg contributions given in \eqref{eq:parallel_D5}, the perturbative contribution becomes
\begin{equation}\label{eq:Stripmin_broken_pert}
    \boxed{\begin{split}
        &\left| \frac{ \cZ_{\rm pert}^{\rm strip} (\vec M, \vec N, K) }{ \cZ_{||}^{\rm left}(\vec M;\bm\emptyset) } \right|^2_{\vec M\to[3,1]_{K'}} = \delta\left( \frac{K}{K' N_1} \right) \delta\left( \frac{\mf t}{\mf q}\frac{K}{K' N_2} \right) \\
        &\qquad\quad \times \; \left| \MM{\frac{\mf t}{\mf q}}^2 \MM{\frac{\mf t^2}{\mf q^2}} \MM{\frac{K'}{M_4}} \MM{\frac{\mf t}{\mf q}\frac{K'}{M_4}} \right|^{-2} \\
        &\quad \times \; \left| \frac{ \MM{\frac{\mf t}{\mf q}\frac{N_4}{N_3}} }{ M(\mf t,\mf q)^2 \times \MM{\frac{\mf q}{\mf t}\frac{K'N_3}{K}} \MM{\frac{\mf q}{\mf t}\frac{K'N_4}{K}} } \right|^2 \times \; \left| \MM{\frac{\mf t}{\mf q}\frac{K}{M_4N_3}} \MM{\frac{\mf t}{\mf q}\frac{K}{M_4N_4}} \right|^{-2} \\[1.5mm]
        &\qquad + \text{ $11 \;$ permutations,}
    \end{split}}
\end{equation}
The first inverse modulus-squared factor is the Goldstone contribution associated with the incoming $[3,1]$ VEV. The two delta functions freeze two out of the four $SU(4)$ fugacities and therefore reduce the symmetry gauged upon gluing from $SU(4)$ to $SU(2)$. Correspondingly, the number of frames is $\frac{|\mathcal W_{SU(4)}|}{|\mathcal W_{SU(2)}|}=\frac{4!}{2!}=12$.

The instantonic contribution simplifies on the same support. In particular, we find
\begin{equation}
    \begin{split}
        \left.\lim_{d_i\rightarrow1}\cZ_{\bm\emptyset,\bm\alpha}^{\rm strip} (\vec M,\vec N,K)\right|_{M_i\rightarrow[3,1]_{K'},\, N_1\rightarrow \frac{K}{K'd_1},\, N_2\rightarrow \frac{\mf t}{\mf q}\frac{K}{K'd_2}} &= \cN_{\alpha_1\emptyset}\left(\frac{\mf t}{\mf q}\right) \; \cN_{\alpha_2\emptyset}\left(\frac{\mf t}{\mf q}\right) \times (\cdots) \\
        &= \delta_{\alpha_1,\emptyset} \delta_{\alpha_2,\emptyset} \times (\cdots) ~,
    \end{split}
\end{equation}
where the second equality follows from \eqref{eq:Ntrivial_identity}. The partitions associated with the two frozen fugacities are therefore empty. For the representative frame, the instantonic contribution becomes
\begin{equation}\label{eq:Strip_31_broken_inst}
    \begin{split}
        &\left.\lim_{d_i\rightarrow1}\cZ_{\bm\emptyset,\bm\alpha}^{\rm strip} (\vec M,\vec N,K)\right|_{M_i\rightarrow[3,1]_{K'},\, N_1\rightarrow \frac{K}{K'd_1},\, N_2\rightarrow \frac{\mf t}{\mf q}\frac{K}{K'd_2}} \\[0.15cm]
        &\quad = \delta_{\alpha_1,\emptyset} \; \delta_{\alpha_2,\emptyset} \; \times \; \prod_{a=3}^4 \Big[ \mf q^{\frac{||\alpha_a||^2}{2}} \widetilde Z_{\alpha_a}(\mf t,\mf q) \Big] \times \; \frac{ \cN_{\alpha_3\emptyset}\left(\frac{\mf t}{\mf q}\frac{K}{M_4N_3}\right) \cN_{\emptyset\alpha_4}\left(\frac{M_4N_4}{K}\right) \prod_{a=3}^4 \cN_{\emptyset\alpha_a}\left(\frac{\mf q}{\mf t}\frac{K'\,N_a}{K}\right) }{ \cN_{\alpha_3\alpha_4}\left(\frac{\mf t}{\mf q}\frac{N_4}{N_3}\right) } ~.
    \end{split}
\end{equation}
Combining these ``perturbative" and ``instantonic" contributions gives
\begin{equation}\label{eq:Strip31_broken}
    \boxed{\begin{split}
        &\left| \frac{ Z^{\rm strip} (\vec M, \vec N, K;\bm\emptyset,\bm\alpha) }{ \cZ_{||}^{\rm left}(\vec M;\bm\emptyset) } \right|^2_{\vec M\to[3,1]_{K'}} = \delta\left( \frac{K}{K'N_1} \right) \times \delta\left( \frac{\mf t}{\mf q}\frac{K}{K'N_2} \right) \\
        &\qquad \times \; \left| \MM{\frac{\mf t}{\mf q}}^2 \MM{\frac{\mf t^2}{\mf q^2}} \MM{\frac{K'}{M_4}} \MM{\frac{\mf t}{\mf q}\frac{K'}{M_4}} \right|^{-2} \\
        &\qquad \times \; \left| \frac{1}{M(\mf t,\mf q)^2} \times \frac{\cM\left( \frac{\mf t}{\mf q} \frac{N_4}{N_3} \right)}{\cN_{\alpha_3\alpha_4}\left( \frac{\mf t}{\mf q} \frac{N_4}{N_3} \right)} \times \prod_{a=3}^4 \frac{\cN_{\emptyset\alpha_a}\left( \frac{\mf q}{\mf t}\frac{N_aK'}{K} \right)}{\cM\left( \frac{\mf q}{\mf t}\frac{N_aK'}{K} \right)} \right|^2 \\[1.5mm]
        &\qquad \times \; \left| \; \delta_{\alpha_1,\emptyset} \; \delta_{\alpha_2,\emptyset} \;\times\; \prod_{b=3}^4 \Big[ \mf q^{\frac{||\alpha_b||^2}{2}} \widetilde Z_{\alpha_b}(\mf t,\mf q) \Big] \frac{ \cN_{\alpha_3\emptyset}\left( \frac{\mf t}{\mf q}\frac{K}{M_4N_3} \right) \cN_{\emptyset\alpha_4}\left( \frac{M_4N_4}{K} \right) }{ \MM{\frac{\mf t}{\mf q}\frac{K}{M_4N_3}} \MM{\frac{M_4N_4}{K}} }  \right|^2 \\[1.5mm]
        &\qquad\quad + \text{ $11$ permutations ,}
    \end{split}}
\end{equation}
This is the $N=4$ specialisation of the general result derived below.

\paragraph{General case: $N\geq4$ :}

The sequential pinching mechanism demonstrated in the $N=4$ example can be understood uniformly from the pole structure of the ``perturbative" part of the strip partition function. For each pair of adjacent masses $M_a,M_{a+1}$ belonging to the Jordan block, the perturbative contribution \eqref{eq:Zstrip_pert} contains, up to factors that remain regular in the degeneration, the contribution
\begin{equation}\label{eq:local_pinching_Nm11}
    \left| \frac{ \MM{\frac{M_a}{M_{a+1}}} }{ \MM{\frac{M_aN_a}{K}}\,\MM{\frac{\mf t}{\mf q}\frac{K}{M_{a+1}N_a}} } \right|^2 ~.
\end{equation}

Using \eqref{eq:Minv_identity}, the second denominator in
\eqref{eq:local_pinching_Nm11} can equivalently be written, within the
modulus square, as $\MM{M_{a+1}N_a/K}$. The first denominator then has a
pole of type $(P)$ at $N_a=\frac{\mf t}{\mf q}\frac{K}{M_a}$ while the second has a pole of type $(\bar P)$ at $N_a=\frac{K}{M_{a+1}}$.
Under the nilpotent specialisation
$M_a/M_{a+1}\to\mf t/\mf q$, these poles approach the same point from
opposite sides of the gluing contour. The numerator vanishes in the
same limit. Taking the residue at the pinched contour produces the
corresponding periodic delta function.

After taking the corresponding residue, the remaining expression contains the same structure for the next adjacent pair in the Jordan block. Iterating this procedure gives the representative sequence of pinchings with support
\begin{equation}\label{eq:local_support_Nm11}
    N_a = \frac{\mf t}{\mf q}\frac{K}{M_a} = \left(\frac{\mf t}{\mf q}\right)^{a+1-\frac{N}{2}}
    \frac{K}{K'd_a} \;\;,\quad a=1,\ldots,N-2 ~.
\end{equation}
More generally, the frozen variables can be chosen among the $N$ parameters $\vec N$, and thus the frames are labelled by ordered $(N-2)$-tuples of distinct indices
\begin{equation}
    (a_1,\dots,a_{N-2}) \;:\quad N_{a_i} = \left(\frac{\mf t}{\mf q}\right)^{i+1-\frac{N}{2}}\frac{K}{K'} \;,\quad a_i\neq a_j \; \text{ for all } i\neq j ~,
\end{equation}
giving $N\times(N-1)\times\cdots\times3=\frac{N!}{2!}=\frac{|\mathcal W_{SU(N)}|}{|\mathcal W_{SU(2)}|}$ frames. 
The $N-2$ delta functions reduce the $SU(N)$ symmetry gauged upon gluing to $SU(2)$, consistent with this Weyl group quotient. After decoupling the external parallel leg contributions along the left edge, we find:
\begin{equation}\label{eq:Strip31_broken_gen}
    \boxed{\begin{split}
        &\left| \frac{ Z^{\rm strip} (\vec M, \vec N, K;\bm\emptyset,\bm\alpha) }{ \cZ_{||}^{\rm left}(\vec M;\bm\emptyset) } \right|^2_{\vec M\to[N-1,1]_{K'}} = \prod_{a=1}^{N-2} \delta\left( \left(\frac{\mf t}{\mf q}\right)^{a+1-\frac{N}{2}}\frac{K}{K' N_a} \right) \\[1.5mm]
        &\quad \times \; \prod_{a=1}^{N-2} \; \left| \; \MM{\frac{\mf t^a}{\mf q^a}}^{N-a-1} \MM{ \left(\frac{\mf t}{\mf q}\right)^{\frac{N}{2}-a} \frac{K'}{M_N} } \; \right|^{-2} \\
        &\quad \times \; \left| \frac{1}{M(\mf t,\mf q)^{N-2}} \times \frac{\cM\left( \frac{\mf t}{\mf q} \frac{N_N}{N_{N-1}} \right)}{\cN_{\alpha_{N-1}\alpha_N}\left( \frac{\mf t}{\mf q} \frac{N_N}{N_{N-1}} \right)} \times \prod_{a=N-1}^N \frac{\cN_{\emptyset\alpha_a}\left( \left(\frac{\mf t}{\mf q}\right)^{1-\frac{N}{2}} \frac{N_aK'}{K} \right)}{\cM\left( \left(\frac{\mf t}{\mf q}\right)^{1-\frac{N}{2}} \frac{N_aK'}{K} \right)} \right|^2 \\[1.5mm]
        &\quad \times \; \left| \; \prod_{a=1}^{N-2} \delta_{\alpha_a,\emptyset} \;\times\; \prod_{b=N-1}^N \Big[ \mf q^{\frac{||\alpha_b||^2}{2}} \widetilde Z_{\alpha_b}(\mf t,\mf q) \Big] \frac{ \cN_{\alpha_{N-1}\emptyset}\left( \frac{\mf t}{\mf q}\frac{K}{M_NN_{N-1}} \right) \cN_{\emptyset\alpha_N}\left( \frac{M_NN_N}{K} \right) }{ \MM{\frac{\mf t}{\mf q}\frac{K}{M_NN_{N-1}}} \MM{\frac{M_NN_N}{K}} }  \right|^2 \\[1.5mm]
        &\qquad + \text{ $\frac{N!}{2!}-1$ permutations ,}
    \end{split}}
\end{equation}
Writing $\vec N'=(N_{N-1},N_N)$ and $\bm\alpha'=(\alpha_{N-1},\alpha_N)$, the different contributions can be organised according to the general broken theory formula \eqref{eq:broken_formula} as
\begin{equation}\label{eq:MinStrip_broken_gen}
    \left.
    \begin{split}
        \bm{\delta}_{\rm pert}^{\rm Nil} &= \prod_{a=1}^{N-2} \delta\left( \left(\frac{\mf t}{\mf q}\right)^{a+1-\frac{N}{2}}\frac{K}{K' N_a} \right) ~, \\
        \bm{\delta}_{\rm pert}^{\rm Ex} &= 1 ~, \\
        \bm{\delta}_{\rm inst} &= \prod_{a=1}^{N-2} \delta_{\alpha_a,\emptyset} ~, \\
        \frac{Z^{\rm residual}}{\cZ_{||}^{\rm residual}} &= \prod_{b=N-1}^N \Big[ \mf q^{\frac{||\alpha_b||^2}{2}} \widetilde Z_{\alpha_b}(\mf t,\mf q) \Big]  \frac{ \cN_{\alpha_{N-1}\emptyset}\left(\frac{\mf t}{\mf q}\frac{K}{M_NN_{N-1}}\right) \cN_{\emptyset\alpha_N}\left(\frac{M_NN_N}{K}\right) }{ \MM{\frac{\mf t}{\mf q}\frac{K}{M_NN_{N-1}}} \MM{\frac{M_NN_N}{K}} } ~, \\
        {\rm GM}_{\rm 5d}^{[N-1,1]} &= \prod_{a=1}^{N-2} \left| \; \MM{\left(\frac{\mf t}{\mf q}\right)^a}^{N-a-1} \MM{\left(\frac{\mf t}{\mf q}\right)^{\frac{N}{2}-a}\frac{K'}{M_N}} \; \right|^{-2} ~, \\
        \Delta^{\rm corr}_{\rm 5d} &= \frac{1}{{\rm GM}^{[N-2,1^2]}_{\rm 5d}\left(\vec N'{}^{-1}{},\sqrt{\frac{\mf q}{\mf t}}\frac{K'}{K};\bm\alpha'\right)} \times \left| \frac{ \cZ_{||}^{\rm right} (\vec N',\bm\alpha') \; \cZ_{||}^{\rm alt \; left} (\vec N',\bm\alpha') }{ M(\mf t,\mf q)^{N-2} \; \cZ_{||}^{\rm alt \; left} (\vec N,\bm\alpha) } \right|^2_{\bm\delta_{\rm pert}^{\rm Nil},\,\bm\delta_{\rm inst}} \\
        &= \left| \frac{1}{M(\mf t,\mf q)^{N-2}} \frac{ \MM{\frac{\mf t}{\mf q}\frac{N_N}{N_{N-1}}} }{ \cN_{\alpha_{N-1}\alpha_N}\left(\frac{\mf t}{\mf q}\frac{N_N}{N_{N-1}}\right) } \prod_{a=N-1}^N \frac{\cN_{\emptyset\alpha_a}\left( \left(\frac{\mf t}{\mf q}\right)^{1-\frac{N}{2}} \frac{N_aK'}{K} \right)}{\cM\left( \left(\frac{\mf t}{\mf q}\right)^{1-\frac{N}{2}} \frac{N_aK'}{K} \right)} \right|^2 ~.
    \end{split}
    \quad\right\vert\;\text{\small Frame 1}
\end{equation}
The external parallel leg contribution associated with the residual theory is simply $\cZ_{||}^{\rm residual} = \MM{\frac{\mf t}{\mf q}\frac{N_N}{N_{N-1}}}/\cN_{\alpha_{N-1}\alpha_N}\left(\frac{\mf t}{\mf q}\frac{N_N}{N_{N-1}}\right)$.

All the perturbative delta functions are of nilpotent type, while $\bm\delta_{\rm pert}^{\rm Ex}=1$. Indeed, the Dirac deltas in \eqref{eq:Strip31_broken_gen} impose a geometric progression constraint associated with a $[N-2,1^2]$ nilpotent VEV. Therefore the incoming $[N-1,1]$ deformation propagates across the strip and induces a $[N-2,1^2]$ nilpotent condition on the opposite $SU(N)$ symmetry.

The frozen fugacities appear in the reverse Weyl order relative to the convention used for the incoming VEV. To express the induced specialisation in that convention, we define
\begin{equation}
    \widehat N_b\equiv N_{N-1-b} \;\;,\quad K_{\rm out} \coloneq \sqrt{\frac{\mf t}{\mf q}}\frac{K}{K'} \;\;,\quad b=1,\ldots,N-2 ~.
\end{equation}
The conditions imposed by the delta functions then become
\begin{equation}
    \widehat N_b = \left(\frac{\mf t}{\mf q}\right)^{\frac{N-1}{2}-b} K_{\rm out} ~.
\end{equation}
This is precisely the standard geometric progression associated with a $[N-2,1^2]$ nilpotent VEV. The reversal of the ordering is a Weyl permutation and therefore does not affect the nilpotent orbit, but must be accounted for when tracking the parameter across successive strips, as we will see in Subsection \ref{subsec:general_NilHW}.

The $N-2$ delta functions break the $SU(N)$ gauge group upon gluing down to $SU(2)$, parameterised by $N_{N-1}$ and $N_N$. The factor $\Delta_{\rm 5d}^{\rm corr}$ reduces the gluing measure from $SU(N)$ to $SU(2)$ and cancels the Goldstone contribution generated by the induced VEV in the test strip.\footnote{The inverse Goldstone factor in \eqref{eq:MinStrip_broken_gen} removes the contribution generated by the induced VEV in the test strip. Since this strip uses the alternate parametrisation, its standard left edge variables are $\vec N'{}^{-1}$, and the parameter of the induced VEV is $K_{\rm out}^{-1}=\sqrt{\mf q/\mf t}\,K'/K$.} The residual theory in the general case remains a single free hypermultiplet in the fundamental representation of $SU(2)$. The general form of the incoming Goldstone contributions ${\rm GM}_{\rm 5d}^{[N-1,1]}$ can once again be easily computed using our prescription in Appendix \ref{app:GM}.

\begin{figure}[h!]
    \centering
    \includegraphics[width=0.66\linewidth]{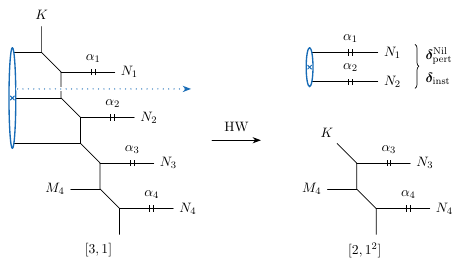}
    \caption{Five-brane realisation of the $[3,1]$ deformation of the length-four strip. After the Hanany--Witten transition, the web consists of a residual asymmetric strip engineering a free $SU(2)$ fundamental hypermultiplet and two semi-infinite D5 branes ending on the same D7 brane. The latter encode the two periodic delta functions imposing the induced $[2,1^2]$ nilpotent condition on the opposite side of the strip, as well as the associated discrete delta functions.}
    \label{fig:Braneweb_31}
\end{figure}

The five-brane interpretation provides a geometric explanation of this structure. The incoming $[N-1,1]$ VEV is implemented by terminating $N-1$ D5 branes on a common D7 brane. After a Hanany--Witten transition, the web decomposes into a residual asymmetric strip and $N-2$ semi-infinite D5 branes ending on the same D7 brane. The asymmetric strip engineers the free $SU(2)$ fundamental hypermultiplet, while the $N-2$ semi-infinite D5 branes account for the $N-2$ periodic delta functions. Since these branes end on the same D7 brane, their positions are correlated according to the geometric progression characteristic of the induced $[N-2,1^2]$ VEV. This explains geometrically why all the delta functions are of nilpotent type.

The different frames correspond to choosing and ordering the $N-2$ frozen D5 branes, or equivalently to choosing the pair of D5 branes that remains dynamical and assigning the distinct frozen values to the remaining branes. This gives precisely $N!/2!$ Weyl-related frames. The $N=4$ realization is shown in Figure \ref{fig:Braneweb_31}.

\subsubsection{3d counterpart and the 5d/3d map}
\label{subsubsubsec:stripminimal_3d}

\begin{figure}[h!]
    \centering
    \includegraphics[width=\linewidth]{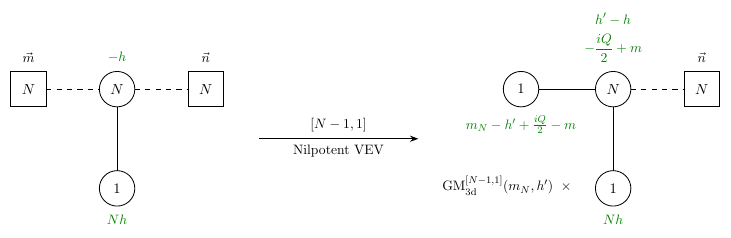}
    \caption{The $S_{1,N}$ magnetic quiver before and after the $[N-1,1]$ nilpotent deformation of one of its full $T[U(N)]$ tails. The deformation partially closes the tail to $T_{[N-1,1]}[U(N)]$ and produces the corresponding Goldstone sector. The $U(N)$ node of the deformed quiver sees $N+1$ effective flavours and is therefore bad.}
    \label{fig:3dStripMin}
\end{figure}

We now study the corresponding deformation of the $S_{1,N}$ magnetic quiver associated with the symmetric length-$N$ strip. The undeformed quiver and its nilpotent deformation are shown in Figure \ref{fig:3dStripMin}. 

The $\cN=2^*$-refined counterpart of the five-dimensional tuning in \eqref{eq:StripVEV_minimal} is
\begin{equation}
    m_j = h'+\left(N-1-2j\right)\left(\frac{iQ}{2}-m\right) \;\;,\quad \text{for } j=1,2,\dots,N-1 ~.
\end{equation}
This corresponds to a monopole VEV that partially closes the $T[U(N)]$ tail to a $T_{[N-1,1]}[U(N)]$ tail producing the quiver on the right of Figure \ref{fig:3dStripMin} together with the corresponding Goldstone contribution. The central $U(N)$ node has only $N+1$ effective flavours and is therefore bad. Since the opposite $T[U(N)]$ tail remains full, the deformed quiver belongs to the broken class described in Subsection \ref{subsec:MQbreakbad}.
 
Applying the electric dualisation algorithm on the deformed quiver gives a partition function of the general form in \eqref{eq:sknbrokengen}. For a representative frame, the different contributions are:\footnote{After reversing the order of the constrained masses, $\bm\delta_{\rm Nil}$ takes the form of the nilpotent specialisation used in Appendix \ref{app:GM}, with parameter $h-h'+iQ-2m$. This fixes the second argument of the induced Goldstone contribution and is consistent with the corresponding 5d Goldstone contribution in \eqref{eq:MinStrip_broken_gen}.}
\begin{equation}\label{eq:3dbroken_stripNm11}
    \left.
    \begin{aligned}
        \PFS{\text{residual}} &=\; \prod_{j=N-1}^{N} \left[ s_b\left(m\pm\left(m_N+n_j-h\right)\right)\right] ~, \\[0.69mm]
        \dd{\rm{Nil}} &=\; \prod_{j=1}^{N-2}\delta\left((N-2-2j)\left(\frac{iQ}{2}-m\right)+n_j+h'-h\right) ~, \\[1.1mm] 
        \dd{\rm{Ex}} &=\;1 ~, \\
        \Delta^{\rm corr}_{\rm 3d} &= \left.\frac{1}{{\rm GM}_{\rm 3d}^{[N-2,1^2]}(\vec{n}',h-h'+iQ-2m)}\frac{\DCB{SU(2)}\left(\vec{n}';m\right)}{\DCB{SU(N)}(\vec{n};m)} \right\vert_{\dd{\text{Nil}}} ~, \\
        &= \prod_{k=N-1}^{N}\;\prod_{l=-\frac{N-2}{2}}^{\frac{N-2}{2}} s_b\left(m+l\,(2m-iQ)\pm\left(m-\frac{iQ}{2}+h'-h+n_k\right)\right) \\
        &\qquad \times \; \prod_{p=1}^{N-2} s_b\left(\frac{iQ}{2}+p\,(2m-iQ)\right)^{N-1-p} \; \prod_{q=1}^{N-3} s_b\left(\frac{iQ}{2}-q\,(2m-iQ)\right)^{N-2-q} ~, \\[1.1mm]
        {\rm GM}_{\rm 3d}^{[N-1,1]} &= \prod_{j=1}^{N-2} \left[ s_b\left( m\pm(2j-1)\left( \frac{iQ}{2}-m \right) \right)^{N-j-1} \right. \\
        &\qquad \times \; \left. s_b\left( m+(N-2j-1)\left(\frac{iQ}{2}-m\right) \pm \left(h'-m_N-\left(\frac{iQ}{2}-m\right)\right) \right) \right] ~,
    \end{aligned}
    \right\vert\;\text{\small Frame 1}
\end{equation}
with $\vec n'=\{n_{N-1},n_N\}$ and the product over $l$ proceeds in unit steps. As in the previous subsection, the residual theory is entirely free. The removal of the overall decoupled $U(1)$, which is implicit in the definition of an interacting magnetic quiver, must therefore be implemented explicitly in applying the electric dualisation algorithm. The expression in \eqref{eq:3dbroken_stripNm11} is understood with this decoupled contribution removed according to the prescription of Section \ref{subsec:sknbreakingbad}. The remaining frames are obtained by permuting the $n_i$, with the same permutation acting on the delta function support, residual contribution, and measure correction. There are $N!/2!$ equivalent frames, in agreement with the breaking of the symmetry gauged upon gluing from $SU(N)$ to $SU(2)$.

We can now compare the 3d result in \eqref{eq:3dbroken_stripNm11} with its 5d counterpart in \eqref{eq:MinStrip_broken_gen} using the 5d/3d dictionary \eqref{eq:dict}. We display the comparison for one representative frame; Weyl symmetry extends the comparison to all remaining frames. A representative frame in the $N=4$ case is summarized in Table \ref{tab:Comp_31}.

\begin{table}[h!]
    \centering
    \setlength{\tabcolsep}{8pt}
    \renewcommand{\arraystretch}{1.4}

    \begin{tabular}{>{\centering\arraybackslash}m{0.46\linewidth} | >{\centering\arraybackslash}m{0.46\linewidth}}
        \textbf{5d} & \textbf{3d} \\ \hline
        $\Delta_{\rm 5d}^{\rm corr}$ & $\Delta_{\rm 3d}^{\rm corr}$ \\
        $\bm\delta_{\rm pert}^{\rm Nil}$ & $\bm\delta_{\rm Nil}$ \\
        \includegraphics[width=\linewidth]{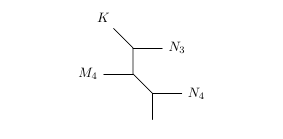} &
        \includegraphics[width=\linewidth]{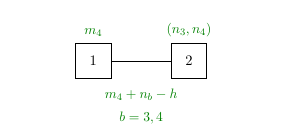}
    \end{tabular}
    \caption{Comparison between representative frames of the 5d and 3d broken theories produced by the $[N-1,1]$ deformation at $N=4$. The delta functions impose the induced nilpotent VEV $\rho'=[N-2,1^2]$ ($[2,1^2]$ here) on the test theory. The residual theory consists of a single free hypermultiplet in the fundamental representation of the unbroken $SU(2)$.}
    \label{tab:Comp_31}
\end{table}

\begin{itemize}
    \item \textit{Delta functions and measure correction}: Unlike the $[2,1^{N-2}]$ case, here we have $\dd{\rm{Ex}}=\bm\delta^{\rm Ex}_{\rm pert}=1$. Instead, the $N-2$ non-trivial constraints belong to the nilpotent sector. At the canonical $\mathcal{N}=4$ value $m=iQ/4$, the 3d constraints become
    \begin{equation}\label{eq:StripMin_3d_freeze}
        n_j = h - h' - (N-2-2j)\left(\frac{iQ}{4}\right) \;\;,\quad j=1,\dots,N-2 ~.
    \end{equation}
    Their 5d counterparts, $\bm\delta^{\rm Nil}_{\rm pert}$, impose: \begin{equation}\label{eq:StripMin_5d_freeze}
        N_j = \left(\frac{\mf t}{\mf q}\right)^{j+1-\frac{N}{2}}\frac{K}{K'} \;\;,\quad j=1,\dots,N-2 ~,
    \end{equation}
    These two sets of conditions impose the same induced nilpotent VEV $\rho'=[N-2,1^2]$ on a test theory and leave a residual $SU(2)$ gauging. This is consistent with the common multiplicity $\frac{|\mathcal W_{SU(N)}|}{|\mathcal W_{SU(2)}|}=\frac{N!}{2!}$ of equivalent frames. Accordingly, both $\Delta_{\rm 3d}^{\rm corr}$ and $\Delta_{\rm 5d}^{\rm corr}$ supply the correction from the original $SU(N)$ measure to the residual $SU(2)$ measure. They also cancel the Goldstone contributions generated by the induced $[N-2,1^2]$ VEV in the test theory. The 5d result additionally contains $\bm\delta_{\rm inst} = \prod_{a=1}^{N-2} \delta_{\alpha_a,\emptyset}$, which trivialises the partitions associated with the frozen variables. This discrete instantonic constraint has no independent counterpart in the three-dimensional partition function. 
    \item \textit{Residual theory}: On both sides, the residual theory is a single free hypermultiplet in the fundamental representation of the unbroken $SU(2)$. Its two components are labelled by the surviving parameters $\vec n'=\{n_{N-1},n_N\}$ in three dimensions and by $\vec N'=\{N_{N-1},N_N\}$ in five dimensions. At $m=iQ/4$, we have the contribution
    $$\prod_{j=N-1}^N s_b\left( \frac{iQ}{4}\pm(m_N+n_j-h) \right) \;\;\text{ in 3d} \;,\;\;\text{and}\quad \prod_{a=N-1}^N \left| \MM{\frac{M_N N_a}{K}} \right|^{-2} \;\;\text{ in 5d ,}$$
    where the latter is after setting external partitions to be empty. The 5d/3d dictionary \eqref{eq:dict} identifies the flavour fugacities of the two components.
    \item \textit{Goldstone modes}: The incoming nilpotent VEV is labelled by the same nilpotent orbit $\rho=[N-1,1]$ in both descriptions, so the corresponding symmetry-breaking patterns agree. The contributions ${\rm GM}_{\rm 3d}^{[N-1,1]}$ and ${\rm GM}_{\rm 5d}^{[N-1,1]}$ follow from the same broken-generator decomposition, using the respective 3d and 5d prescriptions reviewed in Appendix \ref{app:GM}.
\end{itemize}

\subsection{Propagation of nilpotent VEVs across a single strip}
\label{subsec:general_strip}

Having analysed the $[2,1^{N-2}]$ and $[N-1,1]$ examples explicitly, we now formulate the general rule governing the propagation of a nilpotent VEV across a single strip. We first consider hook-type partitions $[L,1^{N-L}]$, for which the complete distributional partition function can be written explicitly, and then extend the propagation rule to an arbitrary partition of $N$ by applying it to each non-trivial Jordan block. This local propagation rule will be applied to a two-strip box in Subsection \ref{subsec:brokenbox} and iterated across a general box in Subsection \ref{subsec:general_NilHW}.

\subsubsection*{Hook-type partitions}

A nilpotent VEV of hook type $\rho=[L,1^{N-L}]$ is implemented by the specialisation
\begin{equation}\label{eq:VEV_hook}
    M_a = \left(\frac{\mf t}{\mf q}\right)^{\frac{L+1}{2}-a}K'\,d_a \;\;,\quad a=1,\dots,L \;\;,\quad \prod_{a=1}^L d_a=1 ~,
\end{equation}
while the remaining parameters $M_{L+1},\ldots,M_N$ are left generic. The parameters $d_a$ regulate the pole collisions. In the limit $d_a\to1$, the modulus-squared refined topological string partition function can be organised according to the general decomposition \eqref{eq:broken_formula}. For $L\geq3$, a representative frame has the following contributions
\begin{equation}\label{eq:Strip_hook_gen}
    \left.
    \begin{split}
        \bm{\delta}_{\rm pert}^{\rm Nil} &= \prod_{a=1}^{L-1} \delta\left(   \left(\frac{\mf t}{\mf q}\right)^{a-\frac{L-1}{2}} \frac{K}{K' N_a} \right) ~, \\
        \bm{\delta}_{\rm pert}^{\rm Ex} &= 1 ~, \\[1.2mm]
        \bm{\delta}_{\rm inst} &= \prod_{a=1}^{L-1} \delta_{\alpha_a,\emptyset} ~, \\
        \frac{ Z^{\rm residual} }{ \cZ_{||}^{\rm residual} } &= \prod_{b=L}^N \Big[ \mf q^{\frac{||\alpha_b||^2}{2}} \widetilde Z_{\alpha_b}(\mf t,\mf q) \Big] \times \prod_{a=L+1}^N \left[ \prod_{b=L}^{a-1} \frac{ \cN_{\alpha_b\emptyset}\left( \frac{\mf t}{\mf q}\frac{K}{M_aN_b} \right) }{ \MM{\frac{\mf t}{\mf q}\frac{K}{M_aN_b}} } \prod_{b=a}^N \frac{ \cN_{\emptyset\alpha_b}\left(\frac{M_aN_b}{K}\right) }{ \MM{\frac{M_aN_b}{K}} } \right] ~, \\[1.2mm]
        {\rm GM}_{\rm 5d}^{[L,1^{N-L}]} &= \left| \prod_{a=1}^{L-1} \MM{\frac{\mf t^a}{\mf q^a}}^{L-a} \prod_{b=L+1}^N \prod_{a=1}^{L-1} \MM{\left(\frac{\mf t}{\mf q}\right)^{\frac{L+1}{2}-a}\frac{K'}{M_b}} \right|^{-2} ~, \\[1.2mm]
        \Delta^{\rm corr}_{\rm 5d} &= \frac{1}{{\rm GM}^{[L-1,1^{N-L+1}]}_{\rm 5d}\left(\vec N'\,{}^{-1},\sqrt{\frac{\mf q}{\mf t}}\frac{K'}{K};\bm\alpha'\right)} \\
        &\quad \times \left| \frac{1}{M(\mf t,\mf q)^{L-1}} \frac{ \cZ_{||}^{\rm right} (\vec N',\bm\alpha') \; \cZ_{||}^{\rm alt \; left} (\vec N',\bm\alpha') }{ \cZ_{||}^{\rm alt \; left} (\vec N,\bm\alpha) } \right|^2_{\bm\delta_{\rm pert}^{\rm Nil},\bm\delta_{\rm inst}} \\
        &= \left| \frac{1}{M(\mf t,\mf q)^{L-1}} \prod_{L\leq a<b\leq N} \frac{ \MM{\frac{\mf t}{\mf q}\frac{N_b}{N_a}} }{ \cN_{\alpha_a\alpha_b}\left( \frac{\mf t}{\mf q}\frac{N_b}{N_a} \right) } \times \prod_{a=L}^N \frac{ \cN_{\emptyset\alpha_a}\left( \left(\frac{\mf t}{\mf q}\right)^{\frac{1-L}{2}} \frac{N_aK'}{K}\right) }{ \MM{ \left(\frac{\mf t}{\mf q}\right)^{\frac{1-L}{2}} \frac{N_aK'}{K}} } \right|^2 ~,
    \end{split}
    \quad\right\vert\;\text{\small Frame 1}
\end{equation}
where $\vec M'=(M_{L+1},\dots,M_N)$, $\vec N'=(N_L,\dots,N_N)$, and $\bm\alpha'=(\alpha_L,\dots,\alpha_N)$. Note that the induced Goldstone contribution cancelled by the inverse factor in $\Delta_{\rm 5d}^{\rm corr}$ uses the alternate parametrisation of the strip. The contribution from external parallel D5 branes of the residual configuration is
\begin{equation}
    \cZ_{||}^{\rm residual} = \prod_{L+1\leq a<b\leq N} \MM{\frac{M_a}{M_b}} \prod_{L\leq a<b\leq N} \frac{ \MM{\frac{\mf t}{\mf q}\frac{N_b}{N_a}} }{ \cN_{\alpha_a\alpha_b}\left(\frac{\mf t}{\mf q}\frac{N_b}{N_a}\right) } ~,
\end{equation}
where we retain the non-trivial partitions $\bm\alpha'$ on the right edge.

The residual theory consists of a free hypermultiplet in the bifundamental representation of $SU(N-L)\times SU(N-L+1)$. It is engineered by the asymmetric strip with $N-L$ D5 branes on the left and $N-L+1$ D5 branes on the right.

The $L-1$ periodic delta functions freeze $L-1$ of the $SU(N)_{\vec N}$ parameters. Upon gluing the Higgsed strip to a test strip, they therefore reduce the gauged symmetry according to $SU(N)_{\vec N} \rightarrow SU(N-L+1)_{\vec N'}$. Correspondingly, the complete answer is a sum of $\frac{|\mathcal W_{SU(N)}|}{|\mathcal W_{SU(N-L+1)}|}=\frac{N!}{(N-L+1)!}$ Weyl-related frames. Before gluing, these frames are distinct distributions supported on different loci in the $\vec N$ parameter space. After gluing to a Weyl-invariant test theory, however, their contributions are related by permutations of the integration variables, with their multiplicity converting the original Weyl normalisation to that of the residual gauge group.

The correction $\Delta_{\rm 5d}^{\rm corr}$ converts the original $SU(N)$ gluing measure into the measure of the residual $SU(N-L+1)$ symmetry. It also removes the Goldstone contribution generated by the nilpotent VEV induced on the test strip. The discrete factor $\bm\delta_{\rm inst}$ simultaneously trivialises the partitions associated with the frozen fugacities.

The support of the perturbative delta functions is
\begin{equation}\label{eq:hook_outgoing_support}
    N_a = \left(\frac{\mf t}{\mf q}\right)^{a-\frac{L-1}{2}} \frac{K}{K'} \;\;,\quad
    a=1,\ldots,L-1 ~.
\end{equation}
The powers of $\mf t/\mf q$ increase with $a$, so the geometric progression appears in the reverse order relative to the incoming VEV convention. To put the outgoing progression in the standard ordering, we define
\begin{equation}\label{eq:Kout_hook}
    \widehat N_b\equiv N_{L-b} \;\;,\quad b=1,\ldots,L-1 \;\;,\quad \text{and} \quad K_{\rm out} \coloneq \sqrt{\frac{\mf t}{\mf q}}\,\frac{K}{K'} ~.
\end{equation}
The constraints in \eqref{eq:hook_outgoing_support} then become
\begin{equation}\label{eq:hook_outgoing_standard}
    \widehat N_b = \left(\frac{\mf t}{\mf q}\right)^{\frac{L}{2}-b} K_{\rm out} \;\;,\quad b=1,\ldots,L-1 ~.
\end{equation}
This is precisely the standard geometric progression associated with a $[L-1,1^{N-L+1}]$ nilpotent VEV. The reversal is only a Weyl permutation of the outgoing parameters and therefore leaves the nilpotent orbit unchanged. It must nevertheless be tracked in the parameter map when the propagation is iterated across several strips.

We can summarise the action of the strip on the partition as
\begin{equation}\label{eq:deltamap_hook_partition}
    \boxed{
    [L,1^{N-L}] \quad\longrightarrow\quad [L-1,1^{N-L+1}]
    }
\end{equation}
and on the corresponding parameters as
\begin{equation}\label{eq:deltamap_hook_strip}
    \boxed{
    \begin{aligned}
        M_a &= \left(\frac{\mf t}{\mf q}\right)^{\frac{L+1}{2}-a}K'\,d_a \;\;,\quad a=1,\dots,L ~, \\
        N_b &= \frac{\mf t}{\mf q}\,\frac{K}{M_b} = \left(\frac{\mf t}{\mf q}\right)^{b-\frac{L-1}{2}}\frac{K}{K'\,d_b} \;\;,\quad b=1,\dots,L-1 ~.
    \end{aligned}
    }
\end{equation}
This relation describes one representative frame. The remaining frames are obtained by Weyl permutations of the variables $\vec N$. At finite regulator values, the reversal also maps $d_a$ to inverse outgoing regulators; this distinction disappears in the nilpotent limit $d_a\to1$. The case $L=2$ obeys the same local propagation rule, but its single delta function belongs to $\delta_{\rm pert}^{\rm Ex}$.\footnote{For $L=2$, the propagated partition is $[1^N]$, corresponding to the trivial nilpotent orbit. Hence no nilpotent condition is imposed on the test strip, and the constraint produced by the pinching is classified as:
$$\bm\delta_{\rm pert}^{\rm Nil}=1 \quad\text{ and }\quad \bm\delta_{\rm pert}^{\rm Ex}=\delta\left( \sqrt{\frac{\mf t}{\mf q}} \frac{K}{K' N_1} \right) ~.$$}

\begin{figure}[h!]
    \centering
    \includegraphics[width=0.66\linewidth]{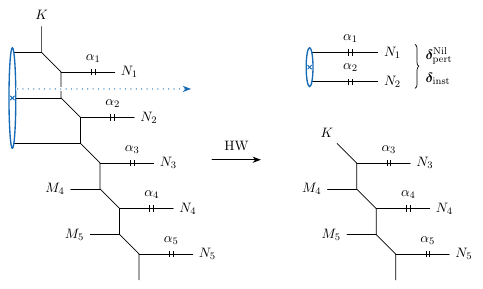}
    \caption{Propagation of a $[3,1^2]$ nilpotent VEV across a length-five strip. Three D5 branes initially end on a common D7 brane. After the Hanany--Witten transition, two semi-infinite D5 branes remain attached to the D7 brane and encode the outgoing $[2,1^3]$ nilpotent condition, while the residual asymmetric strip engineers the free bifundamental sector.}
    \label{fig:Braneweb_311}
\end{figure}

The five-brane web gives an immediate geometric realisation of the rule in \eqref{eq:deltamap_hook_strip}. The incoming $[L,1^{N-L}]$ VEV is realised by terminating $L$ D5 branes on a common D7 brane. Moving this D7 brane through the adjacent NS5 brane by a Hanany--Witten transition removes one of the attached D5 branes. The remaining $L-1$ semi-infinite D5 branes produce the outgoing delta functions and encode the $[L-1,1^{N-L+1}]$ specialisation. The unfrozen D5 branes form the residual asymmetric strip, with $N-L$ external D5 branes on the left that do not participate in the incoming VEV and the $N-L+1$ additional external D5 branes on the right. We illustrate this mechanism for $L=3$ and $N=5$ in Figure \ref{fig:Braneweb_311}.

\subsubsection*{General partitions of $N$}

The local propagation rule \eqref{eq:deltamap_hook_strip} extends naturally to an arbitrary partition: for generic choices of the VEV parameters, the pole collisions associated with different Jordan blocks can be treated independently.

Consider the following partition of $N$:
\begin{equation}\label{eq:general_partition_strip}
    Y_{\rm left} = [r_1,r_2,\ldots,r_s,1^f] \;\;,\quad r_1\geq r_2\geq\cdots\geq r_s\geq2 \;\;,\quad \sum_{i=1}^{s}r_i+f=N ~,
\end{equation}
We associate a distinct centre parameter $K_i'$ with the $i$-th non-trivial Jordan block and label its masses by $M_{i,a}$, with $a=1,\ldots,r_i$. The nilpotent specialisation is
\begin{equation}\label{eq:general_partition_tuning}
    M_{i,a} = \left(\frac{\mf t}{\mf q}\right)^{\frac{r_i+1}{2}-a} K_i'd_{i,a} \;\;,\quad \prod_{a=1}^{r_i}d_{i,a}=1 ~.
\end{equation}

For a representative frame, the $i$-th block produces the constraints
\begin{equation}\label{eq:general_block_propagation}
    N_{i,a} = \frac{\mf t}{\mf q}\frac{K}{M_{i,a}} = \left(\frac{\mf t}{\mf q}\right)^{a-\frac{r_i-1}{2}}
    \frac{K}{K_i'd_{i,a}} \;\;,\quad a=1,\ldots,r_i-1 ~.
\end{equation}
As in the hook case, this progression appears in the opposite Weyl ordering. Defining
\begin{equation}
    \widehat N_{i,b} \coloneq N_{i,r_i-b} \;\;,\quad K_{i,\rm out} \coloneq \sqrt{\frac{\mf t}{\mf q}}\frac{K}{K_i'} \;\;,\quad b=1,\ldots,r_i-1 ~,
\end{equation}
and taking the nilpotent limit, we obtain
\begin{equation}
    \widehat N_{i,b} = \left(\frac{\mf t}{\mf q}\right)^{\frac{r_i}{2}-b} K_{i,\rm out} ~,
\end{equation}
which is the standard progression for a block of size $r_i-1$. 

Consequently, each non-trivial Jordan block loses one box as it propagates across the strip:
\begin{equation}
    [r_i]\longrightarrow[r_i-1] ~.
\end{equation}
Applying this independently to all blocks gives
\begin{equation}\label{eq:general_partition_map_strip}
    \boxed{
    Y_{\rm left} = [r_1,\ldots,r_s,1^f] \quad\longrightarrow\quad Y_{\rm right} = [r_1-1,\ldots,r_s-1,1^{s+f}]
    } ~,
\end{equation}
where the entries on the right-hand side are already in the non-increasing order due to \eqref{eq:general_partition_strip}. The nature of the associated delta functions is also determined block by block:
\begin{itemize}
    \item A block with $r_i\geq3$ produces $r_i-1$ delta functions that collectively impose a non-trivial block of size $r_i-1$ in $Y_{\rm right}$. These belong to $\bm\delta_{\rm pert}^{\rm Nil}$.
    \item A block with $r_i=2$ produces a single delta function. Since its propagated block has size one, it does not impose a non-trivial nilpotent VEV and is therefore of type $\bm\delta_{\rm pert}^{\rm Ex}$.
    \item Each of the original size-one blocks produces no delta function.
\end{itemize}
Altogether, the constraints freeze
\begin{equation}
    \sum_{i=1}^s (r_i-1) = N-s-f
\end{equation}
fugacities. Upon gluing to a test strip, the surviving gauge group is therefore $SU(s+f)$, and the number of Weyl-related frames is $N!/(s+f)!$. The residual strip has $f$ D5 branes on the left and $s+f$ on the right, giving $f(s+f)$ free hypermultiplet components when $f>0$.

\paragraph{The $[3,2]$ example.}

Consider the incoming partition $Y_{\rm left}=[3,2]$. The corresponding nilpotent specialisation is
\begin{equation}\label{eq:VEV_32}
    \begin{aligned}
        M_1 &= \frac{\mf t}{\mf q}K'd_1 \;\;,\quad M_2 = K'd_2 \;\;,\quad M_3 = \frac{\mf q}{\mf t}K'd_3 ~, \\
        M_4 &= \sqrt{\frac{\mf t}{\mf q}}K''d_4 \;\;,\quad M_5 = \sqrt{\frac{\mf q}{\mf t}}K''d_5 ~,
    \end{aligned}
\end{equation}
with $d_1d_2d_3=1$ and $d_4d_5=1$. The size-three block propagates to a size-two block and therefore produces two delta functions of nilpotent type, whereas the size-two block produces one additional delta function of Ex type: $[3,2]\longrightarrow[2,1^3]$.

\begin{figure}[h!]
    \centering
    \includegraphics[width=0.66\linewidth]{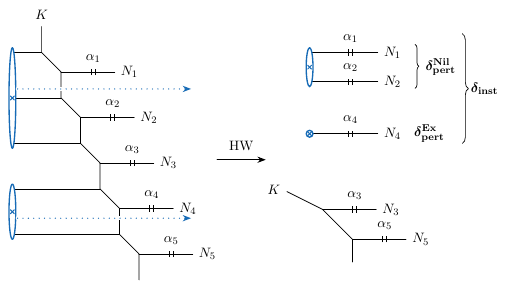}
    \caption{Propagation of a $[3,2]$ nilpotent VEV across a length-five strip. The size-three Jordan block produces two semi-infinite D5 branes and hence the outgoing $[2,1^3]$ nilpotent condition. The size-two block produces one additional semi-infinite D5 brane and therefore an excessive delta function. The resulting distribution is a sum of $60$ Weyl-related frames.}
    \label{fig:Braneweb_32}
\end{figure}

For a representative frame, the distributional partition function takes the general form \eqref{eq:broken_formula}, with
\begin{equation}\label{eq:broken32_gen}
    \left.
    \begin{split}
        \bm{\delta}_{\rm pert}^{\rm Nil} &= \delta\left(   \frac{K}{K' N_1} \right) \, \delta\left(   \frac{\mf t}{\mf q}\frac{K}{K'N_2} \right) ~, \\[1.2mm]
        \bm{\delta}_{\rm pert}^{\rm Ex} &= \delta\left(  \sqrt{\frac{\mf t}{\mf q}}\frac{K}{K'' N_4} \right) ~, \\[1.2mm]
        \bm{\delta}_{\rm inst} &= \delta_{\alpha_1,\emptyset} \; \delta_{\alpha_2,\emptyset} \; \delta_{\alpha_4,\emptyset} ~, \\[1.2mm]
        \frac{Z^{\rm residual}}{\cZ_{||}^{\rm residual}} &= \prod_{b=3,5} \Big[ \mf q^{\frac{||\alpha_b||^2}{2}} \widetilde Z_{\alpha_b}(\mf t,\mf q) \Big] \;  ~, \\
        {\rm GM}_{\rm 5d}^{[3,2]} &= \left| \MM{\frac{\mf t}{\mf q}}^3 \MM{\frac{\mf t^2}{\mf q^2}} \MM{\sqrt{\frac{\mf q}{\mf t}}\frac{K'}{K''}} \MM{\sqrt{\frac{\mf t}{\mf q}}\frac{K'}{K''}}^2 \MM{\sqrt{\frac{\mf t^3}{\mf q^3}}\frac{K'}{K''}} \right|^{-2} ~, \\[1.2mm]
        \Delta^{\rm corr}_{\rm 5d} &= \left[ \frac{1}{{\rm GM}^{[2,1^3]}_{\rm 5d}\left(N_{3,4,5}^{-1},\sqrt{\frac{\mf q}{\mf t}}\frac{K'}{K};\alpha_{3,4,5}\right)} \left| \frac{ \cZ_{||}^{\rm right} (\vec N',\bm\alpha') \; \cZ_{||}^{\rm alt \; left} (\vec N',\bm\alpha') }{ M(\mf t,\mf q)^3 \; \cZ_{||}^{\rm alt \; left} (\vec N,\bm\alpha) } \right|^2 \right]_{\bm\delta_{\rm pert},\bm\delta_{\rm inst}} \\
        &= \left| \frac{1}{M(\mf t,\mf q)^3} \times \frac{1}{\MM{\sqrt{\frac{\mf q}{\mf t}}\frac{K'}{K''}}} \times \frac{ \MM{\frac{\mf t}{\mf q}\frac{N_5}{N_3}} }{ \cN_{\alpha_3\alpha_5}\left( \frac{\mf t}{\mf q}\frac{N_5}{N_3} \right) } \prod_{a=3,5} \frac{ \cN_{\emptyset\alpha_a}\left( \frac{\mf q}{\mf t}\frac{N_aK'}{K} \right) }{ \MM{\frac{\mf q}{\mf t}\frac{N_aK'}{K}} } \right. \\
        &\qquad \times \left. \frac{ \cN_{\alpha_3\emptyset}\left( \sqrt{\frac{\mf t}{\mf q}}\frac{K}{N_3K''} \right) \cN_{\emptyset\alpha_5}\left( \sqrt{\frac{\mf q}{\mf t}}\frac{N_5K''}{K} \right) }{ \MM{\sqrt{\frac{\mf t}{\mf q}}\frac{K}{N_3K''}} \MM{\sqrt{\frac{\mf q}{\mf t}}\frac{N_5K''}{K}} } \right|^2 ~,
    \end{split}
    \quad\right\vert\;\text{\small Frame 1}
\end{equation}
where $\vec N'=(N_3,N_5)$, $\bm\alpha'=(\alpha_3,\alpha_5)$, and $\cZ_{||}^{\rm residual}=\MM{\frac{\mf t}{\mf q}\frac{N_5}{N_3}}/\cN_{\alpha_3\alpha_5}\left( \frac{\mf t}{\mf q}\frac{N_5}{N_3} \right)$ is the external parallel leg contribution associated with the residual configuration.

There are no residual hypermultiplets in this case. After removing the parallel leg contribution, the residual factor consists only of the external partition factor displayed above and becomes one when those partitions are empty. The Dirac deltas $\bm\delta^{\rm Nil}_{\rm pert}$ impose the propagated $[2,1^3]$ VEV associated with the original size-three block. The additional delta function of Ex type originates from the size-two block and freezes one of the parameters that would otherwise remain among the singlets of the propagated partition.\footnote{Accordingly, $N_4$ and $\alpha_4$ are retained among the arguments of the induced Goldstone contribution until the additional constraints imposed by the deltas are applied.} Consequently, only $N_3$ and $N_5$ remain dynamical upon gluing, and the original $SU(5)_{\vec N}$ measure is reduced to an $SU(2)_{\vec N'}$ measure. There are $ \frac{|\mathcal W_{SU(5)}|}{|\mathcal W_{SU(2)}|}= \frac{5!}{2!}=60$
equivalent frames.  

The five-brane interpretation is shown in Figure \ref{fig:Braneweb_32}. The three D5 branes associated with the first Jordan block end on one D7 brane and, after the Hanany--Witten transition, leave two semi-infinite D5 branes producing the nilpotent delta functions. The two D5 branes associated with the second block end on a different D7 brane and leave a single semi-infinite D5 brane, producing the delta function of Ex-type.

We have therefore obtained a local rule for propagating an arbitrary nilpotent VEV across a single strip: each non-trivial Jordan block loses one box, and the resulting geometric progression appears in the opposite Weyl ordering. Size-two blocks produce Ex-type rather than nilpotent delta functions. In Subsection \ref{subsec:brokenbox}, we verify explicitly how two successive applications of this rule describe the propagation of a $[3,1]$ VEV across the $4+SU(4)_0+4$ box. In Subsection \ref{subsec:general_NilHW}, we then iterate the same construction across the general $N+SU(N)_0^{k-1}+N$ box and use it to characterise the good, ugly, and broken regimes in five dimensions.

\subsection{Breaking the box}
\label{subsec:brokenbox}

We now apply the strip analysis of the previous subsections to the $4+SU(4)_0+4$ box theory. This provides the first example in which a nilpotent VEV propagates through more than one strip and produces a broken partition function with an interacting residual sector. We focus here on the $[3,1]$ nilpotent VEV and defer the general analysis to the following subsection.

\subsubsection{5d computation}
\label{subsubsubsec:box_5d}

The topological string partition function of the $4+SU(4)_0+4$ theory is obtained by gluing two length-four strips:
\begin{equation}\label{eq:SU4plus8_mod2}
    \begin{split}
        &\left| Z^{4+SU(4)+4} (\vec M, \vec N, \vec L, K_1, K_2, Q_{\rm inst} ; \bm\beta,\bm\gamma) \right|^2 \\[0.1cm]
        &\hspace{1.5cm} = \left| \, \sum_{\bm\alpha} \prod_{a=1}^4 Q_a^{|\alpha_a|} \; Z^{\rm strip} (\vec M, \vec N, K_1; \bm\beta,\bm\alpha) \; Z^{\rm alt \; strip} (\vec N,\vec L, K_2; \bm\alpha,\bm\gamma) \, \right|^2 \\
        &\hspace{1.5cm} = \left| \, \cZ_{\rm pert}^{\rm strip} (\vec M, \vec N, K_1) \, \right|^2 \; \left| \, \cZ_{\rm pert}^{\rm alt \; strip} (\vec N,\vec L, K_2) \, \right|^2 \\
        &\hspace{3cm} \times \left| \, \sum_{\bm\alpha} \prod_{a=1}^4 Q_a^{|\alpha_a|} \; \cZ_{\bm\beta,\bm\alpha}^{\rm strip} (\vec M, \vec N, K_1) \; \cZ_{\bm\alpha,\bm\gamma}^{\rm alt \; strip} (\vec N,\vec L, K_2) \, \right|^2 ~.
    \end{split}
\end{equation}
We impose a $[3,1]$ nilpotent VEV on the left flavour symmetry by specialising
\begin{equation}\label{eq:31VEV}
    M_1 = \frac{\mf t}{\mf q}K'\,d_1 \;\;,\quad M_2 = K'\,d_2 \;\;,\quad M_3 = \frac{\mf q}{\mf t}K'\,d_3 \;\;,\quad d_1d_2d_3=1 ~,
\end{equation}
and subsequently taking $d_a\rightarrow1$.

The action of this VEV on the first strip is the $N=4$ case of Subsection \ref{subsec:stripminimal}. The ``perturbative" part of this Higgsed strip takes the form
\begin{equation}\label{eq:31SQCD_pert}
    \begin{split}
        &\left| \cZ_{\rm pert}^{\rm strip}(\vec M\vert_{[3,1]}, \vec N, K_1) \right|^2 \, = \; \delta\left( \frac{K_1}{K'N_1} \right) \; \delta\left( \frac{\mf t}{\mf q}\frac{K_1}{K'N_2} \right) \; |M(\mf t,\mf q)|^{-4} \\
        &\hspace{3cm} \times \left| \, \frac{\MM{\frac{\mf q}{\mf t}\frac{K'}{M_4}}}{ \MM{\frac{\mf q}{\mf t}\frac{K'\,N_3}{K_1}} \MM{\frac{\mf q}{\mf t}\frac{K'\,N_4}{K_1}} } \frac{ \MM{\frac{\mf t}{\mf q}\frac{N_4}{N_3}} }{ \MM{\frac{\mf t}{\mf q}\frac{K_1}{M_4N_3}} \MM{\frac{M_4N_4}{K_1}} } \, \right|^2 \\[0.3cm]
        &\hspace{4cm} + \text{ 11 frames} ~.
    \end{split}
\end{equation}
On the support of the displayed frame, the corresponding instantonic contribution is
\begin{equation}\label{eq:31SQCD_inst}
    \begin{split}
        &\left.\cZ_{\bm\emptyset,\bm\alpha}^{\rm strip}(\vec M\vert_{[3,1]}, \vec N, K_1)\right|_{N_1\to \frac{K_1}{K'}, N_2\to\frac{\mf t}{\mf q}\frac{K_1}{K'}} \; = \; \delta_{\alpha_1,\emptyset} \; \delta_{\alpha_2,\emptyset} \times \prod_{b=3}^4 \Big[ \mf q^{\frac{||\alpha_b||^2}{2}} \widetilde Z_{\alpha_b}(\mf t,\mf q) \Big] \\
        &\hspace{2cm} \times \; \cN_{\emptyset\alpha_3}\left(\frac{\mf q}{\mf t}\frac{K'\,N_3}{K_1}\right) \cN_{\emptyset\alpha_4}\left(\frac{\mf q}{\mf t}\frac{K'\,N_4}{K_1}\right) \frac{ \cN_{\alpha_3\emptyset}\left(\frac{\mf t}{\mf q}\frac{K_1}{M_4N_3}\right) \cN_{\emptyset\alpha_4}\left(\frac{M_4N_4}{K_1}\right) }{ \cN_{\alpha_3\alpha_4}\left(\frac{\mf t}{\mf q}\frac{N_4}{N_3}\right) } ~,
    \end{split}
\end{equation}
The other eleven frames are obtained by simultaneously permuting the fugacities $N_a$ and their associated partitions $\alpha_a$.

This Higgsed strip naturally acts on the other strip $Z^{\rm alt \; strip}(\vec N, \vec L, K_2; \bm\alpha,\bm\gamma)$ via the gluing expressed in \eqref{eq:SU4plus8_mod2}. The delta functions generated by the first strip impose a $[2,1^2]$ nilpotent VEV on the variables $\vec N$. Indeed, their support can be written as
\begin{equation}\label{eq:211VEV}
    N_1 = \sqrt{\frac{\mf q}{\mf t}}\,K_{\rm out} \;\;,\quad N_2 = \sqrt{\frac{\mf t}{\mf q}}\,K_{\rm out} \;\;,\quad \alpha_1=\alpha_2=\emptyset ~,
\end{equation}
where
\begin{equation}\label{eq:Kout_box31}
    K_{\rm out} \coloneq \sqrt{N_1N_2} = \sqrt{\frac{\mf t}{\mf q}}\frac{K_1}{K'} ~.
\end{equation}
Thus $K_{\rm out}$ is not an additional parameter, but the centre of the geometric progression generated by the propagation of the original $[3,1]$ VEV through the first strip. The partitions $\alpha_3$ and $\alpha_4$ remain arbitrary and act as spectators for the degeneration.

The propagated $[2,1^2]$ VEV now acts on the second strip. Since this strip uses the alternate parametrisation, the induced VEV is implemented in its standard variables $N_a^{-1}$, with parameter $K_{\rm out}^{-1}$. Applying the $[2,1^{2}]$ result of Subsection \ref{subsubsec:Strip211_5d} then gives\footnote{The induced Goldstone contribution is ${\rm GM}_{\rm 5d}^{[2,1^2]}\left((N_3^{-1},N_4^{-1}),\sqrt{\frac{\mf q}{\mf t}}\frac{K'}{K_1};(\alpha_3,\alpha_4)\right)$. Its inverse appears in the correction $\Delta_{\rm 5d}^{\rm corr}$ supplied by the first strip in \eqref{eq:Strip31_broken_gen}.}
\begin{equation}
    L_1 = \sqrt{\frac{\mf t}{\mf q}}\frac{K_2}{K_{\rm out}} = \frac{K_2K'}{K_1} \;\;,\quad \gamma_1=\emptyset ~.
\end{equation}
After this second propagation, the resulting partition is trivial: $[2,1^2]\to[1^4]$, but one constraint remains supported on the external fugacities $\vec L$.

For concreteness, we now display the complete perturbative and instantonic contributions associated with this representative frame. The remaining frames are obtained by simultaneously permuting the fugacities $\vec L$ and their associated partitions $\gamma_a$. The perturbative contribution is
\begin{equation}\label{eq:211SQCD_pert}
    \begin{split}
        &\left| \cZ_{\rm pert}^{\rm alt \; strip}(\vec N\vert_{[2,1^2]}, \vec L, K_2) \right|^2 \, = \, \delta\left( \frac{K_2K'}{K_1L_1} \right) \times \left| \, \frac{ M(\mf t,\mf q)^{-1} }{ \MM{\frac{K_2K'}{L_2K_1}} } \times \prod_{a=3}^4 \frac{ \MM{\frac{\mf q}{\mf t}\frac{N_aK'}{K_1}} }{ \MM{\frac{K_2K'}{L_aK_1}} } \right. \\
        &\hspace{3cm} \times \; \left. \frac{ \MM{\frac{N_4}{N_3}} \prod_{2\leq a<b\leq4} \MM{\frac{\mf t}{\mf q}\frac{L_a}{L_b}} }{ \prod_{b=3}^4\left[ \prod_{a=2}^{b-1} \MM{\frac{L_aN_b}{K_2}} \prod_{a=b}^4 \MM{\frac{\mf t}{\mf q}\frac{K_2}{L_aN_b}} \right] } \, \right|^2 + \text{ 3 frames} ~,
    \end{split}
\end{equation}
The corresponding instantonic contribution, evaluated on the support of this delta function, is
\begin{equation}\label{eq:211SQCD_inst}
    \begin{split}
        &\left.\cZ_{\{\emptyset,\emptyset,\alpha_3,\alpha_4\},\bm\gamma}^{\rm alt \; strip}(\vec N\vert_{[2,1^2]}, \vec L, K_2)\right|_{L_1\to\frac{K_2K'}{K_1}} = \, \delta_{\gamma_1,\emptyset} \prod_{a=3}^4 \Big[ \mf t^{\frac{||\alpha_a^t||^2}{2}} \widetilde Z_{\alpha_a^t}(\mf q,\mf t) \Big] \prod_{b=2}^4 \Big[ \mf q^{\frac{||\gamma_b||^2}{2}} \widetilde Z_{\gamma_b}(\mf t,\mf q) \Big] \\
        &\qquad \times \; \frac{ \prod_{a=2}^4\cN_{\emptyset\gamma_a}\left(\frac{K_2K'}{L_aK_1}\right) }{ \prod_{a=3}^4\cN_{\emptyset\alpha_a}\left(\frac{\mf q}{\mf t}\frac{N_aK'}{K_1}\right) } \times \frac{ \prod_{b=3}^4\left[\prod_{a=2}^{b-1} \cN_{\gamma_a\alpha_b}\left(\frac{L_aN_b}{K_2}\right) \prod_{a=b}^4 \cN_{\alpha_b\gamma_a}\left(\frac{\mf t}{\mf q}\frac{K_2}{L_aN_b}\right)\right] }{ \cN_{\alpha_3\alpha_4}\left(\frac{N_4}{N_3}\right) \prod_{2\leq a<b\leq4} \cN_{\gamma_a\gamma_b}\left(\frac{\mf t}{\mf q}\frac{L_a}{L_b}\right) } ~,
    \end{split}
\end{equation}
The discrete delta function trivialises the partition associated with the frozen external leg.

The twelve frames produced by the first step differ by Weyl permutations of the internal variables $\vec N$. In the gauge integration implementing the gluing, the two internal delta functions saturate two integrals, and the twelve contributions become equivalent after relabelling the remaining integration variables. Their multiplicity converts the original Weyl normalisation $1/4!$ into $12/4!=1/2!$, as appropriate for the residual $SU(2)$. These branches therefore do not produce distinct distributions in the external parameters. The second propagation instead gives four possible supports for the delta function, corresponding to the four choices of the external parameter $L_a$ that is frozen. Upon gauging $SU(4)_{\vec L}$, the surviving periodic delta in \eqref{eq:211SQCD_pert} reduces the gauge group to $SU(3)$, consistently with the Weyl-group ratio $|\mathcal W_{SU(4)}|/|\mathcal W_{SU(3)}|=4$.

Combining the two strips, the factors associated with the intermediate $[2,1^2]$ Goldstone contribution cancel. We then divide by the contribution of the external parallel D5 branes on the left. The paired refined topological string partition function of the Higgsed box can then be expressed as:\footnote{In writing the result below, we have inserted the two-strip contribution
into the $SU(4)_{\vec N}$ gauge integral, performed the integrations fixed
by the internal delta functions (subject to the conditions in \eqref{eq:211VEV}), and combined the twelve internal Weyl
images. The expression displayed is the resulting integrand for the
remaining $SU(2)$ gauging; the four external frames remain distinct.}
\begin{equation}\label{eq:SU4plus8_broken}
    \boxed{\begin{split}
        &\left| \frac{ Z^{4+SU(4)+4} (\vec M, \vec N, \vec L, K_1, K_2,Q_{\rm inst} ; \bm\emptyset,\bm\gamma) }{ \cZ_{||}^{\rm left}(\vec M;\bm\emptyset) } \right|^2_{\vec M\to[3,1]_{K'}} \\[0.1cm]
        &\quad = \delta\left(\frac{K_2K'}{K_1L_1}\right) \times \left| \MM{\frac{\mf t}{\mf q}}^2\MM{\frac{\mf t^2}{\mf q^2}} \MM{\frac{K'}{M_4}} \MM{\frac{\mf t}{\mf q}\frac{K'}{M_4}} \right|^{-2} \\
        &\qquad \times \; \left| \frac{1}{M(\mf t,\mf q)} \times \prod_{2\leq a<b\leq4} \frac{ \MM{\frac{\mf t}{\mf q}\frac{L_a}{L_b}} }{ \cN_{\gamma_a\gamma_b}\left( \frac{\mf t}{\mf q}\frac{L_a}{L_b} \right) } \frac{ \cN_{\gamma_2\emptyset}\left(\frac{\mf t}{\mf q}\frac{K_1L_2}{K_2K'}\right) }{ \MM{\frac{\mf t}{\mf q}\frac{K_1L_2}{K_2K'}} } \times \prod_{a=3,4} \frac{ \cN_{\emptyset\gamma_a}\left(\frac{K_2K'}{L_aK_1}\right) }{ \MM{\frac{K_2K'}{L_aK_1}} } \right|^2 \\[0.1cm]
        &\qquad \times \left| \, \delta_{\gamma_1,\emptyset} \times \prod_{b=2}^4 \Big[ \mf q^{\frac{||\gamma_b||^2}{2}} \widetilde Z_{\gamma_b}(\mf t,\mf q) \Big] \times \sum_{\alpha_3,\alpha_4} \prod_{a=3}^4 Q_a^{|\alpha_a|} \; \prod_{a=3}^4 \Big[ \mf q^{\frac{||\alpha_a||^2}{2}} \mf t^{\frac{||\alpha_a^t||^2}{2}} \widetilde Z_{\alpha_a}(\mf t,\mf q) \widetilde Z_{\alpha_a^t}(\mf q,\mf t) \Big] \right. \\
        &\qquad\quad \times \; \frac{ \MM{\frac{N_4}{N_3}} \MM{\frac{\mf t}{\mf q}\frac{N_4}{N_3}} }{ \cN_{\alpha_3\alpha_4}\left(\frac{\mf t}{\mf q}\frac{N_4}{N_3}\right) \cN_{\alpha_3\alpha_4}\left(\frac{N_4}{N_3}\right) } \frac{ \cN_{\alpha_3\emptyset}\left(\frac{\mf t}{\mf q}\frac{K_1}{M_4N_3}\right) \cN_{\emptyset\alpha_4}\left(\frac{M_4N_4}{K_1}\right) }{ \MM{\frac{\mf t}{\mf q}\frac{K_1}{M_4N_3}} \MM{\frac{M_4N_4}{K_1}} } \\
        &\qquad\quad \left. \times \; \prod_{b=3}^4 \left[ \prod_{a=2}^{b-1} \frac{\cN_{\gamma_a\alpha_b}\left(\frac{L_aN_b}{K_2}\right)}{\MM{\frac{L_aN_b}{K_2}}} \prod_{a=b}^4 \frac{\cN_{\alpha_b\gamma_a}\left(\frac{\mf t}{\mf q}\frac{K_2}{L_aN_b}\right)}{\MM{\frac{\mf t}{\mf q}\frac{K_2}{L_aN_b}}} \right] \right|^2 \; + \text{ 3 frames.}
    \end{split}}
\end{equation}

\begin{figure}[t!]
    \centering
    \includegraphics[width=0.75\linewidth]{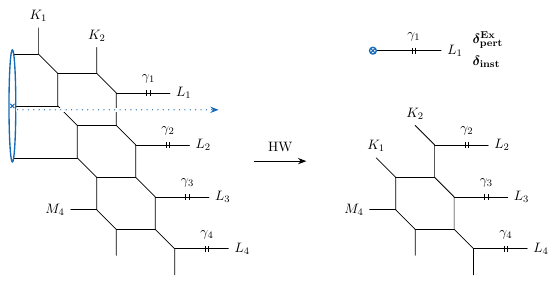}
    \caption{Hanany--Witten realisation of the $[3,1]$ nilpotent VEV in the $4+SU(4)_0+4$ box theory. After the Hanany--Witten moves, the residual asymmetric web engineers the interacting $1+SU(2)_0+3$ theory, while the isolated semi-infinite D5 brane accounts for the remaining delta function and the
    associated trivialisation of the external partition.}
    \label{fig:Braneweb_Box31}
\end{figure}

Since this is a broken theory, comparing this expression with the general formula in \eqref{eq:broken_formula} gives:
\begin{equation}\label{eq:SQCD_broken}
    \left.
    \begin{split}
        \bm{\delta}_{\rm pert}^{\rm Nil} &= 1 ~, \\
        \bm{\delta}_{\rm pert}^{\rm Ex} &= \delta\left(\frac{K_2K'}{K_1L_1}\right) ~, \\
        \bm{\delta}_{\rm inst} &= \delta_{\gamma_1,\emptyset} ~, \\
        \frac{Z^{\rm residual}}{\cZ_{||}^{\rm residual}} &= \prod_{b=2}^4 \Big[ \mf q^{\frac{||\gamma_b||^2}{2}} \widetilde Z_{\gamma_b}(\mf t,\mf q) \Big] \sum_{\alpha_3,\alpha_4} \prod_{a=3}^4 Q_a^{|\alpha_a|} \; \prod_{a=3}^4 \Big[ \mf q^{\frac{||\alpha_a||^2}{2}} \mf t^{\frac{||\alpha_a^t||^2}{2}} \widetilde Z_{\alpha_a}(\mf t,\mf q) \widetilde Z_{\alpha_a^t}(\mf q,\mf t) \Big] \\
        &\quad \times \frac{ \MM{\frac{N_4}{N_3}} \MM{\frac{\mf t}{\mf q}\frac{N_4}{N_3}} }{ \cN_{\alpha_3\alpha_4}\left(\frac{\mf t}{\mf q}\frac{N_4}{N_3}\right) \cN_{\alpha_3\alpha_4}\left(\frac{N_4}{N_3}\right) } \times \frac{\cN_{\alpha_3\emptyset}\left(\frac{\mf t}{\mf q}\frac{K_1}{M_4N_3}\right) \cN_{\emptyset\alpha_4}\left(\frac{M_4N_4}{K_1}\right)}{\MM{\frac{\mf t}{\mf q}\frac{K_1}{M_4N_3}} \MM{\frac{M_4N_4}{K_1}}} \\
        &\quad \times \; \prod_{b=3}^4 \left[ \prod_{a=2}^{b-1} \frac{\cN_{\gamma_a\alpha_b}\left(\frac{L_aN_b}{K_2}\right)}{\MM{\frac{L_aN_b}{K_2}}} \prod_{a=b}^4 \frac{\cN_{\alpha_b\gamma_a}\left(\frac{\mf t}{\mf q}\frac{K_2}{L_aN_b}\right)}{\MM{\frac{\mf t}{\mf q}\frac{K_2}{L_aN_b}}} \right] ~, \\
        {\rm GM}_{\rm 5d}^{[3,1]} &= \left| \MM{\frac{\mf t}{\mf q}}^2\MM{\frac{\mf t^2}{\mf q^2}} \MM{\frac{K'}{M_4}} \MM{\frac{\mf t}{\mf q}\frac{K'}{M_4}} \right|^{-2} ~, \\
        \Delta^{\rm corr}_{\rm 5d} &= \left| \frac{1}{M(\mf t,\mf q)} \frac{ \cZ_{||}^{\rm alt\;right}(\vec L',\bm\gamma') \cZ_{||}^{\rm left}(\vec L',\bm\gamma') }{ \cZ_{||}^{\rm left}(\vec L,\bm\gamma) } \right|^2 \\
        &= \left| \frac{1}{M(\mf t,\mf q)} \times \prod_{2\leq a<b\leq4} \frac{ \MM{\frac{\mf t}{\mf q}\frac{L_a}{L_b}} }{ \cN_{\gamma_a\gamma_b}\left( \frac{\mf t}{\mf q}\frac{L_a}{L_b} \right) } \prod_{a=2}^4 \frac{ \cN_{\emptyset\gamma_a}\left(\frac{K_2K'}{L_aK_1}\right) }{ \MM{\frac{K_2K'}{L_aK_1}} } \right|^2 ~. \\
    \end{split}
    \quad\right\vert\;\text{\small Frame 1}
\end{equation}
Here $\vec L'=(L_2,L_3,L_4)$ and $\bm\gamma'=(\gamma_2,\gamma_3,\gamma_4)$. The contribution from the external parallel D5 legs of the residual five-brane web is\footnote{The contributions from the parallel NS5 and $(1,-1)$ branes in the residual five-brane web (see Figure \ref{fig:Braneweb_Box31}) are inherited from the original top and bottom parallel leg factors. As in Subsections \ref{subsec:good} and \ref{subsec:ugly}, the Hanany--Witten transitions change the charges of the relevant external legs while preserving their separation. These contributions therefore remain included on both sides of Eqs.\ (\ref{eq:SU4plus8_broken}) and (\ref{eq:SQCD_broken}) and must be decoupled when extracting the physical index of the isolated residual theory.}
\begin{equation}
    \cZ_{||}^{\rm residual} = \prod_{2\leq a<b\leq4} \frac{ \MM{\frac{\mf t}{\mf q}\frac{L_a}{L_b}} }{ \cN_{\gamma_a\gamma_b}\left( \frac{\mf t}{\mf q}\frac{L_a}{L_b} \right) } ~.
\end{equation}

The residual contribution in \eqref{eq:SQCD_broken} can be identified with the partition function of the $1+SU(2)_0+3$ gauge theory: the sum over $\alpha_3,\alpha_4$ describes the residual $SU(2)_{(N_3,N_4)}$ gauge group, while $M_4$ parameterises the flavour on the left and $L_2,L_3,L_4$ the three flavours on the right. To compare with the magnetic quiver description, it is convenient to use the Hanany--Witten transitions discussed in Subsection \ref{subsubsec:5dugly}, which transport the D7 brane associated with $L_4$ from the right to the left of the web. This gives an equivalent presentation of the interacting theory, with its flavours distributed equally on both sides of the web.

The mass of the transported flavour is determined directly by matching its contribution before and after the move. Denoting the corresponding fugacity in the symmetric frame by $R_2$, we require
\begin{equation}
    \frac{R_2N_a}{K_1} = \frac{N_aL_4}{K_2} \;\;,\quad a=3,4 ~,
\end{equation}
which gives $R_2=L_4K_1/K_2$. The interacting residual theory therefore admits the symmetric $$2+SU(2)_0+2$$ presentation, with flavour fugacities
\begin{equation}
    \vec R = \left\{ M_4,\frac{L_4K_1}{K_2} \right\} \quad \text{ on the left }, \quad \vec S = \{ L_2,L_3 \} \quad \text{ on the right} ~.
\end{equation}
As in the ugly case, we bring these parameters to the canonical $SU(2)$ parametrisations before comparing with the magnetic quiver. The frozen values $N_1=K_1/K'$ and $N_2=\mf t/\mf q\,K_1/K'$ imply that $N_3N_4\neq1$. Likewise, the two components of each residual flavour vector do not have unit product: the original $SU(4)$ constraints give $M_4=K'^{-3}$ and $L_2L_3=(L_1L_4)^{-1}$. We therefore write
\begin{equation}\label{eq:box_NRS}
    \begin{split}
        \underline{\vec N}{}' &\coloneq \frac{\vec N'}{c_N} \;\;,\quad c_N = \sqrt{\frac{\mf q}{\mf t}}\frac{K'}{K_1} ~, \\
        \underline{\vec R} &\coloneq \frac{\vec R}{c_R} \;\;,\quad c_R = \sqrt{\frac{L_4K_1}{K'^3K_2}} \;\;,\quad \underline{\vec S} \coloneq \frac{\vec S}{c_S} \;\;,\quad c_S = \sqrt{\frac{K_1}{K'K_2L_4}} ~.
    \end{split}
\end{equation}
where $\vec N'=(N_3,N_4)$, and we redefine
\begin{equation}\label{eq:box_K}
    \underline{K}_1 \coloneq \frac{K_1}{c_Nc_R} = \sqrt{\frac{\mf t}{\mf q}\,\frac{K_1^3K_2K'}{L_4}} \;\;,\quad \underline{K}_2 \coloneq \frac{K_2}{c_Nc_S} = \sqrt{\frac{\mf t}{\mf q}\,\frac{K_1K_2^3L_4}{K'}} ~.
\end{equation}
These simultaneous rescalings preserve the matter arguments and impose the canonical unit-product constraints $\underline{N}_3\underline{N}_4=\underline{R}_1\underline{R}_2=\underline{S}_1\underline{S}_2=1$. The residual theory is thus the $2+SU(2)_0+2$ box in the canonical parametrisation with parameters $(\underline{\vec R},\underline{\vec S},\underline{K}_1,\underline{K}_2)$, with the associated strong-coupling magnetic quiver $S_{2,2}$.

The Goldstone contribution in \eqref{eq:SQCD_broken} is determined entirely by the original $[3,1]$ nilpotent orbit and its embedding into $SU(4)$. It is therefore unaffected by the subsequent propagation of the VEV through the second strip and coincides with the contribution already encountered in Subsection \ref{subsec:stripminimal}. It can be computed independently using the prescription reviewed in Appendix \ref{app:GM}.

The five-brane interpretation of the result is shown in Figure~\ref{fig:Braneweb_Box31}. Starting from the configuration on the left, a sequence of Hanany--Witten moves produces the residual asymmetric web together with an isolated semi-infinite D5 brane. The residual web engineers the interacting $1+SU(2)_0+3$ theory, whereas the isolated D5 brane accounts for the remaining distributional
constraint: the periodic delta function $\bm\delta_{\rm pert}^{\rm Ex}$ fixing one of the external flavour fugacities, together with the discrete constraint $\bm\delta_{\rm inst}$ that trivialises the corresponding partition. Since only one D5 brane terminates on the associated D7 brane, it cannot form a non-trivial Jordan block or, equivalently, define a non-trivial geometric progression. The remaining periodic delta function is therefore of the Ex-type. The resulting web provides a geometric realisation of the propagation
\begin{equation}
    [3,1]\longrightarrow[2,1^2]\longrightarrow[1^4] ~,
\end{equation}
where although the outgoing partition is trivial, the Ex-type delta function $\bm\delta_{\rm pert}^{\rm Ex}$ survives.

\subsubsection{3d counterpart and the 5d/3d map}
\label{subsubsubsec:box_3d}

\begin{figure}[h!]
    \centering
    \includegraphics[width=\linewidth]{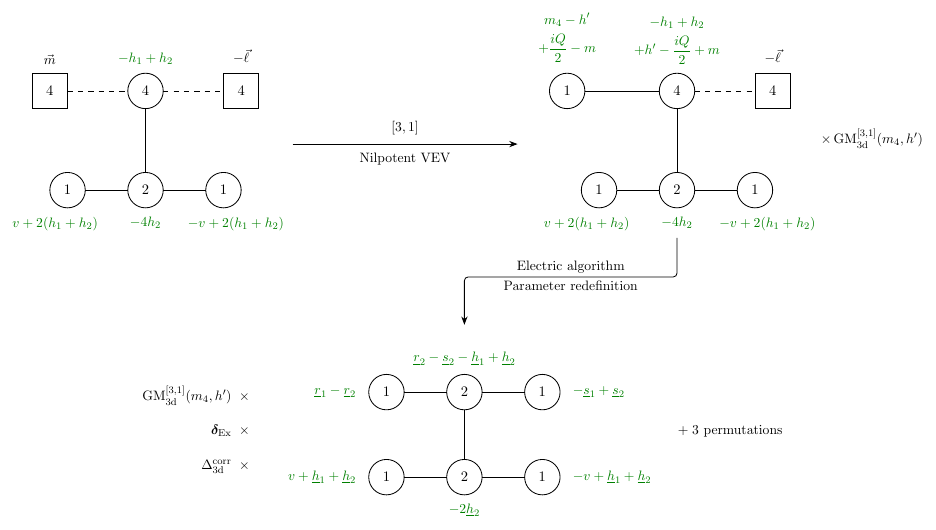}
    \caption{The $[3,1]$ nilpotent deformation of $S_{2,4}$. Partial closure of the left $T[U(4)]$ tail produces a bad quiver. Applying the electric dualisation algorithm gives four Weyl-related frames, each containing the interacting residual quiver $S_{2,2}$ in the canonical parametrisation, together with an Ex-type delta function and the accompanying measure correction factor.}
    \label{fig:Quiver_Box31}
\end{figure}

We now study the same deformation in the strong coupling 3d magnetic quiver $S_{2,4}$ associated with the 5d $4+SU(4)_0+4$ theory, shown at the top left of Figure \ref{fig:Quiver_Box31}. Using the $\cN=2^*$-refined version of the 5d/3d dictionary, the $[3,1]$ nilpotent VEV is implemented by
\begin{equation}
    m_1=h'+\frac{iQ}{2}-m \;\;,\quad m_2=h'-\frac{iQ}{2}+m \;\;,\quad m_3=h'-\frac{3iQ}{2}+3m ~.
\end{equation}
The associated monopole VEV partially closes the left $T[U(4)]$ tail to $T_{[3,1]}[U(4)]$ and produces the Goldstone contribution ${\rm GM}_{\rm 3d}^{[3,1]}$. The resulting quiver, shown at the top right of Figure \ref{fig:Quiver_Box31}, contains a bad central $U(4)$ node that sees six effective flavours. Applying the electric algorithm gives four Weyl-related frames, each containing the interacting residual theory shown at the bottom of Figure \ref{fig:Quiver_Box31}.

The partition function of this broken theory takes the general form \eqref{eq:sknbrokengen}, where for the first frame, we identify
\begin{equation}\label{eq: ex3d3}
\left.
\begin{aligned}
    \PFS{\text{residual}} &=\; \text{(Quiver in the bottom line of Figure \ref{fig:Quiver_Box31})} ~, \\[0.69mm]
    \dd{\rm{Nil}} &=\; 1 ~, \\[1.1mm]
    \dd{\rm{Ex}} &=\; \delta\left(-l_1+h'-h_1+h_2+m-\frac{iQ}{2}\right) ~, \\[2.1mm]
    \Delta^{\text{corr}}_{3d} &= \left.\frac{\DCB{SU(3)}\left(\vec{l}';m\right)}{\DCB{SU(4)}(\vec{l};m)} \right\vert_{\dd{\text{Ex}}} ~, \\
    &=\; s_b\left(2m-\frac{iQ}{2}\right) \times \prod_{j=2}^{4} \left[s_b\left(\frac{iQ}{2}\pm\left(-l_j+h'-h_1+h_2+m-\frac{iQ}{2}\right)\right) \right. \\
    &\qquad \times\; \left. s_b\left(2m-\frac{iQ}{2}\pm\left(-l_j+h'+m-\frac{iQ}{2}-h_1+h_2\right)\right)\right] ~, \\[1.1mm]
    {\rm GM}_{\rm 3d}^{[3,1]} &=\; s_b\left(2m-\frac{iQ}{2}\right)^2 s_b\left(\frac{iQ}{2}\right)^2 s_b\left(\frac{iQ}{2}\pm\left(h'+m-\frac{iQ}{2}-m_4\right)\right) \\
    &\qquad \times\; s_b\left(2m-\frac{iQ}{2}\pm\left(h'+m-\frac{iQ}{2}-m_4\right)\right) s_b\left(-2m+\frac{3iQ}{2}\right) s_b\left(4m-\frac{3iQ}{2}\right) ~,
\end{aligned}
\right\vert\;\text{\small Frame 1}
\end{equation}
where $\vec l'=(l_2,l_3,l_4)$. As in Section \ref{subsec:sknbreakingbad}, the overall $U(1)$ is understood to have been removed from the displayed interacting quiver.\footnote{As a consistency check, the FI parameter of this $U(1)$ vanishes on the support of $\bm\delta_{\rm Ex}$ in the inherited parametrisation.}

To compare the residual theory with its five-dimensional counterpart, we bring its inherited parameters to the canonical $S_{2,2}$ parametrisation. Writing the inherited tail parameters as $\vec r=(m_4,h_1-h_2+l_4)$ and $\vec s=(l_2,l_3)$, we define the following traceless parametrisation of the two $T[U(2)]$ tails:
\begin{equation}\label{eq:box_rund_sund}
    \begin{split}
        \underline r_i &\coloneq r_i - \frac{l_4+h_1-h_2-3h'+3(iQ/2-m)}{2} ~, \\[1mm]
        \underline s_i &\coloneq s_i - \frac{h_1-h_2-h'-l_4+iQ/2-m}{2} ~,
    \end{split}
\end{equation}
for $i=1,2$. Thus $\underline r_1+\underline r_2=\underline s_1+\underline s_2=0$. We also redefine
\begin{equation}\label{eq:box_hund}
    \underline{h}_1 = \frac32\,h_1 + \frac{h_2+h'-l_4-iQ/2+m}{2} \;\;,\quad
    \underline{h}_2 = \frac32\,h_2 + \frac{h_1-h'+l_4+iQ/2-m}{2} ~.
\end{equation}
These simultaneous redefinitions lead to the parametrisation of the residual quiver displayed in Figure \ref{fig:Quiver_Box31}, and are precisely the additive counterparts of Eqs.\ (\ref{eq:box_NRS}) and (\ref{eq:box_K}) at $m=iQ/4$.

We can now compare the three-dimensional broken theory in \eqref{eq: ex3d3} with its five-dimensional counterpart in \eqref{eq:SQCD_broken}. We display the comparison for one representative frame; Weyl symmetry extends it to the remaining three frames. This example is qualitatively distinct from the examples considered so far since the residual theory is interacting. Accordingly, we match the parametrisation of the interacting residual theories, along with the delta functions, the measure corrections, and the Goldstone sectors. This comparison is summarised in Table \ref{tab:Comp_Box31}.

\begin{table}[h!]
    \centering
    \setlength{\tabcolsep}{8pt}
    \renewcommand{\arraystretch}{1.4}

    \begin{tabular}{>{\centering\arraybackslash}m{0.46\linewidth} | >{\centering\arraybackslash}m{0.46\linewidth}}
        \textbf{5d} & \textbf{3d} \\ \hline
        $\Delta_{\rm 5d}^{\rm corr}$ & $\Delta_{\rm 3d}^{\rm corr}$ \\
        $\bm\delta_{\rm pert}^{\rm Ex}$ & $\bm\delta_{\rm Ex}$ \\
        \includegraphics[width=0.75\linewidth]{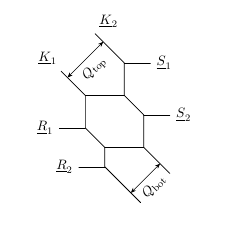} &
        \includegraphics[width=0.9\linewidth]{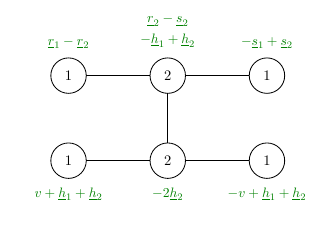}
    \end{tabular}
    \caption{Comparison between representative frames of the 5d and 3d broken theories produced by the $[3,1]$ deformation of the $4+SU(4)_0+4$ box. The Ex-type delta function reduces the external $SU(4)$ gauging down to $SU(3)$, while the outgoing nilpotent partition is trivial. The interacting residual sector is described by the canonical $2+SU(2)_0+2$ box in five dimensions and its strong-coupling magnetic quiver $S_{2,2}$ in three dimensions.}
    \label{tab:Comp_Box31}
\end{table}

\begin{itemize}
    \item \textit{Delta functions and measure correction}: After the second propagation, the outgoing partition is trivial, so $\dd{\rm{Nil}}=\bm\delta^{\rm Nil}_{\rm pert}=1$. The excessive deltas nevertheless survive, with support
    \begin{equation}
        l_1 = h'-h_1+h_2-\frac{iQ}{2}+m \;\;\text{ in 3d} \;,\;\;\text{and}\quad L_1 = \frac{K_2K'}{K_1} \;\;\text{ in 5d} ~.
    \end{equation}
    At $m=iQ/4$, these conditions are related by the 5d/3d dictionary \eqref{eq:dict}. When the external $SU(4)$ symmetry is gauged, they freeze one Cartan and leave an $SU(3)$ gauging. The four possible choices of the frozen parameter account for the multiplicity $\frac{|\mathcal W_{SU(4)}|}{|\mathcal W_{SU(3)}|}=4$, while $\Delta_{\rm 3d}^{\rm corr}$ and $\Delta_{\rm 5d}^{\rm corr}$ convert the original $SU(4)$ gauging measures into the residual $SU(3)$ ones. There is no induced Goldstone sector to remove because $\rho'=[1^4]$. In five dimensions, the accompanying constraint $\bm\delta_{\rm inst} = \delta_{\gamma_1,\emptyset}$ also trivialises the partition on the frozen external leg; it has no independent three-dimensional counterpart.
    
    \item \textit{Residual theory}: The residual theory in this case is interacting and coincides with the $1+SU(2)_0+3$ gauge theory
    or equivalently with  the $2+SU(2)_0+2$ box gauge theory with parameters $(\underline{\vec R},\underline{\vec S},\underline K_1,\underline K_2)$, following a Hanany--Witten transition. Its strong-coupling magnetic quiver is $S_{2,2}$, which is precisely the interacting residual quiver obtained via the electric algorithm in three dimensions. At $m=iQ/4$, the 5d/3d dictionary \eqref{eq:dict} identifies these parameters with $(\underline{\vec r},\underline{\vec s},\underline h_1,\underline h_2)$ reproducing the FI parameters of the three-dimensional residual quiver. Therefore, the canonical parametrisation of the residual magnetic quiver reproduces exactly the parameter reshuffling and rescaling of the residual five-dimensional theory.
    
    \item \textit{Goldstone modes}: The Goldstone sector is fixed by the original $[3,1]$ VEV of $SU(4)$. In both three and five dimensions, the Goldstone contribution of the induced $[2,1^2]$ VEV is cancelled by the inverse Goldstone factor supplied by the first strip. Accordingly, ${\rm GM}_{\rm 3d}^{[3,1]}$ and ${\rm GM}_{\rm 5d}^{[3,1]}$ coincide with the respective contributions in the $[3,1]$ strip example of Subsection \ref{subsec:stripminimal}. Their common broken-generator content is then consistently translated into the two partition functions using our prescription in Appendix \ref{app:GM}.
\end{itemize}

Therefore, the five- and three-dimensional theories agree on all the data characterising the broken theory: the distributional support, the reduction of the gauged symmetry, the interacting residual theory and its parametrisation, and the Goldstone sector associated with the original nilpotent VEV.

\subsection{General broken box theories and the Hanany-Witten propagation rule}
\label{subsec:general_NilHW}

In this section, we derive the general propagation rule for a nilpotent VEV through the $N+SU(N)_0^{k-1}+N$ box theory. We regard the box as a sequence of $k$ symmetrical length-$N$ strips and transport the D7 branes associated with the Jordan blocks of the nilpotent orbit across successive NS5 branes. At each Hanany--Witten move, one of the D5 branes ending on each transported D7 brane is annihilated. Consequently, after crossing $p$ strips, a Jordan block of size $r$ is reduced to one of size $r-p$, provided that $r>p$. 

This procedure determines both the partition propagated to each successive strip and its exact parametrisation in terms of the original five-dimensional variables.\footnote{Hanany--Witten transitions of topological string partition functions, including the removal of decoupled parallel-brane sectors and the induced K\"ahler parameter maps, were recently studied also in  \cite{Bourget:2026hsg}.} If no D5 brane remains attached to a transported D7 brane after crossing all $k$ strips, the final index remains an ordinary function. By contrast, if some D5 branes remain attached to a transported D7 brane after crossing all $k$ strips, the corresponding constraints act on the external parameters and the index becomes distributional. This gives a direct five-brane criterion for when the nilpotent VEV renders the box theory broken.

We will show that both the five-dimensional box theory and its strong-coupling magnetic quiver $S_{k,N}$ become broken when the largest Jordan block of the nilpotent VEV exceeds the number of strips in the box. Their respective partition functions then develop the same distributional structure as a sum of symmetry-related frames.

\subsubsection*{Hook-type nilpotent VEVs}

We begin with a hook-type partition $[L,1^{N-L}]$, with $2\leq L\leq N$, for which the single-strip result \eqref{eq:Strip_hook_gen} can be applied iteratively. As discussed in Subsection \ref{subsec:general_strip}, crossing a single strip reduces the size of the non-trivial Jordan block by one, with the exact parameter specialisation given in \eqref{eq:deltamap_hook_strip}. In the five-brane web, this corresponds to moving the D7 brane supporting the Jordan block across the adjacent NS5 brane, thereby annihilating one of the $L$ D5 branes ending on it; see Figure \ref{fig:Braneweb_311}.

Consider the specialisation
\begin{equation}
\label{eq:VEV_hook_2}
    M_a = \left(\frac{\mf t}{\mf q}\right)^{\frac{L+1-2a}{2}} K'd_a \;\;,\quad a=1,\dots,L \;\;,\quad \prod_{a=1}^{L}d_a=1 ~.
\end{equation}
The parameters $d_a$ regulate the degeneration. We keep them generic while deriving the support of the delta functions and take the nilpotent limit $d_a\to1$ only at the end.

After $p$ Hanany--Witten moves, with $1\leq p\leq k$ and $p<L$, the perturbative and instantonic delta functions in the nilpotent limit are
\begin{equation}\label{eq:VEVprop_hook}
    \begin{split}
        \bm\delta_{\rm pert}^{(p)} &= \prod_{a=1}^{L-p} \delta\left( (N_a^{(p)})^{-1} \left(\frac{\mf t}{\mf q}\right)^{\frac{L-p+1-2a}{2}(-1)^p} \nu^{(p)}d_a^{(-1)^p} \right) ~, \\
        \bm\delta_{\rm inst}^{(p)} &= \prod_{a=1}^{L-p} \delta_{\alpha_a^{(p)},\emptyset} ~,
    \end{split}
\end{equation}
where
\begin{equation}\label{eq:nup_defn}
    \nu^{(p)} = \left(\frac{\mf t}{\mf q}\right)^{\frac{1-(-1)^p}{4}} \left[ \frac{ K'\prod_{j=1}^{\lfloor p/2\rfloor}K_{2j} }{ \prod_{j=1}^{\lceil p/2\rceil}K_{2j-1} } \right]^{(-1)^p} ~.
\end{equation}
In particular, $\nu^{(0)}=K'$. These centres obey the recurrence
\begin{equation}
    \nu^{(p)} = \sqrt{\frac{\mf t}{\mf q}}\,\frac{K_p}{\nu^{(p-1)}} ~.
\end{equation}
The factors of $(-1)^p$ encode the inversion of the geometric progression at each successive propagation step, with the alternating strip parametrisations fixed by \eqref{eq:alt_strip}. This reverses the ordering of the parameters within the Jordan block but leaves the associated partition unchanged. Once $p\geq L$, no constraint from this block propagates further.

Since each Hanany--Witten move reduces the size of the Jordan block by one, the qualitative outcome is determined entirely by the comparison between $L$ and the number $k$ of strips. The propagation may terminate before the transported D7 brane reaches the opposite boundary, precisely on the final strip, or leave surviving D5 branes at the opposite boundary. These possibilities correspond, respectively, to $L<k$, $L=k$, and $L>k$, which we now discuss in turn.

\begin{figure}[h!]
    \centering
    \includegraphics[width=\linewidth]{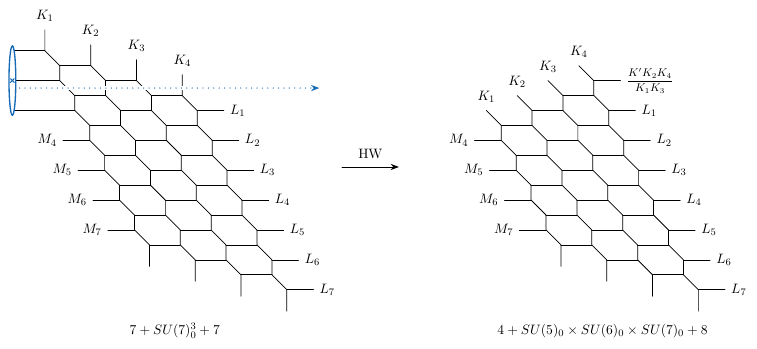}
    \caption{Propagation of a hook-type nilpotent VEV with $L<k$, illustrated for $N=7$, $k=4$, and $L=3$. The D5 branes ending on the transported D7 brane are exhausted before the D7 brane reaches the opposite boundary. The last local constraint is integrated over and generates an additional fundamental hypermultiplet for the first rank-$N$ gauge node.}
    \label{fig:gen_good}
\end{figure}

\noindent\underline{\textbf{Case $L<k$}}: In this case, the single constraint is produced after crossing $p=L-1$ strips, while the transported D7 brane is still inside the box. The constraint is therefore integrated over in the gluing to the next strip, and no condition is imposed on the external $SU(N)$ parameters at the opposite boundary. The final index is consequently non-distributional, and the resulting configuration is good. We illustrate this situation in Figure \ref{fig:gen_good} for $N=7$, $k=4$, and $L=3$.
The resulting index takes the form

\begin{equation}
    \begin{split}
        &\cI^{N+SU(N)_0^{k-1}+N} (\vec M,\vec L,K_i,\vec Q_{\rm inst}) \Big\vert_{\vec M\to[L,1^{N-L}]_{K'}} \\[1mm]
        &\quad = \cI^{(N-L)+SU(N-L+1)_0\times\cdots\times SU(N-1)_0\times(SU(N)_0+1\,{\rm F})\times SU(N)_0^{k-L-1}+N} (\vec M',\vec L,K_i,K',\vec Q_{\rm inst}) \\
        &\qquad\qquad \times \; {\rm GM}_{\rm 5d}^{[L,1^{N-L}]} (\vec M',K';\bm\emptyset) ~.
    \end{split}
\end{equation}
Here $\vec M'=(M_{L+1},\dots,M_N)$ and $(SU(N)_0+1\,{\rm F})$ denotes the first rank-$N$ gauge node, which contains one additional fundamental hypermultiplet. This is the situation studied in Subsection \ref{subsec:good} for $L=2$ and $k=3$.

\begin{figure}[h!]
    \centering
    \includegraphics[width=\linewidth]{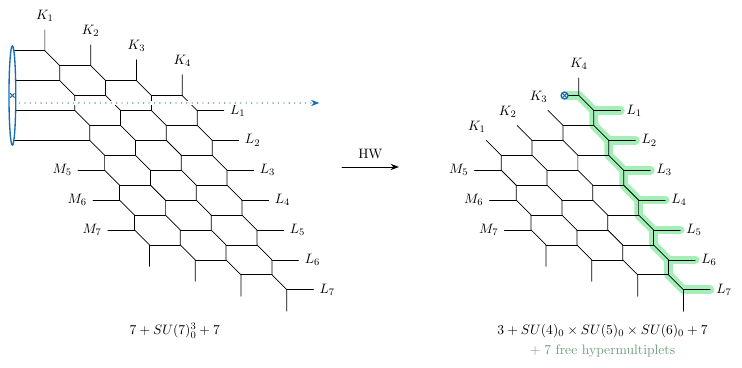}
    \caption{Propagation of a hook-type nilpotent VEV with $L=k$, illustrated for $N=7$ and $k=L=4$. The last D5 brane reaches the final strip and produces one decoupled $SU(N)$ fundamental hypermultiplet, while no delta function remains supported on the external parameters.}
    \label{fig:gen_ugly}
\end{figure}

\noindent\underline{\textbf{Case $L=k$}}: When $L=k$, a single delta is produced after crossing the first $k-1$ strips. The gauge integration attaching the final strip saturates this constraint and leaves a decoupled fundamental hypermultiplet of the right $SU(N)$ flavour symmetry. Thus, the final index remains non-distributional but now contains a decoupled free sector. We illustrate this configuration in Figure~\ref{fig:gen_ugly} for $N=7$ and $k=L=4$.
The index becomes
\begin{equation}
    \begin{split}
        &\cI^{N+SU(N)_0^{k-1}+N} (\vec M,\vec L,K_i,\vec Q_{\rm inst}) \Big\vert_{\vec M\to[L,1^{N-L}]_{K'}}
        \\[1mm]
        &\quad = \cI^{(N-L)+SU(N-L+1)_0\times\cdots\times SU(N-1)_0+N} (\vec M',\vec L,K_i,K',\vec Q_{\rm inst}) \\
        &\qquad\qquad \times \; \cI^{\rm free}(K_1,\dots,K_k,K',\vec L) \times {\rm GM}_{\rm 5d}^{[L,1^{N-L}]} (\vec M',K';\bm\emptyset) ~,
    \end{split}
\end{equation}
where the free fundamental hypermultiplet contributes
\begin{equation}
    \cI^{\rm free}(K_1,\dots,K_k,K',\vec L) = \prod_{a=1}^{N} \left| \MM{\frac{\nu^{(k-1)}L_a}{K_k}} \right|^{-2} ~,
\end{equation}
with $\nu^{(k-1)}$ defined in \eqref{eq:nup_defn}. This reproduces the ugly configuration studied in Subsection \ref{subsec:ugly} for $L=k=2$.

\noindent\underline{\textbf{Case $L>k$}}: The situation changes qualitatively when $L>k$. In this case, the transported D7 brane reaches the opposite boundary of the box before all the D5 branes ending on it have been annihilated. Setting $p=k$, we find $L-k$ constraints acting directly on the external $SU(N)$ mass parameters. The final index is therefore distributional, and the resulting configuration is broken.

For $L\geq k+2$, at least two D5 branes reach the opposite boundary and impose a non-trivial outgoing nilpotent VEV:
\begin{equation}
    L\geq k+2 \;: \quad \bm\delta_{\rm pert}^{\rm Nil}\neq1 \;\;,\quad \bm\delta_{\rm pert}^{\rm Ex}=1 \;\;,\quad \rho'_{(k)}=[L-k,1^{N-L+k}] ~.
\end{equation}
For $L=k+1$, only one D5 brane survives and the outgoing partition is trivial:
\begin{equation}
    L=k+1 \;:\quad \bm\delta_{\rm pert}^{\rm Nil}=1 \;\;,\quad \bm\delta_{\rm pert}^{\rm Ex}\neq1 \;\;,\quad \rho'_{(k)}=[1^N] ~.
\end{equation}
The latter is precisely the situation studied in the previous subsection for $N=4$, $L=3$, and $k=2$.

\begin{figure}[h!]
    \centering
    \includegraphics[width=\linewidth]{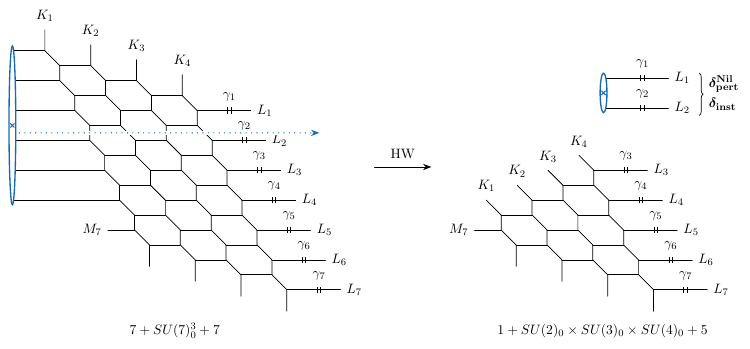}
    \caption{Propagation of a hook-type nilpotent VEV with $L>k$, illustrated for $N=7$, $k=4$, and $L=6$. Two D5 branes remain attached to the transported D7 brane after it reaches the opposite boundary. They impose a non-trivial outgoing nilpotent VEV and render the index distributional.}
    \label{fig:gen_broken}
\end{figure}

The five-brane origin of this distinction is transparent. If a single D5 brane remains attached to the transported D7 brane, it fixes one external parameter and produces the single delta function $\bm\delta_{\rm pert}^{\rm Ex}$. If several D5 branes survive, their parameters form a geometric progression and collectively impose a smaller nilpotent VEV at the opposite boundary. We illustrate the latter situation in Figure \ref{fig:gen_broken} for $N=7$, $k=4$, and $L=6$.

We now collect the different contributions to the modulus square of the refined topological string partition function for a broken theory obtained from a hook-type VEV. The result contains
\begin{equation}
    \frac{N!}{(N-L+k)!} \nn
\end{equation}
Weyl-related frames. As in Subsection \ref{subsec:brokenbox}, the multiplicities generated at internal interfaces are absorbed into the residual Weyl normalisations upon gluing; only the distinct external supports label the final frames.

For $L\geq k+2$, the perturbative and instantonic delta functions are
\begin{equation}
\label{eq:Strip_hook_gen_deltas}
    \begin{split}
        \bm\delta_{\rm pert}^{\rm Nil} &= \prod_{a=1}^{L-k} \delta\left( \left(\frac{\mf t}{\mf q}\right)^{\frac{L-k+1-2a}{2}(-1)^k}\frac{\nu^{(k)}}{L_a} \right) \;\;,\quad \bm\delta_{\rm pert}^{\rm Ex}=1 ~, \\
        \bm\delta_{\rm inst} &= \prod_{a=1}^{L-k} \delta_{\gamma_a,\emptyset} ~.
    \end{split}
\end{equation}
For $L=k+1$, the same local propagation rule gives instead
\begin{equation}
    \bm\delta_{\rm pert}^{\rm Nil}=1 \;\;,\quad \bm\delta_{\rm pert}^{\rm Ex} = \delta\left(\frac{\nu^{(k)}}{L_1}\right) \;\;,\quad \bm\delta_{\rm inst} = \delta_{\gamma_1,\emptyset} ~.
\end{equation}
Upon gluing the broken configuration to a test strip, these constraints reduce the original $SU(N)_{\vec L}$ gauging to $SU(N-L+k)_{\vec L'}$ with surviving fugacities
\begin{equation}
    \vec L'=(L_{L-k+1},L_{L-k+2},\dots,L_{N-1},L_N) ~.
\end{equation}
The corresponding measure is produced by the correction
\begin{equation}\label{eq:Strip_hook_gen_deltacorr}
    \begin{split}
        \Delta^{\rm corr}_{\rm 5d} &= \left| \frac{1}{M(\mf t,\mf q)^{L-k}} \prod_{L-k+1\leq a<b\leq N} \frac{ \MM{ \frac{\mf t}{\mf q} \left(\frac{L_b}{L_a}\right)^{(-1)^{k-1}} } }{ \cN_{\gamma_a\gamma_b}\left( \frac{\mf t}{\mf q} \left(\frac{L_b}{L_a}\right)^{(-1)^{k-1}} \right) } \right. \\
        &\hspace{3cm} \left. \times \; \prod_{a=L-k+1}^{N} \frac{ \cN_{\emptyset\gamma_a}\left( \left(\frac{\mf t}{\mf q}\right)^{\frac{k-L+1}{2}} \left(\frac{L_a}{\nu^{(k)}}\right)^{(-1)^{k-1}} \right) }{ \MM{ \left(\frac{\mf t}{\mf q}\right)^{\frac{k-L+1}{2}} \left(\frac{L_a}{\nu^{(k)}}\right)^{(-1)^{k-1}} } } \right|^2 ~.
    \end{split}
\end{equation}
For $L\geq k+2$, this correction also removes the Goldstone contribution generated by the outgoing nilpotent VEV on the test strip. Its parity dependence follows from the alternating strip parametrisations. For $k=1$, it reproduces the correction in \eqref{eq:Strip_hook_gen}; for $k=2$, $L=3$, and $N=4$ it reproduces the correction in \eqref{eq:SQCD_broken}.

The Goldstone modes associated with the original $[L,1^{N-L}]$ nilpotent VEV contribute
\begin{equation}
    {\rm GM}_{\rm 5d}^{[L,1^{N-L}]} (\vec M',K';\bm\emptyset) = \left| \prod_{a=1}^{L-1}
    \MM{\frac{\mf t^a}{\mf q^a}}^{L-a} \prod_{b=L+1}^{N} \prod_{a=1}^{L-1} \MM{ \left(\frac{\mf t}{\mf q}\right)^{\frac{L+1}{2}-a} \frac{K'}{M_b} } \right|^{-2} ~.
\end{equation}
The intermediate Goldstone contributions cancel against the inverse factors supplied by the corrections in the preceding strips, leaving only the contribution associated with the original VEV. Finally, for $k\geq2$, the residual interacting theory is
\begin{equation}
    (N-L)+SU(N-L+1)_0\times\cdots\times SU(N-L+k-1)_0+(N-L+k) ~. \nn
\end{equation}
For $k=1$, the residual sector is instead the free hypermultiplet in the bifundamental representation of $SU(N-L)\times SU(N-L+1)$ described in Subsection \ref{subsec:general_strip}.

Together, these contributions take the general form \eqref{eq:broken_formula} and reproduce the modulus square of the refined topological string partition function of the Higgsed box theory. All the hook-type examples studied in the preceding subsections follow as special cases of these formulas.

\subsubsection*{General nilpotent VEVs}

We now extend the propagation argument to a general nilpotent VEV labelled by a partition
\begin{equation}\label{eq:gen_part}
    \rho=[r_1,r_2,\ldots,r_s] \;\;,\quad r_1\geq r_2\geq\cdots\geq r_s \;\;,\quad \sum_{i=1}^{s}r_i=N ~.
\end{equation}
As in the single-strip analysis of Subsection \ref{subsec:general_strip}, each Jordan block of size $r_i$ is represented by a D7 brane with $r_i$ D5 branes ending on it. We label the masses belonging to the $i$-th block by $M_{i,a}$, with $a=1,\ldots,r_i$, and impose
\begin{equation}
    M_{i,a} = \left(\frac{\mf t}{\mf q}\right)^{\frac{r_i+1}{2}-a} K'_i d_{i,a} \;\;,\quad \prod_{a=1}^{r_i}d_{i,a}=1 ~.
\end{equation}
As before, the $d_{i,a}$ are kept generic while deriving the distributional support and are set to one only after the degeneration has been resolved.

The distributional support propagates block by block according to the hook-type rule. After crossing $p$ strips, a surviving block with $r_i>p$ has centre
\begin{equation}
    \nu_i^{(p)} = \left(\frac{\mf t}{\mf q}\right)^{ \frac{1-(-1)^p}{4}} \left[ \frac{ K'_i\prod_{j=1}^{\lfloor p/2\rfloor}K_{2j} }{ \prod_{j=1}^{\lceil p/2\rceil}K_{2j-1} } \right]^{(-1)^p} ~,
\end{equation}
while the size of the block is reduced from $r_i$ to $r_i-p$. For a representative frame, let $N_{i,a}^{(p)}$ denote the parameters assigned to the surviving part of the $i$-th block. The corresponding constraints in the nilpotent limit, for $1\leq p\leq k$, are
\begin{equation}
    \begin{split}
        \bm\delta_{\rm pert}^{(p)} &= \prod_{i:\,r_i>p} \prod_{a=1}^{r_i-p} \delta\left(
        \left(\frac{\mf t}{\mf q}\right)^{\frac{r_i-p+1-2a}{2}(-1)^p} \frac{\nu_i^{(p)}d_{i,a}^{(-1)^p}}{N_{i,a}^{(p)}} \right) ~, \\
        \bm\delta_{\rm inst}^{(p)} &= \prod_{i:\,r_i>p} \prod_{a=1}^{r_i-p} \delta_{\alpha_{i,a}^{(p)},\emptyset} ~.
    \end{split}
\end{equation}
This immediately gives the general classification:

\begin{itemize}
    \item If $r_1<k$, all the constraints are exhausted inside the box. The final index is an ordinary function and the configuration is {\it good}.
    \item If $r_1=k$, no constraint reaches the opposite boundary, but every     block with $r_i=k$ produces a decoupled fundamental hypermultiplet of the right $SU(N)$ flavour symmetry. The resulting configuration is of {\it ugly} type.
    \item If $r_1>k$, at least one constraint reaches the opposite boundary. The final index is distributional and the configuration is {\it broken}. Blocks with $r_i=k$ may still contribute additional free hypermultiplets.
\end{itemize}
The first two possibilities have the same five-brane interpretation as the hook-type configurations shown in Figures \ref{fig:gen_good} and \ref{fig:gen_ugly}. The broken case occurs whenever at least one D7 brane reaches the opposite side of the box with one or more D5 branes still ending on it.

For a representative frame, we assign to the $i$-th block with $r_i>k$ a subset of the external parameters, denoted by $\{L_{i,a}\}_{a=1}^{r_i-k}$. The remaining frames are obtained by Weyl permutations of the original $SU(N)_{\vec L}$ parameters. Upon gluing to a test strip, these constraints generically reduce the external $SU(N)_{\vec L}$ gauging to the residual gauge group $SU\left(N-\sum_{i:r_i>k}(r_i-k)\right)$. The number of Weyl-related frames is therefore $N!/\left(N-\sum_{i:r_i>k}(r_i-k)\right)!$. When several outgoing blocks have equal size, their centres are generically distinct, so exchanging their assignments to the external fugacities gives distinct frames.

\begin{figure}[h!]
    \centering
    \includegraphics[width=\linewidth]{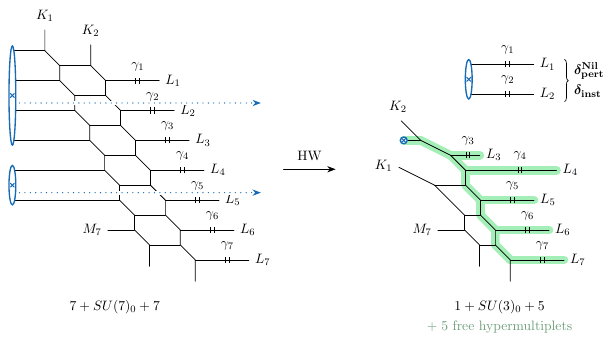}
    \caption{Propagation of the general nilpotent VEV $\rho=[4,2,1]$ through the $N=7$, $k=2$ box. The block of size four produces a non-trivial outgoing nilpotent VEV, while the block of size two gives an additional free fundamental hypermultiplet of the surviving $SU(5)$ flavour symmetry.}
    \label{fig:gen_broken_gen}
\end{figure}

At the opposite boundary, blocks with $r_i\geq k+2$ impose non-trivial nilpotent delta functions, while blocks with $r_i=k+1$ produce $\bm\delta_{\rm pert}^{\rm Ex}$. Explicitly,
\begin{equation}
    \begin{split}
        \bm\delta_{\rm pert}^{\rm Nil} &= \prod_{i:\,r_i\geq k+2} \prod_{a=1}^{r_i-k} \delta\left( \left(\frac{\mf t}{\mf q}\right)^{\frac{r_i-k+1-2a}{2}(-1)^k} \frac{\nu_i^{(k)}}{L_{i,a}}  \right) ~, \\
        \bm\delta_{\rm pert}^{\rm Ex} &= \prod_{i:\,r_i=k+1} \delta\left( \frac{\nu_i^{(k)}}{L_{i,1}} \right) ~, \\
        \bm\delta_{\rm inst} &= \prod_{i:\,r_i>k} \prod_{a=1}^{r_i-k} \delta_{\gamma_{i,a},\emptyset} ~.
    \end{split}
\end{equation}
As an illustration, consider $N=7$, $k=2$, and $\rho=[4,2,1]$, shown in Figure \ref{fig:gen_broken_gen}.
The block of size four reaches the opposite boundary with two attached D5 branes. It therefore imposes the outgoing $[2,1^5]$ nilpotent specialisation and freezes two external fugacities. The block of size two is exhausted on the final strip and produces a decoupled fundamental hypermultiplet of the surviving $SU(5)$ flavour symmetry, while the block of size one does not generate any non-trivial propagation. The residual interacting theory is $1+SU(3)_0+5$, and the distribution contains $7!/5!=42$ Weyl-related frames.

We conclude that a general nilpotent VEV therefore produces a broken box configuration precisely when the size of its largest Jordan block exceeds the number of strips:
\begin{equation}
    r_1>k \qquad\Longleftrightarrow\qquad \text{broken configuration} ~. \nn
\end{equation}
This criterion agrees precisely with the magnetic-quiver analysis of Section \ref{sec:review_3d}. There, we considered a nilpotent VEV on one $T[U(N)]$ tail of $S_{k,N}$, keeping the other tails undeformed. The resulting theory is broken precisely when the central $U(N)$ node becomes bad. This is the three-dimensional counterpart of the deformation of the box theory studied in this section. The central $U(N)$ node becomes bad precisely when $r_1>k,$ in exact agreement with the five-brane web criterion derived above.

We therefore find a precise parallel between five and three dimensions: the theories become broken at the same threshold, and their superconformal index and $S_b^3$ partition function develop the same distributional structure, organised as a sum of symmetry-related frames.

\section{Two-sided nilpotent Higgsings: bad theories}
\label{sec:multiplenil_bad}

So far, we have studied nilpotent VEVs for only one of the two $SU(N)$ flavour moment maps of the $N+SU(N)_0^{k-1}+N$ box theory. Depending on the partition, these deformations can lead to good, ugly, or broken theories. In the broken case, the partition function becomes a distributional sum of frames related by global-symmetry transformations.

We now turn on nilpotent VEVs for both $SU(N)$ flavour moment maps. For sufficiently large VEVs, these simultaneous deformations can make the index a distributional sum of \emph{inequivalent} frames. We refer to the resulting five-dimensional configuration as a bad theory, in direct analogy with bad three-dimensional $\mathcal N=4$ theories. Using the 5d/3d dictionary, we match the delta functions, residual interacting theories, free sectors, and parameter maps appearing in the two descriptions.

Our main example is the $4+SU(4)_0^3+4$ box theory deformed by two $[3,1]$ nilpotent VEVs, one for each $SU(4)$ flavour symmetry. In Subsection \ref{subsec:bad5d}, we propagate the two VEVs through the box by applying the Hanany--Witten propagation rules developed in Subsections \ref{subsec:general_strip} and \ref{subsec:general_NilHW}. The two propagated constraints eventually act on the same set of internal Coulomb parameters and the resolution of this collision produces two inequivalent classes of contributions. In Subsection \ref{subsec:bad3d}, we reproduce precisely the same two classes by applying the electric dualisation algorithm to the magnetic quiver $S_{4,4}$ with the corresponding two nilpotent VEVs. Finally, Subsection \ref{subsec:bad_triangulations} identifies the corresponding residual brane webs through two inequivalent tessellations of the same generalised toric diagram.

\subsection{5d analysis and brane interpretation}
\label{subsec:bad5d}

We begin with the refined topological string partition function \eqref{eq:Ztop_SUNkm12N} for $N=k=4$. The two $[3,1]$ nilpotent VEVs are implemented by the specialisations:
\begin{equation}\label{eq:bad_31_31_vevs}
    \begin{aligned}
        M_1 & =\frac{\mf t}{\mf q}K'd'_1 \;\;,& M_2 &= K'd'_2 \;\;,& M_3 &= \frac{\mf q}{\mf t}K'd'_3 ~, \\
        L_1 &= \frac{\mf t}{\mf q}K''d''_1 \;\;,& L_2 &= K''d''_2 \;\;,& L_3 &= \frac{\mf q}{\mf t}K''d''_3 ~,
    \end{aligned}
\end{equation}
where $d'_1d'_2d'_3=d''_1d''_2d''_3=1$. As before, the regulators are kept independent and generic until all relevant pole collisions have been resolved; only then do we take $d'_a,d''_a\to1$.

\begin{figure}[h!]
    \centering
    \includegraphics[width=\linewidth]{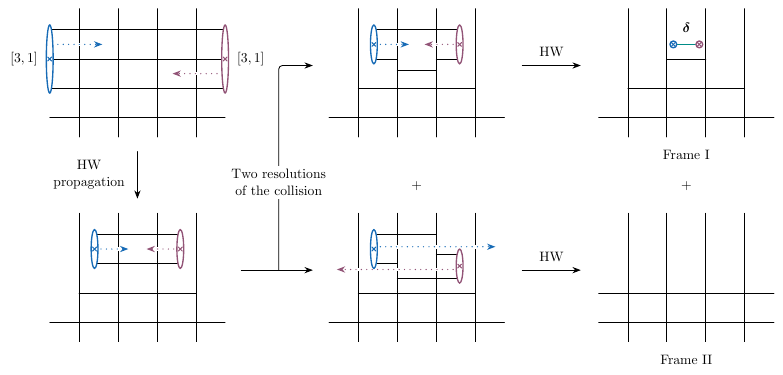}
    \caption{Propagation and collision of the two $[3,1]$ nilpotent VEVs in the $4+SU(4)_0^3+4$ box theory. When the two propagated constraints act on the same set of Coulomb parameters $\vec N^{(2)}$, the collision admits two inequivalent resolutions, corresponding to Frames I and II. Internal $(1,1)$ five-brane segments are contracted only to simplify the schematic drawing.}
    \label{fig:badbrane}
\end{figure}

The propagation of the two VEVs is summarised schematically in Figure \ref{fig:badbrane}. The internal $(1,1)$ five-brane segments have been contracted for clarity. The figure should therefore be regarded as a visualisation of the propagation and the two possible resolutions of the collision; a geometric description in terms of tessellated generalised toric polygons is given in Subsection \ref{subsec:bad_triangulations}.

Since both partitions are of hook type, we may apply the propagation rule in \eqref{eq:VEVprop_hook}. After each VEV has crossed two strips, the constraints coming from the left and from the right act on the same set of variables, $\vec N^{(2)}$ and $\bm\alpha^{(2)}$. Introducing
\begin{equation}\label{eq:bad_AB}
    A \coloneq \frac{K_2K'}{K_1} \;\;,\quad B \coloneq \frac{K_3K''}{K_4} ~,
\end{equation}
the propagated contributions take the form
\begin{equation}\label{eq:leftrigh_deltas}
    \delta\left(\frac{N_i^{(2)}}{A}\right) \delta_{\alpha_i^{(2)},\emptyset} ~.~ \delta\left(\frac{N_j^{(2)}}{B}\right) \delta_{\alpha_j^{(2)},\emptyset} \;\;,\quad i,j=1,\dots,4 ~.
\end{equation}
The ordered pair $(i,j)$ labels sixteen contributions. Evaluating them explicitly, we find that the result depends only on whether the two constraints act on the same Coulomb parameter, $i=j$, or on two distinct Coulomb parameters, $i\neq j$. The sixteen contributions therefore organise into two inequivalent classes: four with $i=j$ and twelve with $i\neq j$. We refer to these two classes as Frames I and II.

Both resolutions contain Goldstone modes associated with the two incoming $[3,1]$ VEVs. For brevity, we denote their contributions by
\begin{equation}
    {\rm GM}_{\rm 5d,\,L}^{[3,1]} \coloneq {\rm GM}_{\rm 5d}^{[3,1]}(M_4,K';\bm\emptyset) \;\;,\quad {\rm GM}_{\rm 5d,\,R}^{[3,1]} \coloneq {\rm GM}_{\rm 5d}^{[3,1]}(L_4,K'';\bm\emptyset) ~.
\end{equation}
These are the Goldstone contributions already encountered in Subsection \ref{subsec:stripminimal}, evaluated in the original external parameters. Since these are common to both frames, we factor them out when computing the two frame contributions below.

\paragraph{Frame I: $i=j$}: When the two constraints freeze the same Coulomb parameter, \eqref{eq:leftrigh_deltas} gives\footnote{We use the identity
\begin{equation}\label{eq:bad_delta_compatibility}
    \delta\left(\frac{N_i^{(2)}}{A}\right)\delta\left(\frac{N_i^{(2)}}{B}\right) = \delta\left(\frac{A}{B}\right)\delta\left(\frac{N_i^{(2)}}{A}\right)
\end{equation}
understood under the $N_i^{(2)}$ contour integral.}
\begin{equation}\label{eq:bad_frameI_deltas}
    \delta\left(\frac{N_i^{(2)}}{A}\right) \delta\left(\frac{A}{B}\right) \delta_{\alpha_i^{(2)},\emptyset} = \delta\left(\frac{K_1N_i^{(2)}}{K_2K'}\right) \delta\left(\frac{K_2K_4K'}{K_1K_3K''}\right) \delta_{\alpha_i^{(2)},\emptyset} ~.
\end{equation}
The internal delta function reduces $SU(4)_{\vec N^{(2)}}$ to $SU(3)$, while the discrete delta accompanies this freezing by trivialising one internal partition corresponding to the frozen Coulomb modulus. There are four such frames consistently with the ratio of the orders of the Weyl groups of the original $SU(4)$ and the residual $SU(3)$. The residual interacting theory is therefore
\begin{equation}
    1+SU(2)_0\times\big(SU(3)_0+2\big)\times SU(2)_0+1 ~. \nn
\end{equation}
The remaining periodic delta function imposes $A=B$ on the external parameters. In the web, the two D7 branes transported into the box must be aligned before the finite D5 brane stretched between them can be moved away from the remaining configuration. The required tuning of their relative position is the geometric counterpart of the remaining delta $\delta\left(A/B\right)$.

To express the contribution of Frame I to the index in a form particularly well suited to the 5d/3d comparison, we proceed as follows. The natural parameters of the residual theory $1+SU(2)_0\times\big(SU(3)_0+2\big)\times SU(2)_0+1$ are related to those of the original $4+SU(4)_0^3+4$ box theory by a non-trivial change of variables. This map can  be derived by explicitly tracking the parameters through the sequence of Hanany-Witten moves leading to the residual five-brane web.

There is, however, a more convenient way to describe the same residual theory. It can be obtained from the canonically parametrised $3+SU(3)_0^3+3$ box theory by turning on a $[2,1]$ nilpotent VEV for each of its two flavour moment maps. Each of these VEVs partially Higgses one of the two terminal $SU(3)$ gauge nodes down to $SU(2)$, thereby producing precisely the residual theory displayed above. These are the minimal nilpotent VEVs studied in Subsection \ref{subsec:good}. As shown there, the Higgsing is accompanied by a definite reparametrisation of the five-dimensional variables. Applying this reparametrisation at both ends of the $3+SU(3)_0^3+3$ box reproduces the parametrisation obtained by following the original flow from the $4+SU(4)_0^3+4$ theory with two $[3,1]$ VEVs to the residual theory in Frame I.

The description in terms of the $3+SU(3)_0^3+3$ box theory with two $[2,1]$ VEVs is therefore not an additional deformation, but a convenient alternative presentation of the same residual theory. Its advantage is that it uses the canonical parametrisation of a box theory and can consequently be mapped directly to the corresponding three-dimensional magnetic quiver. Notice, however, that this presentation introduces the Goldstone contributions associated with the two auxiliary $[2,1]$ VEVs; these must therefore be divided out when writing the contribution of Frame I.

With this clarification, we rescale the residual parameters so that they satisfy the appropriate $SU(3)$ conditions.\footnote{As in Subsection \ref{subsec:ugly}, the retained Coulomb parameters are rescaled as
\begin{equation}
    \underline{\vec N}^{(p)}{}' = \frac{\vec N^{(p)}{}'}{c_N^{(p)}} \;\;,\text{ with }\; (c_N^{(1)})^3 = \frac{\mf q}{\mf t}\,\frac{K'}{K_1} \;\;,\quad (c_N^{(2)})^3 = \frac{K_1}{K_2K'} \;\;,\quad (c_N^{(3)})^3 = \frac{K''}{K_4} ~.
\end{equation}
Here $\vec N^{(p)}{}'$ denotes the three-component vector retained at each interface in the auxiliary $SU(3)$ box. The flavour parameters are normalised similarly: $\underline{\vec M}'=(M_1,M_2,M_4)/c_M$ and $\underline{\vec L}'=(L_2,L_3,L_4)/c_L$, with $c_M^3=M_1M_2M_4$ and $c_L^3=L_2L_3L_4$. The corresponding rescaling factors are absorbed into the abelian fugacities.} We denote the index of the residual theory by
\begin{equation}\label{eq:bad_frameI_index}
    \cI_{\rm res}^{\rm I} \coloneq \frac{ \left. \cI^{3+SU(3)_0^3+3} \left(\underline{\vec M}',\underline{\vec L}',\underline{\vec K}^{\rm I},\vec Q_{\rm inst}\right) \right|_{\underline{\vec M}'\to[2,1]_{K'},\,\underline{\vec L}'\to[2,1]_{K''}} }{ {\rm GM}_{\rm 5d}^{[2,1]}(\underline M_3',\underline K';\bm\emptyset) \; {\rm GM}_{\rm 5d}^{[2,1]}(\underline L_3',\underline K'';\bm\emptyset) }
\end{equation}
The two $[2,1]$ specialisations and the map to the original parameters are given by
\begin{equation}\label{eq:bad_frameI_params}
    \begin{split}
        \underline{\vec M}' &= \left( \sqrt{\frac{\mf t}{\mf q}}\,\underline K' ,\; \sqrt{\frac{\mf q}{\mf t}}\,\underline K' ,\; \underline K'^{\,-2} \right) \;\;,\quad \underline{\vec L}' = \left( \sqrt{\frac{\mf t}{\mf q}}\,\underline K'' ,\; \sqrt{\frac{\mf q}{\mf t}}\,\underline K'' ,\; \underline K''^{\,-2} \right) ~, \\
        \underline{\vec K}^{\rm I} &= \left( \frac{K_1^{\frac43}}{\big(K'^{\,3}M_4\big)^{\frac13}} \,,\; \left(\frac{\mf t}{\mf q}\,K_2^4\right)^{\frac13} \,,\; \left(\frac{K_2K_3^4K_4K'}{K_1K''}\right)^{\frac13} \,,\; \left(\frac{\mf t}{\mf q}\,\frac{K_4^4}{K''^3L_4}\right)^{\frac13} \right) ~, \\
        \underline K' &= \left(\sqrt{\frac{\mf t}{\mf q}}\,\frac{K'}{M_4}\right)^{\frac13} \;\;,\quad \underline K'' = \left(\sqrt{\frac{\mf q}{\mf t}}\,\frac{K''}{L_4}\right)^{\frac13} ~. \\
    \end{split}
\end{equation}
Combining this with the two Goldstone contributions associated with the initial nilpotent VEVs, Frame I therefore contributes
\begin{equation}{\rm GM}_{\rm 5d,\,L}^{[3,1]} \; {\rm GM}_{\rm 5d,\,R}^{[3,1]} \times \delta\left(\frac AB\right)\;\cI_{\rm res}^{\rm I} 
\end{equation}
to the original index.

\paragraph{Frame II: $i\neq j$.} When the two constraints freeze distinct Coulomb parameters, their product remains
\begin{equation}\label{eq:bad_frameII_deltas}
    \delta\left(\frac{N_i^{(2)}}{A}\right) \delta\left(\frac{N_j^{(2)}}{B}\right) \delta_{\alpha_i^{(2)},\emptyset} \delta_{\alpha_j^{(2)},\emptyset} \;\;,\quad i\neq j ~.
\end{equation}
The two internal delta functions reduce $SU(4)_{\vec N^{(2)}}$ to $SU(2)$ and are both integrated in the gluing. Consequently, no delta function remains supported on an external fugacity. The residual interacting theory is the $2+SU(2)_0^3+2$ box theory. The twelve contributions in this class satisfy $\frac{|W_{SU(4)}|}{|W_{SU(2)}|} = \frac{4!}{2!} = 12$, consistently with the reduction of the gauge group from $SU(4)$ to $SU(2)$.

After normalising the residual parameters to satisfy the $SU(2)$ conditions, we denote the index of the residual theory in Frame II by
\begin{equation}\label{eq:bad_frameII_index}
    \cI_{\rm res}^{\rm II} \coloneq \cI^{2+SU(2)_0^3+2} \left(\underline{\vec M}',\underline{\vec L}',\underline{\vec K}^{\rm II},\vec Q_{\rm inst}\right) ~.
\end{equation}
The residual parameters are
\begin{equation}\label{eq:bad_frameII_params}
    \begin{aligned}
        \underline{\vec M}' &=\left( \sqrt{\frac{M_4K_2K_4}{K_1K_3K''}} \,,\; \sqrt{\frac{K_1K_3K''}{M_4K_2K_4}} \right) \;\;,\quad \underline{\vec L}' =\left( \sqrt{\frac{L_4K_1K_3}{K_2K_4K'}} \,,\; \sqrt{\frac{K_2K_4K'}{L_4K_1K_3}} \right) ~, \\
        \underline{\vec K}^{\rm II} &= \left( \sqrt{\frac{\mf t}{\mf q}\,\frac{K_1^3K_2K_4}{M_4K_3K'^2K''}} \,,\; \sqrt{\frac{\mf t}{\mf q}\,\frac{K_1K_2^3K_3K''}{K_4K'}} \,,\; \sqrt{\frac{\mf t}{\mf q}\,\frac{K_2K_3^3K_4K'}{K_1K''}} \,,\; \sqrt{\frac{\mf t}{\mf q}\,\frac{K_1K_3K_4^3}{L_4K_2K'K''^2}} \right) ~.
    \end{aligned}
\end{equation}
This frame also contains two decoupled free hypermultiplets. Their combined contribution can be written as
\begin{equation}\label{eq:bad_free_index}
    \cI_{\rm free} \coloneq \left| \MM{\frac{\underline M'_1\underline K_1\underline K_3} {\underline L'_1\underline K_2\underline K_4}}
      \MM{\frac{\mf t}{\mf q} \frac{\underline M'_1\underline K_1\underline K_3} {\underline L'_1\underline K_2\underline K_4}} \right|^{-2} = \left| \MM{\frac AB} \MM{\frac{\mf t}{\mf q}\,\frac AB} \right|^{-2} ~.
\end{equation}
Including the common Goldstone factors from the original VEVs, the contribution of Frame II to the original index is
$$ {\rm GM}_{\rm 5d,\,L}^{[3,1]} \; {\rm GM}_{\rm 5d,\,R}^{[3,1]} \times \cI_{\rm free}\;\cI_{\rm res}^{\rm II} ~. $$

The multiplicities four and twelve combine with the Weyl normalisation $1/4!$ of the original middle gauge node to give $1/3!$ and $1/2!$, respectively. They are thus already accounted for in the canonical measures of the two residual indices. Therefore, the complete expression for the superconformal index of the $4+SU(4)_0^3+4$ box theory with two $[3,1]$ nilpotent VEVs is
\begin{equation}\label{eq:bad_complete_index}
    \boxed{
    \left. \cI^{4+SU(4)_0^3+4} \right|_{[3,1],\,[3,1]} = {\rm GM}_{\rm 5d,\,L}^{[3,1]} \; {\rm GM}_{\rm 5d,\,R}^{[3,1]} \left[ \delta\left(\frac AB\right) \cI_{\rm res}^{\rm I} + \cI_{\rm free} \, \cI_{\rm res}^{\rm II} \right] ~.
    }
\end{equation}
The two terms contain inequivalent residual theories, and only Frame I is supported on an external periodic delta function. This is precisely the structure expected of a bad theory.

The propagation rule gives a convenient shorthand for this computation, but collisions involving more than two independently supported delta functions require additional care. In such cases one should retain generic regulators and resolve the underlying pole collisions before taking the degeneration limit, rather than manipulate products of already-formed distributions. At the end of Subsection \ref{subsec:bad3d}, we compare the general collision threshold with the three-dimensional badness criterion and find an exact agreement.

\subsection{3d counterpart and the 5d/3d map}
\label{subsec:bad3d}

\begin{figure}[h!]
    \centering
    \includegraphics[width=\linewidth]{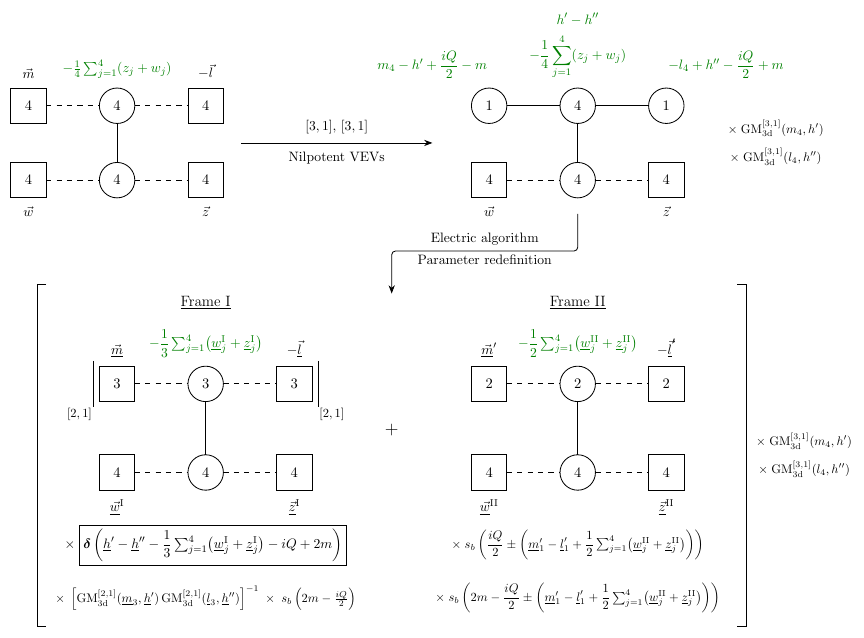}
    \caption{Simultaneous $[3,1]$ nilpotent deformations of the two $T[U(4)]$ tails of $S_{4,4}$ render the upper $U(4)$ node bad. The electric dualisation algorithm produces two inequivalent frames. Frame I contains an external delta function and an interacting residual sector presented as $S_{4,3}$, with a $[2,1]$ VEV on each of its $T[U(3)]$ tails. Frame II contains the interacting residual sector $S_{4,2}$, together with a decoupled free sector. The Goldstone contributions associated with the original pair of $[3,1]$ VEVs multiply both frames.}
    \label{fig:quiver_bad}
\end{figure}

We now study the same pair of nilpotent VEVs in the strong-coupling magnetic quiver $S_{4,4}$. In the $\cN=2^*$ refinement of the 5d/3d dictionary, the two $[3,1]$ specialisations are
\begin{equation}\label{eq:VEV31_31_new}
    \begin{aligned}
        m_1&=h'+\frac{iQ}{2}-m \;\;, &m_2&=h'-\frac{iQ}{2}+m \;\;, &m_3&=h'-\frac{3iQ}{2}+3m ~, \\
        l_1&=h''+\frac{iQ}{2}-m \;\;, &l_2&=h''-\frac{iQ}{2}+m ~, &l_3&=h''-\frac{3iQ}{2}+3m ~.
    \end{aligned}
\end{equation}
As shown in Figure \ref{fig:quiver_bad}, they partially close the two $T[U(4)]$ tails to $T_{[3,1]}[U(4)]$. Each partially closed tail contributes one effective flavour to the upper $U(4)$ node, while the lower $U(4)$ node contributes four. The upper node therefore sees
\begin{equation}
    N_f^{\rm eff} = 1+1+4=6 < 2N_c-1 = 7 ~,
\end{equation}
and is consequently bad. Applying the electric dualisation algorithm produces the sum of two inequivalent frames displayed in Figure \ref{fig:quiver_bad}. This shows directly that the simultaneous VEVs in \eqref{eq:VEV31_31_new} render $S_{4,4}$ bad rather than broken. The Goldstone contributions generated by the original two VEVs are common to both frames and are retained in the original parameters, as in the five-dimensional calculation.

Frame I contains an external Dirac delta function, a decoupled factor, and an interacting residual sector given by the strong-coupling magnetic quiver of the five-dimensional $$1+SU(2)_0\times\big(SU(3)_0+2\big)\times SU(2)_0+1$$ theory. As in the five-dimensional analysis, it is convenient to represent this interacting quiver as the $S_{4,3}$ theory with a $[2,1]$ VEV on each of its two $T[U(3)]$ tails. This presentation provides a natural parametrisation of the residual theory that can be compared directly with its five-dimensional counterpart.

Frame II has no external delta function. Its interacting sector is $S_{4,2}$, the strong-coupling magnetic quiver of the $2+SU(2)_0^3+2$ box theory, and is accompanied by a decoupled free sector. Since both the residual sectors are interacting, throughout this discussion their magnetic quivers are understood with the overall decoupled $U(1)$ removed, as in Subsection \ref{subsec:sknbreakingbad}.

To compare with the five-dimensional result, we bring the parameters of both frames to the canonical box parametrisation. For Frame I, this uses the $S_{4,3}$ presentation with two $[2,1]$ VEVs described above. The required parameter redefinitions for Frame I are\footnote{In addition to these, the parameters of the two auxiliary $[2,1]$ specialisations are $\underline h' = \frac43h'$, $\underline h'' = \frac43h''-\frac13(iQ-2m)$.}
\begin{equation}\label{eq:bad_3d_underline_I}
    \begin{aligned}
        \underline{\vec m} &= \left( \frac43 h', \frac43 h'-iQ+2m, -\frac83 h'+iQ-2m \right) \;\;,\quad \underline{\widetilde h}^{\rm I}_j = \frac43\widetilde h_j-\frac16(iQ-2m)+\delta_{j,3}\frac \zeta3 ~, \\
        \underline{\vec l} &= \left( \frac43 h''-\frac13(iQ-2m), \frac43 h''-\frac43(iQ-2m), -\frac83 h''+\frac53(iQ-2m) \right) ~, \\
        \underline w^{\rm I}_j &= w_j-\frac14(iQ-2m)+\delta_{j,3}\frac \zeta2 \;\;,\quad \underline z^{\rm I}_j = z_j-\frac14(iQ-2m)+\delta_{j,3}\frac \zeta2 ~.
    \end{aligned}
\end{equation}
For Frame II they are
\begin{equation}\label{eq:bad_3d_underline_II}
    \begin{aligned}
        \underline{\vec m}' &= \pm\Big( -\frac12(h_1-h_2+h_3-h_4+3h'+h'')+iQ-2m \Big) ~, \\
        \underline{\vec l}' &= \pm\Big( -\frac12(-h_1+h_2-h_3+h_4+h'+3h'')+iQ-2m \Big) ~, \\
        \underline{\widetilde h}^{\rm II}_j &= 2\widetilde h_j+\frac \zeta2 \;\;,\quad \underline{\vec w}\,{}^{\rm II} = \vec w+\frac \zeta2 \;\;,\quad \underline{\vec z}\,{}^{\rm II} = \vec z+\frac \zeta2 ~.
    \end{aligned}
\end{equation}
Here $j=1,\dots,4$ and we define
\begin{equation}
    \zeta \coloneq h'-h''-\sum_{j=1}^4 \widetilde h_j = h'-h''-h_1+h_2-h_3+h_4 ~.
\end{equation} At the canonical $R$ charge assignment $m=iQ/4$, these redefinitions are the additive counterparts of the five-dimensional parameter maps in Eqs.\ (\ref{eq:bad_frameI_params}) and (\ref{eq:bad_frameII_params}). The flavour vectors are traceless in both frames, while the equal shifts of $w_j$ and $z_j$ preserve their differences, in agreement with the 5d maps leaving the instanton parameters unchanged in Eqs.\ (\ref{eq:bad_frameI_index}) and (\ref{eq:bad_frameII_index}).

We can now compare the two terms produced by the electric dualisation algorithm with the two frames found in the five-dimensional distributional calculation.

\begin{itemize}
    \item \textit{Frame I}:
    In the original variables, this frame contains the Dirac delta function
    \begin{equation}\label{eq:bad_3d_delta_frameI}
        \delta(\zeta) = \delta\left( h'-h''-h_1+h_2-h_3+h_4 \right) ~.
    \end{equation}
    At $m=iQ/4$, the 5d/3d dictionary gives $\frac{K_2K_4K'}{K_1K_3K''} = e^{-\beta\zeta}$. Thus the condition $\zeta=0$ exponentiates to $A/B=1$, matching the support of the five-dimensional periodic delta function
    \begin{equation}
        \delta\left(\frac AB\right) = \delta\left( \frac{K_2K_4K'}{K_1K_3K''} \right) ~.
    \end{equation}
    The interacting sector is $S_{4,3}$ with two $[2,1]$ nilpotent VEVs, precisely the magnetic quiver of the alternative presentation of the residual five-dimensional theory as the $3+SU(3)_0^3+3$ box with two $[2,1]$ VEVs.

    The ratio of Goldstone contributions has the same origin on the two sides. The numerator contains the Goldstone modes generated by the original pair of $[3,1]$ VEVs, whereas the denominator removes the Goldstone factors associated with the two auxiliary $[2,1]$ VEVs used to present the residual theory in canonical box variables. Dividing by these auxiliary contributions therefore gives the interacting sector corresponding to $\cI_{\rm res}^{\rm I}$ in \eqref{eq:bad_frameI_index}.

    \item \textit{Frame II}: Neither the three- nor the five-dimensional contribution contains an external delta function. The interacting sector is $S_{4,2}$, which is the strong-coupling magnetic quiver of the $2+SU(2)_0^3+2$ residual box theory. The traceless mass vectors and the shifts of $\underline{\widetilde h}^{\rm II}_j$, $\underline{\vec w}^{\rm II}$, and $\underline{\vec z}^{\rm II}$ in \eqref{eq:bad_3d_underline_II} map to the rescaled five-dimensional fugacities in \eqref{eq:bad_frameII_params}. Finally, the free sector multiplying $S_{4,2}$ reproduces the two decoupled free hypermultiplets appearing in \eqref{eq:bad_free_index} under the 5d/3d dictionary \eqref{eq:dict} at $m=iQ/4$.
\end{itemize}

The electric dualisation algorithm therefore reproduces the complete structure found in five dimensions: one frame is supported on an external delta function and contains the residual interacting theory presented through the $N=3$ box with two minimal nilpotent VEVs, while the other has no external delta function and contains the residual $N=2$ box theory together with two free hypermultiplets. Since these two frames are inequivalent, the $S_b^3$ partition function has precisely the structure expected of a bad theory.

We can also compare the thresholds for general simultaneous non-trivial nilpotent VEVs. Let $r_L$ and $r_R$ denote the largest Jordan blocks of the two partitions imposed on the $T[U(N)]$ tails of $S_{k,N}$. The partially closed tails contribute $N-r_L$ and $N-r_R$ effective flavours to the central $U(N)$ node, while the transverse $U(k)$ node contributes $k$. Thus
\begin{equation}
    N_f^{\rm eff}=k+2N-r_L-r_R ~.
\end{equation}
The central node is good for $r_L+r_R\leq k$, ugly for $r_L+r_R=k+1$, and bad for
\begin{equation}
    r_L+r_R>k+1 ~.
\end{equation}
On the five-dimensional side, at an internal interface labelled by $p$, the left and right VEVs have crossed $p$ and $k-p$ strips, respectively. The two propagated constraints collide when
\begin{equation}
    p<r_L \;\;,\quad k-p<r_R \;\;,\quad p\in\{1,\dots,k-1\} ~.
\end{equation}
Such an interface exists precisely when $r_L+r_R>k+1$, in agreement with the condition for the central three-dimensional node to become bad. Thus the five-dimensional propagation rule and the magnetic-quiver flavour count identify the same threshold for overlapping constraints. In the examples studied above, resolving their collision yields a sum of inequivalent frames, matching the structure of the corresponding $S_b^3$ partition function. We expect this correspondence to extend to more general simultaneous nilpotent VEVs.

\subsection{Geometric realisation of bad theories: inequivalent tessellations}
\label{subsec:bad_triangulations}

The five-dimensional computation of Subsection \ref{subsec:bad5d} shows that the superconformal index of the $4+SU(4)_0^3+4$ box theory with two $[3,1]$ nilpotent VEVs is a sum of two inequivalent contributions, corresponding to Frames I and II. We now show that this distinction admits a direct geometric interpretation in the generalised toric diagram of the deformed theory: the two frames arise from two inequivalent tessellations of the same diagram into elementary triangles and trapezoids.

\begin{figure}[h!]
    \centering
    \includegraphics[width=\linewidth]{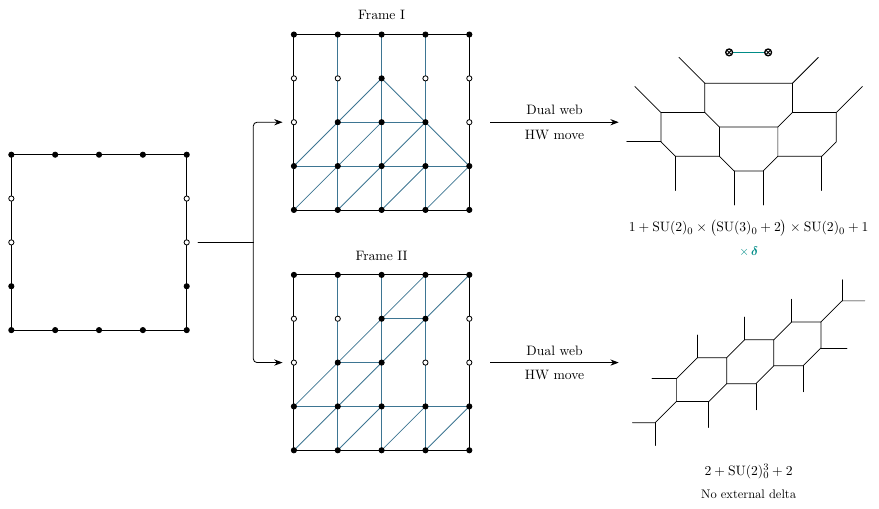}
    \caption{The generalised toric diagram on the left records the infinite-coupling data of the $4+SU(4)_0^3+4$ box theory with two $[3,1]$ nilpotent VEVs, encoded by the white dots on its two vertical edges. The middle panels show two admissible tessellations, whose dual five-brane webs give the configurations on the right after Hanany--Witten transitions. The upper web realises the interacting $1+SU(2)_0\times(SU(3)_0+2)\times SU(2)_0+1$ theory of Frame I. The finite D5 brane suspended between the two D7 branes requires them to be aligned, as encoded by the external periodic delta function. The lower web realises the interacting $2+SU(2)_0^3+2$ box theory of Frame II, which has no external delta function.}
    \label{fig:bad_triangulations}
\end{figure}

The generalised toric diagram corresponding to the theory we consider is shown on the left of Figure \ref{fig:bad_triangulations}. Its boundary data encode the deformed theory at infinite coupling with each set of white dots recording three external D5 branes ending on a common D7 brane. Resolving the interior into elementary triangles and trapezoids gives tessellations compatible with these boundary conditions and the generalised $s$-rule of \cite{Benini:2009gi}.

For the polygon at hand, the admissible tessellations fall into two inequivalent classes under flop transitions and Hanany--Witten moves. Representatives of these classes are displayed in the middle of Figure \ref{fig:bad_triangulations}. Every other admissible tessellation can be related to one of these by such moves, but the two representatives cannot be related by any such moves. They therefore describe genuinely different resolutions of the same generalised toric diagram, and hence two inequivalent $(p,q)$ five-brane webs. 

\begin{itemize}
    \item \textbf{Upper tessellation $\longleftrightarrow$ Frame I}. The resulting web engineers the interacting theory $1+SU(2)_0\times\big(SU(3)_0+2\big)\times SU(2)_0+1$. It also contains a finite D5 brane stretched between the two D7 branes transported inside the box, shown in cyan in Figure \ref{fig:bad_triangulations}. This segment can be detached from the rest of the web only if the two D7 branes are aligned along the D5 direction. The required tuning of their relative position is precisely the geometric counterpart of the external periodic delta function\footnote{Related configurations, including a five-brane segment joining two seven-branes, and their geometric interpretation are discussed in \cite{Alexeev:2024bko}.}
    \begin{equation}
        \delta\left(\frac{K_2K_4K'}{K_1K_3K''}\right) ,
    \end{equation}
    which multiplies Frame I in \eqref{eq:bad_frameI_index}.
    \item \textbf{Lower tessellation $\longleftrightarrow$ Frame II}. The resulting web engineers the $2+SU(2)_0^3+2$ box theory, appearing in \eqref{eq:bad_frameII_index}. No relative alignment of the two incoming D7 branes is required, consistently with the absence of an external delta function in this frame.
\end{itemize}
The two inequivalent frames of the index are therefore in one-to-one correspondence with the two equivalence classes of admissible tessellations of the generalised toric diagram.\footnote{The appearance of inequivalent tessellations of a generalised toric polygon has also been discussed in \cite{CarrenoBolla:2024fxy}.}

This gives a $(p,q)$ five-brane web interpretation of the distinction between the broken and bad five-dimensional theories defined through the structure of the superconformal index:
\begin{itemize}
    \item For a \emph{broken} theory, the tessellations of the generalised toric polygons describing the symmetry-related frames belong to a single equivalence class under flop and Hanany--Witten transitions. The excess of white dots shows up in the web as semi-infinite D5 branes that remain attached to the transported seven-branes after the Hanany--Witten moves. Each such D5 brane ending on a seven-brane is the brane avatar of a delta function, as discussed in Section \ref{sec:Higgsing}. All frames of the index arise from this single equivalence class of webs.
    \item For a \emph{bad} theory, the inequivalent frames are represented by distinct classes of admissible tessellations of the infinite-coupling generalised toric polygon that cannot be related by flops or Hanany--Witten transitions. Each class gives its own residual web, and hence its own residual theory and delta functions. In Frame I, the D5 brane suspended between the two seven-branes can be detached only once they are aligned. This condition is the brane realisation of the external periodic delta function in \eqref{eq:bad_complete_index}. In Frame II, no such alignment is required, and correspondingly no external delta function appears.
\end{itemize}

\acknowledgments

We thank Antoine Bourget, Noppadol Mekareeya, Elli Pomoni, Diego Rodríguez-Gómez, Matteo Sacchi, and Gabi Zafrir for discussions and comments on the draft. We are especially grateful to Riccardo Comi and Andrea Sangiovanni for discussions and collaboration.

\appendix

\section{Special functions}
\label{app:special_func}

In this appendix, we collect the special function identities used to obtain the distributional limits in this paper. Our normalisations agree with those of \cite[Appendix A]{one}.

\subsubsection*{Definitions and conventions}

The exponentiated $\Omega$-deformation parameters are
\begin{equation}\label{eq:Omegadef_Region}
    \mf q=e^{-\beta\epsilon_1} \;,\; \mf t=e^{\beta\epsilon_2} \quad\text{such that}\quad |\mf t|>1 \;,\; |\mf q|<1 ~.
\end{equation}
For an integer partition $\lambda$, let $\lambda^t$ denote its transpose and $l(\lambda)$ its number of non-zero rows. We use
\begin{equation}
    |\lambda| \coloneq \sum_{i=1}^{l(\lambda)} \lambda_i \;,\quad ||\lambda||^2 \coloneq \sum_{i=1}^{l(\lambda)} \lambda_i^2 ~,
\end{equation}
and label a box by its row and column, $(i,j)\in\lambda$.

The perturbative factors in the refined topological string partition functions can be expressed through
\begin{equation}\label{eq:Mfun_def}
    \MM{U} \equiv \MM{U;\mf t, \mf q} \coloneq \prod_{i,j=1}^\infty (1-U\,\mf t^{-i}\mf q^j) ~,
\end{equation}
The partition-dependent factors entering the instanton contributions are defined by the finite products
\begin{equation}\label{eq:ZNfun_def}
    \begin{split}
        \tilde{Z}_\nu(\mathfrak{t}, \mathfrak{q}) \coloneq& \prod_{i=1}^{l(\nu)} \prod_{j=1}^{\nu_i} \left(1 - \mathfrak{t}^{\nu_j^t - i + 1} \mathfrak{q}^{\nu_i - j}\right)^{-1} ~, \\
        \cN_{\lambda\mu}(U) \equiv& \cN_{\lambda\mu}(U;\mf t,\mf q) \\
        =& \prod_{(i,j)\in\lambda} (1-U\,\mf t^{\mu_j^t-i}\mf q^{\lambda_i-j+1}) \prod_{(i,j)\in\mu} (1-U\,\mf t^{-\lambda_j^t+i-1}\mf q^{-\mu_i+j}) ~.
    \end{split}
\end{equation}
We will often suppress the explicit dependence on the deformation parameters and abbreviate these functions as $\MM{U}$ and $\cN_{\lambda\mu}(U)$.

\subsubsection*{Modulus square and some identities}

To assemble the superconformal index, the refined topological string partition function is paired with a second copy in which the fugacities and deformation parameters are inverted. We denote this operation by
\begin{equation}\label{eq:mod_square}
    \Big|\, f(U_1,U_2,\dots;\mf t,\mf q) \,\Big|^2 = f(U_1,U_2,\dots;\mf t,\mf q) \, f(U_1^{-1},U_2^{-1},\dots;\mf t^{-1},\mf q^{-1}) ~,
\end{equation}
where the second factor is understood by analytic continuation. For the perturbative functions, we have the identity
\begin{equation}
    \MM{U^{-1},\mf t^{-1},\mf q^{-1}} = \prod_{i,j=1}^\infty (1-U^{-1}\mf t^{1-i}\mf q^{j-1}) = \MM{\frac{\mf t}{\mf q}U^{-1},\mf t,\mf q} ~.
\end{equation}
Consequently, the paired function has the equivalent expressions
\begin{equation}\label{eq:Minv_identity}
    \left| \MM{U} \right|^2 = \MM{U;\mf t,\mf q} \MM{\frac{\mf t}{\mf q}U^{-1};\mf t,\mf q} = \left| \MM{\frac{\mf t}{\mf q}U^{-1}} \right|^2 ~.
\end{equation}
For the partition-dependent factors, inversion also exchanges the two partitions:
\begin{equation}
    \cN_{\lambda\mu}\left( U^{-1};\mf t^{-1},\mf q^{-1} \right) = \cN_{\mu\lambda}\left( \frac{\mf t}{\mf q}U^{-1};\mf t,\mf q \right) ~.
\end{equation}
The corresponding paired expression is therefore
\begin{equation}\label{eq:Ninv_identity}
    \begin{split}
        \left| \cN_{\lambda\mu}(U) \right|^2 &= \cN_{\lambda\mu}(U;\mf t,\mf q) \; \cN_{\lambda\mu}\left( U^{-1};\mf t^{-1},\mf q^{-1} \right) \\
        &= \cN_{\lambda\mu}(U;\mf t,\mf q) \; \cN_{\mu\lambda}\left(\frac{\mf t}{\mf q}U^{-1};\mf t,\mf q\right) ~.
    \end{split}
\end{equation}

\subsubsection*{Zeroes and degeneration limits}

For generic deformation parameters in \eqref{eq:Omegadef_Region}, the zeroes of the paired MacMahon function fall into the two families
\begin{equation}\label{eq:Msq_zeroes}
    (\bar P) :\; U = \mf t^{-n_1}\mf q^{n_2} \quad,\quad (P) :\; U = \mf t^{n_1+1} \mf q^{-n_2-1} \quad\text{ where } n_1,n_2\in\mathbb Z_{\geq0} ~. 
\end{equation}
When this function occurs in a denominator, these families determine the poles relevant to contour pinching in the superconformal index. In particular, the member $U=1$ of the first family is a simple zero. Removing its vanishing factor defines the Cartan contribution
\begin{equation}\label{eq:Mstr_def}
    | M(\mf t,\mf q) |^2 = \lim_{U\rightarrow 1} \frac{ \left| \MM{U;\mf t,\mf q} \right|^2 }{ 1-U^{-1} } = | \MM{\mf q^{-1};\mf t,\mf q} |^2 ~.
\end{equation}
Finally, the following specialisations impose a discrete constraint
\begin{equation}\label{eq:Ntrivial_identity}
    \cN_{\lambda\emptyset} \left(\frac{\mf t}{\mf q}\right) = \cN_{\emptyset\lambda} (1) = \delta_{\lambda,\emptyset} ~.
\end{equation}
In the main text, these identities force the corresponding partition to be empty, thereby trivialising the associated instanton sum.

\section{$\mathbf{S}^3_b$ partition function conventions}
\label{app:3dstuff}

In this appendix we collect the tools we use to perform the 3d calculations. We work with the $S^3_b$ partition function, whose general form was derived in \cite{Hama:2011ea}, adopting the conventions specified below.

\subsubsection*{Axial mass and charge assignments}

We write $Q=b+b^{-1}$, where $b$ is the squashing parameter of $S^3_b$, and describe the theories in $\cN=2^*$ language \cite{Tong:2000ky}. Specifically, we introduce a real mass ${\rm Re}(m)$ associated with the  $U(1)_A$ axial symmetry, identified as the commutant within the $\cN=4$ symmetry $SU(2)\times SU(2)$ of the $U(1)_R$ subgroup which remains manifest using the $\cN=2$ formalism. Its mixing with the reference $R$ symmetry $R_0$ is given by
\begin{equation}
    R = R_0 + r\,q_A ~,
\end{equation}
where $r$ is the mixing coefficient and $q_A$ is the axial charge. To keep track of this mixing, it is convenient to package the parameters in the following holomorphic combination
\begin{equation}\label{eq:massm}
    m\;=\;\text{Re}(m)+\frac{iQ}{2}\,r ~.
\end{equation}
The charges under the two abelian symmetries must then be distributed compatibly with the $\cN=4$ superpotential, which for a single  gauge node takes the form
\begin{equation}
    \mathcal W = {\rm Tr} \left[ \Phi \,Q\,\widetilde Q  \right] ~,
\end{equation}
where $\Phi$ is the adjoint chiral multiplet in the $\cN=4$ vector multiplet while $Q,\widetilde Q$ are the two $\cN=2$ chiral multiplets of the $\cN=4$ hypermultiplet charged under the gauge node. We collect the charges, both for a generic 3d $\cN=4$ theory and its 3d mirror dual, in Table \ref{chargeN2star}. At the level of the partition function, the $R$ symmetry charges in the electric and magnetic theories are related by mirror symmetry as follows:
\begin{equation}
    m \;\longrightarrow\; \frac{iQ}{2}-m ~.
\end{equation}
The magnetic quivers used in this paper follow the charge assignments for magnetic theories given in Table \ref{chargeN2star}. The canonical $R$ symmetry assignment is recovered at $r=1/2$: the hypermultiplet chirals have $R$ charge $1/2$, while $\Phi$ has $R$ charge one. At vanishing real axial mass, this corresponds to $m=iQ/4$.

\begin{table}[h]
\centering
\begin{tabular}{c|c|c|c}
{Electric theory} & $Q$ & $\widetilde{Q}$ & $\Phi$ \\ \hline
$U(1)_{A}$    & $-1$ & $-1$ & $2$ \\
$U(1)_{R_0}$  & $1$  & $1$  & $0$
\end{tabular}
\qquad\qquad
\begin{tabular}{c|c|c|c}
{Magnetic theory} & $Q$ & $\widetilde{Q}$ & $\Phi$ \\ \hline
$U(1)_{A}$    & $1$ & $1$ & $-2$ \\
$U(1)_{R_0}$  & $0$ & $0$ & $2$
\end{tabular}
\caption{Charge assignments of chiral fields in $\mathcal{N}=2^*$
description. The left table details the conventions we use for a generic
theory, while the right one describes those associated with its mirror. These two are connected via the
exchange of $SU(2)_H \leftrightarrow SU(2)_C$.}
\label{chargeN2star}
\end{table}

\subsubsection*{Monopole charges}

Consider a unitary quiver with gauge groups $U(N_a)$, bifundamental hypermultiplets, and flavour nodes. For a bare monopole with magnetic charges $\{\mathfrak m_i^{(a)}\}$, define \cite{Gaiotto:2008ak}
\begin{equation}\label{eq:GWdim}
    \Delta(\mathfrak{m}) = \frac{1}{2} \sum_{\langle a,b\rangle} \sum_{i=1}^{N_a} \sum_{j=1}^{N_b} \big|\mathfrak{m}^{(a)}_i - \mathfrak{m}^{(b)}_j\big| - \sum_a\sum_{i<j}^{N_a} \big|\mathfrak{m}^{(a)}_i - \mathfrak{m}^{(a)}_j\big| ~.
\end{equation}
The first sum runs over the bifundamental links, counting each link once. With the magnetic assignments of Table \ref{chargeN2star}, the total charge of the monopole can be defined as
\begin{equation}\label{eq:monopolecharge}
    \mathcal{Q}(\mathfrak{m}) = \big(iQ-2m\big)\,\Delta(\mathfrak{m}) + \sum_a\xi_a\sum_{i=1}^{N_a}\mathfrak{m}^{(a)}_i ~,
\end{equation}
with $\xi_a$ the FI parameter for the $U(N_a)$ node, specifying the charges of the monopole relative to the Cartan subgroup within the topological symmetry.

Here $\Delta(\mathfrak m)$ is the scaling dimension of the monopole operator computed using the canonical ultraviolet $\cN=4$ assignment. The factor $iQ-2m$ is also the total $R$ symmetry and axial contribution to the charge of a Coulomb-branch moment map. For the canonical $\mathcal{N}=4$ $R$ charge assignment, $m=\frac{iQ}{4}$, this prefactor reduces to $iQ/2$, so \eqref{eq:monopolecharge} becomes $iQ/2$ times the canonical monopole $R$ charge, together with the FI contribution. In order to acquire a VEV, the monopole of magnetic charge $\mathfrak{m}$ must have zero total charge:
$$\mathcal{Q}(\mathfrak{m})=0 ~.$$
This is the charge-neutrality condition used in the main text to determine the corresponding parameter specialisations.

\subsubsection*{Sphere partition function}

For a theory with gauge group $G$, chiral multiplets in gauge and flavour representations $\mathcal{R}$ and $\mathcal{F}$, and chiral $R$ charge $r$, the associated partition function is given by the following ${\rm rank}(G)$-dimensional integral over $\mathbb{R}^{{\rm rank}(G)}$:
\begin{equation}\label{conventionpf}
    \mathcal{Z}(\eta,\vec{\nu})\;=\;\int \! \text{d}\vec{u}\; \PFS{\text{cl}}(\eta)\, \frac{\displaystyle\prod_{\vec{\lambda}\,\in\,\mathcal{R}} \; \prod_{\vec{\omega}\,\in\,\mathcal{F}} s_b \left[\frac{iQ}{2}\left(1-r\right)-\vec{\lambda}(\vec{u})-\vec{\omega}(\vec{\nu})\right]}{\displaystyle\prod_{\vec{\alpha}\,\in\,\Delta_G} s_b\left[\frac{iQ}{2}-\vec{\alpha}(\vec{u})\right]} ~,
\end{equation}
The products run over weights $\vec\lambda,\vec\omega$ of the matter representations and the roots $\vec\alpha$ of $G$. The variables $\vec u$ parameterise the gauge Cartan, while $\vec\nu$ denotes the real flavour masses. The integration measure is
\begin{equation}
    \text{d}\vec{u}\;=\;\frac{1}{\left|\mathcal W_G\right|}\, \prod_{a=1}^{{\rm rank}(G)}\text{d}u_a \;,
\end{equation}
where $\left|\mathcal W_G\right|$ is the order of the Weyl group of $G$. For a $U(N)$ gauge node with vanishing Chern--Simons level, the classical contribution is the following FI factor with parameter $\eta$:
\begin{equation}\label{classicalprefactorpf}
    \PFS{\text{cl}}(\eta ) \;=\; \exp\left[\,2\pi i\,\eta\sum_{a=1}^{{\rm rank}(G)}u_a\,\right] ~.
\end{equation}
For a product of unitary gauge groups, these factors are multiplied over the nodes, with an independent FI parameter for each node.

The double-sine function appearing in the one-loop determinants is defined by the infinite product
\begin{equation}\label{sbdef}
    s_b(x) = \prod_{n_1,\,n_2\,\geq\,0} \frac{n_1 b+n_2 b^{-1}+\frac{Q}{2}-i\,x}{n_1 b+n_2 b^{-1}+\frac{Q}{2}+i\,x} ~.
\end{equation}

Since we work in the $\cN=2^*$ setup, we record the $U(N)$ measures by grouping the $\mathcal{N}=2$ vector and adjoint contributions together into a single object. Whenever the $U(N)$ node gauges a Higgs branch (flavour) symmetry for the magnetic theory, we obtain:
\begin{equation}\label{HBmeasure}
    \Delta^{\text{HB}}_{N}(\vec{u};m)\;=\; \frac{\displaystyle\prod_{i,j=1}^{N} s_b\!\left(2m-\frac{iQ}{2}+u_i-u_j\right)}{\displaystyle\prod_{i\neq j}^{N} s_b\left(\frac{iQ}{2}+u_i-u_j\right)} ~,
\end{equation}
in agreement with the assignments in Table \ref{chargeN2star}. If, conversely, this $U(N)$ node gauges a Coulomb branch (topological) symmetry, then the appropriate measure is
\begin{equation}\label{eq:CBmeasure}
    \Delta^{\text{CB}}_{N}(\vec{u};m)\;=\; \frac{\displaystyle\prod_{i,j=1}^{N} s_b\left(-2m+\frac{iQ}{2}+u_i -u_j\right)} {\displaystyle\prod_{i\neq j}^{N} s_b\left(\frac{iQ}{2}+u_i-u_j\right)} ~.
\end{equation}
The labels HB and CB refer to the magnetic charge convention. Their roles are exchanged in the electric convention: Higgs branch gauging uses $\Delta_N^{\rm CB}$, while Coulomb branch gauging uses $\Delta_N^{\rm HB}$.

\begin{figure}[h!]
    \centering
    \includegraphics[width=0.66\linewidth]{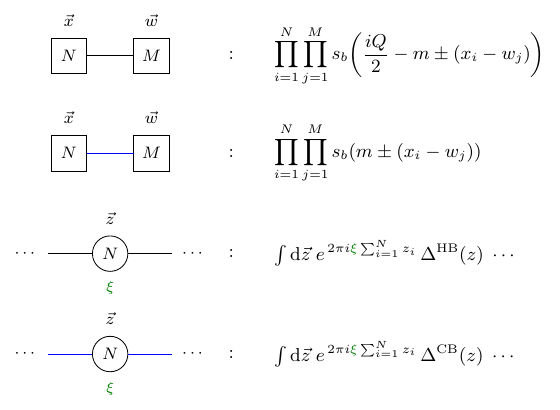}
    \caption{Local rules for constructing the sphere partition function from a quiver. The upper rows show hypermultiplet contributions in the magnetic (black) and electric (blue) conventions; the lower rows show the corresponding gauging prescriptions.}
    \label{fig:conventionsquiver}
\end{figure}

In our quiver diagrams, circles denote gauge groups and squares denote flavour groups. Hypermultiplets with magnetic charge assignments are drawn in black, while those with electric assignments are drawn in \textcolor{blue}{blue}. Figure \ref{fig:conventionsquiver} summarises these conventions and the associated partition function contributions.

\subsubsection*{The $T[U(N)]$ quiver tail}

\begin{figure}[h!]
    \centering
    \includegraphics[width=0.75\linewidth]{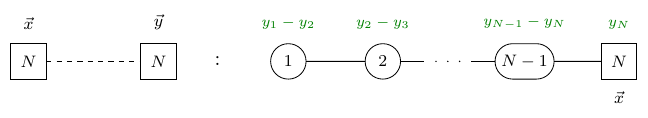}
    \caption{The linear $T[U(N)]$ quiver tail and the corresponding compact dashed-line notation.}
    \label{fig:tsuNappendix1}
\end{figure}

Many of the theories considered here are constructed by gluing $T[U(N)]$ quiver tails. Figure \ref{fig:tsuNappendix1} shows the linear quiver description of this building block and the compact dashed-line notation used in the main text.
The partition function of this theory can be obtained using the rules above. We denote it by
$$ \PFS{}^{T[U(N)]}(\vec{X},\vec{Y},m) ~,$$
where $\vec{X}$ parametrises the $U(N)_{\vec X}$ Higgs-branch global symmetry and $\vec{Y}$ parametrises the $U(N)_{\vec Y}$ enhanced Coulomb-branch global symmetry. Relative to the $T[SU(N)]$ description of \cite{Gaiotto:2008ak}, our $T[U(N)]$ convention includes the background contact term $e^{2\pi iY_N \sum_{j=1}^N X_j}$ for the flavour symmetry.

Mirror symmetry swaps the Higgs and Coulomb branches alongside their corresponding Cartans, while transforming the axial mass according to $m\to iQ/2-m$. At the level of the partition function, this gives
$$\PFS{}^{T[U(N)]}(\vec{X},\vec{Y},m) = \PFS{}^{T[U(N)]}\left(\vec{Y},\vec{X},\frac{iQ}{2}-m\right) ~.$$

\section{Goldstone modes from nilpotent VEVs}
\label{app:GM}

In this appendix, we review how to determine the Goldstone contributions associated with the nilpotent VEVs considered in the main text. The symmetry-breaking data determine the Goldstone directions, while their contributions to the three- and five-dimensional partition functions are obtained using the corresponding special functions.

\paragraph{Three-dimensional Goldstone modes:} We begin with the 3d  $\mathcal N=2^*$ case. A convenient way to impose a VEV for an operator $O$ is to introduce a Lagrange-multiplier chiral multiplet $\Lambda$ through the superpotential\footnote{This deformation preserves only $\mathcal N=2$ supersymmetry manifestly; in the examples considered here, the theory can enhance back to $\mathcal N=4$ at the infrared fixed point.}
\begin{equation}
    \mathcal W=\operatorname{Tr}\left[\Lambda\left(J-O\right)\right],
\end{equation}
whose F-term equation imposes $O=J$. Here $J$ is the nilpotent Jordan matrix associated with the partition $\rho$. At the level of the $S_b^3$ partition function, the nilpotent VEV translates into relations among the real-mass parameters and the parameters entering the trial $R$ symmetry. These relations follow by requiring every component of $O$ acquiring a VEV to have vanishing total charge. For example, consider the $SU(4)$ nilpotent VEV given by the partition $[2,1^2]$, represented by
\begin{equation}
    J = \begin{pmatrix}  0 & 0 & 0 & 0 \\  1 & 0 & 0 & 0 \\  0 & 0 & 0 & 0 \\  0 & 0 & 0 & 0  \end{pmatrix} \;\Longrightarrow\; \langle O_{21}\rangle \neq 0 ~.
\end{equation}

The moment map operator has trial $R$ charge $R_0=2$ and axial charge -2, following the conventions for the Coulomb branch moment maps of the magnetic quivers considered here. We denote the $SU(4)$ mass parameters by $m_1,m_2,m_3,m_4$ (with $\sum_{i=1}^4 m_i=0$). The total charge of $O_{ij}$ is therefore $$m_i-m_j+iQ-2m ~,$$ where $iQ/2$ multiplies the trial $R$ charge, while the complexified axial mass $m$ incorporates the mixing specified in Appendix \ref{app:3dstuff}.

Requiring $O_{21}$ to have zero total charge gives
\begin{equation}
    m_2 = m_1-iQ+2m ~.
\end{equation}
We parameterize the $U(1)$ factor surviving the VEV by setting
\begin{equation}
    m_1 = h' \;\;,\quad m_2 = h'-iQ+2m ~.
\end{equation}
More generally, the entries belonging to each Jordan block are constrained to form an arithmetic progression in additive variables, with successive entries differing by $-iQ+2m$, or a geometric progression in multiplicative variables.

The mass specialisation extends directly to a general hook partition $[L,1^{N-L}]$: vanishing total charge of the components $O_{a+1,a}$ fixes the successive differences to $m_{a+1}-m_a=-iQ+2m$. In the convention used throughout the main text, we parameterise the resulting arithmetic progression as
\begin{equation}
    m_a = h'+\left(\frac L2-a\right)(iQ-2m) \;\;,\quad a=1,\dots,L ~.
\end{equation}
The centre of the block is therefore $h'-iQ/2+m$, independently of its length.

We now return to the $[2,1^{N-2}]$ VEV and determine its Goldstone contribution. The Goldstone directions are tangent to the nilpotent orbit: they are generated by fluctuations $\delta O=[T,J]$, with generators $T$ satisfying $[T,J]=0$ belonging to the stabiliser. For the $[2,1^2]$ example, the commuting matrices take the form
\begin{equation}
    T_{\rm preserved} =
    \begin{pmatrix}
        t_{11} & 0 & 0 & 0 \\
        t_{21} & t_{11} & t_{23} & t_{24} \\
        t_{31} & 0 & t_{33} & t_{34} \\
        t_{41} & 0 & t_{43} & t_{44}
    \end{pmatrix} ~,
\end{equation}
where $2t_{11}+t_{33}+t_{44}=0$. A convenient choice of representatives for the broken directions is
\begin{equation}
    T_{\rm broken} = T - T_{\rm preserved} =
    \begin{pmatrix}
        0 & t_{12} & t_{13} & t_{14} \\
        0 & t_{22}-t_{11} & 0 & 0 \\
        0 & t_{32} & 0 & 0 \\
        0 & t_{42} & 0 & 0
    \end{pmatrix} ~.
\end{equation}
The six independent parameters in $T_{\rm broken}$ label the Goldstone directions. Their combined contribution can be written as a product over these independent entries, assigning the factor
$$s_b\left(2m-\frac{iQ}{2}+m_i-m_j\right)$$
to each entry $(i,j)$. Applying the VEV conditions $m_1=h'$ and $m_2=h'-iQ+2m$, we obtain
\begin{equation}
    {\rm GM}_{\rm 3d}^{[2,1^2]}(\vec m',h') = s_b\left(m\pm\left(\frac{iQ}{2}-m\right)\right) \prod_{j=3}^4 s_b\left(m\pm\left(m-\frac{iQ}{2}+h'-m_j\right)\right) ~.
\end{equation}
where $\vec m'=(m_3,m_4)$ are the residual parameters.

\paragraph{Five-dimensional Goldstone modes:} In the 5d $\cN=1$ case, the $[2,1^2]$ VEV is implemented by
\begin{equation}
    M_1 = \sqrt{\frac{\mf t}{\mf q}}\,K' \;\;,\quad M_2 = \sqrt{\frac{\mf q}{\mf t}}\,K' \quad\text{and}\quad \alpha_1=\alpha_2=\emptyset ~,
\end{equation}
where $M_1,\dots,M_4$ are the $SU(4)$ fugacities and $\alpha_1,\dots,\alpha_4$ the associated external partitions. The Goldstone contribution to the five-dimensional superconformal index is obtained by associating to each entry $(i,j)$ of $T_{\rm broken}$ the factor
$$\frac{\cN_{\alpha_i\alpha_j}\left(\frac{M_i}{M_j}\right)}{\MM{\frac{M_i}{M_j}}} ~,$$
and imposing the VEV conditions. For the present $[2,1^2]$ example, this gives
\begin{equation}
    \frac{ \cN_{\emptyset\emptyset}\left(\frac{\mf t}{\mf q}\right) \cN_{\emptyset\alpha_3}\left(\sqrt{\frac{\mf t}{\mf q}}\frac{K'}{M_3}\right) \cN_{\emptyset\alpha_4}\left(\sqrt{\frac{\mf t}{\mf q}}\frac{K'}{M_4}\right) }{ \MM{\frac{\mf t}{\mf q}} \MM{\sqrt{\frac{\mf t}{\mf q}}\frac{K'}{M_3}} \MM{\sqrt{\frac{\mf t}{\mf q}}\frac{K'}{M_4}} } \times \frac{ \cN_{\emptyset\emptyset}\left(1\right) \cN_{\alpha_3\emptyset}\left(\sqrt{\frac{\mf t}{\mf q}}\frac{M_3}{K'}\right) \cN_{\alpha_4\emptyset}\left(\sqrt{\frac{\mf t}{\mf q}}\frac{M_4}{K'}\right) }{ \MM{1} \MM{\sqrt{\frac{\mf t}{\mf q}}\frac{M_3}{K'}} \MM{\sqrt{\frac{\mf t}{\mf q}}\frac{M_4}{K'}} }
\end{equation}
The two lines pair under the simultaneous inversion using Eqs.\ (\ref{eq:Minv_identity}) and (\ref{eq:Ninv_identity}). Using also $\cN_{\emptyset\emptyset}(U)=1$, we obtain
\begin{equation}
    {\rm GM}^\rho_{\rm 5d}(\vec M',K';\bm\alpha') = \left| \frac{ \cN_{\emptyset\emptyset}\left(\frac{\mf t}{\mf q}\right) \cN_{\emptyset\alpha_3}\left(\sqrt{\frac{\mf t}{\mf q}}\frac{K'}{M_3}\right) \cN_{\emptyset\alpha_4}\left(\sqrt{\frac{\mf t}{\mf q}}\frac{K'}{M_4}\right) }{ \MM{\frac{\mf t}{\mf q}} \MM{\sqrt{\frac{\mf t}{\mf q}}\frac{K'}{M_3}} \MM{\sqrt{\frac{\mf t}{\mf q}}\frac{K'}{M_4}} } \right|^2 ~. 
\end{equation}
Thus the contributions associated with the broken directions combine into complete modulus-squared factors for the five-dimensional Goldstone multiplets. Here $\vec M'=(M_3,M_4)$ and $\bm\alpha'=(\alpha_3,\alpha_4)$ are the fugacities and partitions untouched by the VEV, while $K'$ parameterises the non-trivial Jordan block.

More generally, for $\rho=[L,1^{N-L}]$, the length-$L$ Jordan block imposes the geometric progression
\begin{equation}
    M_a = \left(\frac{\mf t}{\mf q}\right)^{\frac{L+1}{2}-a}K' \;\;,\quad \alpha_a=\emptyset \;\;,\quad a=1,\dots,L ~.
\end{equation}
The same computation then gives
\begin{equation}
    {\rm GM}_{\rm 5d}^{[L,1^{N-L}]}(\vec M',K',\bm\alpha') = \left| \prod_{a=1}^{L-1} \frac{1}{\MM{\frac{\mf t^a}{\mf q^a}}^{L-a}} \prod_{b=L+1}^N \prod_{a=1}^{L-1} \frac{ \cN_{\emptyset\alpha_b}\left(\left(\frac{\mf t}{\mf q}\right)^{\frac{L+1}{2}-a}\frac{K'}{M_b}\right) }{ \MM{\left(\frac{\mf t}{\mf q}\right)^{\frac{L+1}{2}-a}\frac{K'}{M_b}} } \right|^2 ~,
\end{equation}
where $\vec M'=(M_{L+1},\dots,M_N)$ and $\bm\alpha'=(\alpha_{L+1},\dots,\alpha_N)$. This analysis can be easily generalised to arbitrary partitions of $N$.

\bibliographystyle{JHEP}
\bibliography{ref}

@article{Hanany:1996ie,
    author = "Hanany, Amihay and Witten, Edward",
    title = "{Type IIB superstrings, BPS monopoles, and three-dimensional gauge dynamics}",
    eprint = "hep-th/9611230",
    archivePrefix = "arXiv",
    reportNumber = "IASSNS-HEP-96-121",
    doi = "10.1016/S0550-3213(97)00157-0",
    journal = "Nucl. Phys. B",
    volume = "492",
    pages = "152--190",
    year = "1997"
}

@article{BourgetHananyMiketa2020,
  author        = {Bourget, Antoine and Hanany, Amihay and Miketa, Dominik},
  title         = {Quiver origami: discrete gauging and folding},
  journal       = {JHEP},
  volume        = {01},
  pages         = {086},
  year          = {2021},
  doi           = {10.1007/JHEP01(2021)086},
  eprint        = {2005.05273},
  archivePrefix = {arXiv},
  primaryClass  = {hep-th}
}

@article{BourgetGrimmingerMartoneZafrir2021,
  author        = {Bourget, Antoine and Grimminger, Julius F. and
                   Martone, Mario and Zafrir, Gabi},
  title         = {Magnetic quivers for rank 2 theories},
  journal       = {JHEP},
  volume        = {03},
  pages         = {208},
  year          = {2022},
  doi           = {10.1007/JHEP03(2022)208},
  eprint        = {2110.11365},
  archivePrefix = {arXiv},
  primaryClass  = {hep-th}
}

@article{Bourget:2023wlb,
    author = "Bourget, Antoine and Collinucci, Andr{\'e}s and Schafer-Nameki, Sakura",
    title = "{Generalized toric polygons, T-branes, and 5d SCFTs}",
    eprint = "2301.05239",
    archivePrefix = "arXiv",
    primaryClass = "hep-th",
    doi = "10.21468/SciPostPhys.18.3.079",
    journal = "SciPost Phys.",
    volume = "18",
    number = "3",
    pages = "079",
    year = "2025"
}

@article{Aharony:1997bh,
    author = "Aharony, Ofer and Hanany, Amihay and Kol, Barak",
    title = "{Webs of (p,q) five-branes, five-dimensional field theories and grid diagrams}",
    eprint = "hep-th/9710116",
    archivePrefix = "arXiv",
    reportNumber = "IASSNS-HEP-97-113, RU-97-81, SU-ITP-97-40",
    doi = "10.1088/1126-6708/1998/01/002",
    journal = "JHEP",
    volume = "01",
    pages = "002",
    year = "1998"
}

@article{Bergman:2013aca,
    author = "Bergman, Oren and Rodr{\'\i}guez-G{\'o}mez, Diego and Zafrir, Gabi",
    title = "{5-Brane Webs, Symmetry Enhancement, and Duality in 5d Supersymmetric Gauge Theory}",
    eprint = "1311.4199",
    archivePrefix = "arXiv",
    primaryClass = "hep-th",
    doi = "10.1007/JHEP03(2014)112",
    journal = "JHEP",
    volume = "03",
    pages = "112",
    year = "2014"
}

@article{CarrenoBolla:2024fxy,
    author = "Carre{\~n}o Bolla, Ignacio and Franco, Sebasti{\'a}n and Rodr{\'\i}guez-G{\'o}mez, Diego",
    title = "{The 5d tangram: brane webs, 7-branes and primitive T-cones}",
    eprint = "2411.01510",
    archivePrefix = "arXiv",
    primaryClass = "hep-th",
    doi = "10.1007/JHEP05(2025)175",
    journal = "JHEP",
    volume = "05",
    pages = "175",
    year = "2025"
}

@article{Terashima:2012ra,
    author = "Terashima, Seiji",
    title = "{Supersymmetric gauge theories on $S^4$ x $S^1$}",
    eprint = "1207.2163",
    archivePrefix = "arXiv",
    primaryClass = "hep-th",
    reportNumber = "YITP-12-55",
    doi = "10.1103/PhysRevD.89.125001",
    journal = "Phys. Rev. D",
    volume = "89",
    number = "12",
    pages = "125001",
    year = "2014"
}

@article{Awata:2005fa,
    author = "Awata, Hidetoshi and Kanno, Hiroaki",
    title = "{Instanton counting, Macdonald functions and the moduli space of D-branes}",
    eprint = "hep-th/0502061",
    archivePrefix = "arXiv",
    doi = "10.1088/1126-6708/2005/05/039",
    journal = "JHEP",
    volume = "05",
    pages = "039",
    year = "2005"
}

@article{one,
    author = "Garavaglia, Sebastiano and Pasquetti, Sara and Singh, Palash",
    title = "From pq-webs to magnetic quivers and back I: dictionary and deformations",
    journal = "to appear"
}

@article{Bourget:2026hsg,
    author = "Bourget, Antoine and Cheng, Peng and Lamouret, Quentin",
    title = "{BPS Invariants for Generalized Toric Calabi-Yau Threefolds}",
    eprint = "2607.29451",
    archivePrefix = "arXiv",
    primaryClass = "hep-th",
    month = "7",
    year = "2026"
}

@article{Comi:2026lcj,
    author = "Comi, Riccardo and Hwang, Chiung and Marino, Fabio",
    title = "{Algorithmic Dualization of Unitary Circular Quivers}",
    eprint = "2607.01327",
    archivePrefix = "arXiv",
    primaryClass = "hep-th",
    reportNumber = "USTC-ICTS/PCFT-26-34, UWThPh 2026-6",
    month = "7",
    year = "2026"
}

@article{DeWolfe:1999hj,
    author = "DeWolfe, Oliver and Hanany, Amihay and Iqbal, Amer and Katz, Emanuel",
    title = "{Five-branes, seven-branes and five-dimensional E(n) field theories}",
    eprint = "hep-th/9902179",
    archivePrefix = "arXiv",
    reportNumber = "MIT-CTP-2831",
    doi = "10.1088/1126-6708/1999/03/006",
    journal = "JHEP",
    volume = "03",
    pages = "006",
    year = "1999"
}

@article{Spiridonov:2014cxa,
    author = "Spiridonov, Vyacheslav P. and Vartanov, G. S.",
    title = "{Vanishing superconformal indices and the chiral symmetry breaking}",
    eprint = "1402.2312",
    archivePrefix = "arXiv",
    primaryClass = "hep-th",
    doi = "10.1007/JHEP06(2014)062",
    journal = "JHEP",
    volume = "06",
    pages = "062",
    year = "2014"
}

@article{Mitev:2014jza,
    author = "Mitev, Vladimir and Pomoni, Elli and Taki, Masato and Yagi, Futoshi",
    title = "{Fiber-Base Duality and Global Symmetry Enhancement}",
    eprint = "1411.2450",
    archivePrefix = "arXiv",
    primaryClass = "hep-th",
    reportNumber = "DESY-14-212, HU-Mathematik-2014-32, HU-EP-14-48, KIAS-P14066",
    doi = "10.1007/JHEP04(2015)052",
    journal = "JHEP",
    volume = "04",
    pages = "052",
    year = "2015"
}

@article{Katz:1996fh,
    author = "Katz, Sheldon H. and Klemm, Albrecht and Vafa, Cumrun",
    title = "{Geometric Engineering of Quantum Field Theories}",
    eprint = "hep-th/9609239",
    archivePrefix = "arXiv",
    doi = "10.1016/S0550-3213(97)00282-4",
    journal = "Nucl. Phys. B",
    volume = "497",
    pages = "173--195",
    year = "1997"
}

@article{Hayashi:2013qwa,
    author = "Hayashi, Hirotaka and Kim, Hee-Cheol and Nishinaka, Takahiro",
    title = "{Topological strings and 5d $T_N$ partition functions}",
    eprint = "1310.3854",
    archivePrefix = "arXiv",
    primaryClass = "hep-th",
    doi = "10.1007/JHEP06(2014)014",
    journal = "JHEP",
    volume = "06",
    pages = "014",
    year = "2014"
}

@article{Cremonesi:2015lsa,
    author = "Cremonesi, Stefano and Ferlito, Giulia and Hanany, Amihay and Mekareeya, Noppadol",
    title = "{Instanton Operators and the Higgs Branch at Infinite Coupling}",
    eprint = "1505.06302",
    archivePrefix = "arXiv",
    primaryClass = "hep-th",
    doi = "10.1007/JHEP04(2017)042",
    journal = "JHEP",
    volume = "04",
    pages = "042",
    year = "2017"
}

@article{Ferlito:2016grh,
    author = "Ferlito, Giulia and Hanany, Amihay",
    title = "{A tale of two cones: the Higgs Branch of Sp(n) theories with 2n flavours}",
    eprint = "1609.06724",
    archivePrefix = "arXiv",
    primaryClass = "hep-th",
    month = "9",
    year = "2016"
}

@article{Ferlito:2017xdq,
    author = "Ferlito, Giulia and Hanany, Amihay and Mekareeya, Noppadol and Zafrir, Gabi",
    title = "{3d Coulomb branch and 5d Higgs branch at infinite coupling}",
    eprint = "1712.06604",
    archivePrefix = "arXiv",
    primaryClass = "hep-th",
    doi = "10.1007/JHEP07(2018)061",
    journal = "JHEP",
    volume = "07",
    pages = "061",
    year = "2018"
}

@article{Cabrera:2018jxt,
    author = "Cabrera, Santiago and Hanany, Amihay and Yagi, Futoshi",
    title = "{Tropical Geometry and Five Dimensional Higgs Branches at Infinite Coupling}",
    eprint = "1810.01379",
    archivePrefix = "arXiv",
    primaryClass = "hep-th",
    doi = "10.1007/JHEP01(2019)068",
    journal = "JHEP",
    volume = "01",
    pages = "068",
    year = "2019"
}

@article{vanBeest:2020kou,
    author = "van Beest, Marieke and Bourget, Antoine and Eckhard, Jonathan and Sch{\"a}fer-Nameki, Sakura",
    title = "{(Symplectic) Leaves and (5d Higgs) Branches in the Poly(go)nesian Tropical Rain Forest}",
    eprint = "2008.05577",
    archivePrefix = "arXiv",
    primaryClass = "hep-th",
    doi = "10.1007/JHEP11(2020)124",
    journal = "JHEP",
    volume = "11",
    pages = "124",
    year = "2020"
}

@article{Hayashi:2014hfa,
    author = "Hayashi, Hirotaka and Zoccarato, Gianluca",
    title = "{Exact partition functions of Higgsed 5d $T_N$ theories}",
    eprint = "1409.0571",
    archivePrefix = "arXiv",
    primaryClass = "hep-th",
    doi = "10.1007/JHEP01(2015)093",
    journal = "JHEP",
    volume = "01",
    pages = "093",
    year = "2015"
}

@article{Leung:1997tw,
    author = "Leung, Naichung Conan and Vafa, Cumrun",
    title = "{Branes and toric geometry}",
    eprint = "hep-th/9711013",
    archivePrefix = "arXiv",
    reportNumber = "HUTP-97-A090",
    doi = "10.4310/ATMP.1998.v2.n1.a4",
    journal = "Adv. Theor. Math. Phys.",
    volume = "2",
    pages = "91--118",
    year = "1998"
}

@article{Alexeev:2024bko,
    author = {Alexeev, Valery and Arg{\"u}z, H{\"u}lya and Bousseau, Pierrick},
    title = "{Non-toric Brane Webs, Calabi{\textendash}Yau 3-Folds, and 5d SCFTs}",
    eprint = "2410.04714",
    archivePrefix = "arXiv",
    primaryClass = "hep-th",
    doi = "10.1007/s00220-025-05461-9",
    journal = "Commun. Math. Phys.",
    volume = "406",
    number = "11",
    pages = "287",
    year = "2025"
}

@article{Nekrasov:2003rj,
    author = "Nekrasov, Nikita and Okounkov, Andrei",
    title = "{Seiberg-Witten theory and random partitions}",
    eprint = "hep-th/0306238",
    archivePrefix = "arXiv",
    reportNumber = "ITEP-TH-36-03, PUDM-2003, IHES-P-03-43",
    doi = "10.1007/0-8176-4467-9_15",
    journal = "Prog. Math.",
    volume = "244",
    pages = "525--596",
    year = "2006"
}

@article{Aharony:1997ju,
    author = "Aharony, Ofer and Hanany, Amihay",
    title = "{Branes, superpotentials and superconformal fixed points}",
    eprint = "hep-th/9704170",
    archivePrefix = "arXiv",
    reportNumber = "RU-97-25, IASSNS-HEP-97-38",
    doi = "10.1016/S0550-3213(97)00472-0",
    journal = "Nucl. Phys. B",
    volume = "504",
    pages = "239--271",
    year = "1997"
}

@article{Iqbal:2012xm,
    author = "Iqbal, Amer and Vafa, Cumrun",
    title = "{BPS Degeneracies and Superconformal Index in Diverse Dimensions}",
    eprint = "1210.3605",
    archivePrefix = "arXiv",
    primaryClass = "hep-th",
    doi = "10.1103/PhysRevD.90.105031",
    journal = "Phys. Rev. D",
    volume = "90",
    number = "10",
    pages = "105031",
    year = "2014"
}

@article{Kim:2012gu,
    author = "Kim, Hee-Cheol and Kim, Sung-Soo and Lee, Kimyeong",
    title = "{5-dim Superconformal Index with Enhanced En Global Symmetry}",
    eprint = "1206.6781",
    archivePrefix = "arXiv",
    primaryClass = "hep-th",
    reportNumber = "KIAS-P12033",
    doi = "10.1007/JHEP10(2012)142",
    journal = "JHEP",
    volume = "10",
    pages = "142",
    year = "2012"
}

@article{Nekrasov:2002qd,
    author = "Nekrasov, Nikita A.",
    title = "{Seiberg-Witten prepotential from instanton counting}",
    eprint = "hep-th/0206161",
    archivePrefix = "arXiv",
    reportNumber = "ITEP-TH-22-02, IHES-P-04-22",
    doi = "10.4310/ATMP.2003.v7.n5.a4",
    journal = "Adv. Theor. Math. Phys.",
    volume = "7",
    number = "5",
    pages = "831--864",
    year = "2003"
}

@article{Aganagic:2003db,
    author = "Aganagic, Mina and Klemm, Albrecht and Marino, Marcos and Vafa, Cumrun",
    title = "{The Topological vertex}",
    eprint = "hep-th/0305132",
    archivePrefix = "arXiv",
    reportNumber = "CALT-68-2439, HUTP-03-A032, HU-EP-03-24, CERN-TH-2003-111",
    doi = "10.1007/s00220-004-1162-z",
    journal = "Commun. Math. Phys.",
    volume = "254",
    pages = "425--478",
    year = "2005"
}

@article{Iqbal:2007ii,
    author = "Iqbal, Amer and Kozcaz, Can and Vafa, Cumrun",
    title = "{The Refined topological vertex}",
    eprint = "hep-th/0701156",
    archivePrefix = "arXiv",
    doi = "10.1088/1126-6708/2009/10/069",
    journal = "JHEP",
    volume = "10",
    pages = "069",
    year = "2009"
}

@article{Iqbal:2004ne,
    author = "Iqbal, Amer and Kashani-Poor, Amir-Kian",
    title = "{The Vertex on a strip}",
    eprint = "hep-th/0410174",
    archivePrefix = "arXiv",
    reportNumber = "SMS-0402, SLAC-PUB-10804, SU-ITP-04-39",
    doi = "10.4310/ATMP.2006.v10.n3.a2",
    journal = "Adv. Theor. Math. Phys.",
    volume = "10",
    number = "3",
    pages = "317--343",
    year = "2006"
}

@article{Taki:2007dh,
    author = "Taki, Masato",
    title = "{Refined Topological Vertex and Instanton Counting}",
    eprint = "0710.1776",
    archivePrefix = "arXiv",
    primaryClass = "hep-th",
    reportNumber = "UT-07-32",
    doi = "10.1088/1126-6708/2008/03/048",
    journal = "JHEP",
    volume = "03",
    pages = "048",
    year = "2008"
}

@article{Isachenkov:2014eya,
    author = "Isachenkov, Mikhail and Mitev, Vladimir and Pomoni, Elli",
    title = "{Toda 3-Point Functions From Topological Strings II}",
    eprint = "1412.3395",
    archivePrefix = "arXiv",
    primaryClass = "hep-th",
    reportNumber = "DESY-14-236, HU-MATHEMATIK-35-2014, HU-EP-14-57",
    doi = "10.1007/JHEP08(2016)066",
    journal = "JHEP",
    volume = "08",
    pages = "066",
    year = "2016"
}

@article{Comi:2025zwu,
    author = "Comi, Riccardo and Garavaglia, Sebastiano and Giacomelli, Simone and Pasquetti, Sara and Singh, Palash",
    title = "{Breaking bad theories of class $ \mathcal{S} $}",
    eprint = "2508.21071",
    archivePrefix = "arXiv",
    primaryClass = "hep-th",
    doi = "10.1007/JHEP05(2026)075",
    journal = "JHEP",
    volume = "05",
    pages = "075",
    year = "2026"
}

@article{Gaiotto:2012xa,
    author = "Gaiotto, Davide and Rastelli, Leonardo and Razamat, Shlomo S.",
    title = "{Bootstrapping the superconformal index with surface defects}",
    eprint = "1207.3577",
    archivePrefix = "arXiv",
    primaryClass = "hep-th",
    doi = "10.1007/JHEP01(2013)022",
    journal = "JHEP",
    volume = "01",
    pages = "022",
    year = "2013"
}

@article{Kim:2012uz,
    author = "Kim, Hee-Cheol and Kim, Jungmin and Kim, Seok and Lee, Kanghoon",
    title = "{Vortices and 3 dimensional dualities}",
    eprint = "1204.3895",
    archivePrefix = "arXiv",
    primaryClass = "hep-th",
    reportNumber = "KIAS-P11078, SNUTP11-011",
    month = "4",
    year = "2012"
}

@article{Benini:2009gi,
    author = "Benini, Francesco and Benvenuti, Sergio and Tachikawa, Yuji",
    title = "{Webs of five-branes and N=2 superconformal field theories}",
    eprint = "0906.0359",
    archivePrefix = "arXiv",
    primaryClass = "hep-th",
    reportNumber = "PUPT-2303",
    doi = "10.1088/1126-6708/2009/09/052",
    journal = "JHEP",
    volume = "09",
    pages = "052",
    year = "2009"
}

@article{Hama:2011ea,
    author = "Hama, Naofumi and Hosomichi, Kazuo and Lee, Sungjay",
    title = "{SUSY Gauge Theories on Squashed Three-Spheres}",
    eprint = "1102.4716",
    archivePrefix = "arXiv",
    primaryClass = "hep-th",
    reportNumber = "DAMTP-2011-6, YITP-11-1",
    doi = "10.1007/JHEP05(2011)014",
    journal = "JHEP",
    volume = "05",
    pages = "014",
    year = "2011"
}

@article{Benini:2010uu,
    author = "Benini, Francesco and Tachikawa, Yuji and Xie, Dan",
    title = "{Mirrors of 3d Sicilian theories}",
    eprint = "1007.0992",
    archivePrefix = "arXiv",
    primaryClass = "hep-th",
    reportNumber = "MIFPA-10-27, PUTP-2344",
    doi = "10.1007/JHEP09(2010)063",
    journal = "JHEP",
    volume = "09",
    pages = "063",
    year = "2010"
}

@article{Giacomelli:2023zkk,
    author = "Giacomelli, Simone and Hwang, Chiung and Marino, Fabio and Pasquetti, Sara and Sacchi, Matteo",
    title = "{Probing bad theories with the dualization algorithm. Part I}",
    eprint = "2309.05326",
    archivePrefix = "arXiv",
    primaryClass = "hep-th",
    reportNumber = "USTC-ICTS/PCFT-23-24, DMUS-MP-23/12",
    doi = "10.1007/JHEP04(2024)008",
    journal = "JHEP",
    volume = "04",
    pages = "008",
    year = "2024"
}

@article{Giacomelli:2024laq,
    author = "Giacomelli, Simone and Hwang, Chiung and Marino, Fabio and Pasquetti, Sara and Sacchi, Matteo",
    title = "{Probing bad theories with the dualization algorithm. Part II.}",
    eprint = "2401.14456",
    archivePrefix = "arXiv",
    primaryClass = "hep-th",
    doi = "10.1007/JHEP07(2024)165",
    journal = "JHEP",
    volume = "07",
    pages = "165",
    year = "2024"
}

@article{Gaiotto:2008ak,
    author = "Gaiotto, Davide and Witten, Edward",
    title = "{S-Duality of Boundary Conditions In N=4 Super Yang-Mills Theory}",
    eprint = "0807.3720",
    archivePrefix = "arXiv",
    primaryClass = "hep-th",
    doi = "10.4310/ATMP.2009.v13.n3.a5",
    journal = "Adv. Theor. Math. Phys.",
    volume = "13",
    number = "3",
    pages = "721--896",
    year = "2009"
}

@article{Yaakov:2013fza,
    author = "Yaakov, Itamar",
    title = "{Redeeming Bad Theories}",
    eprint = "1303.2769",
    archivePrefix = "arXiv",
    primaryClass = "hep-th",
    doi = "10.1007/JHEP11(2013)189",
    journal = "JHEP",
    volume = "11",
    pages = "189",
    year = "2013"
}

@article{Assel:2017jgo,
    author = "Assel, Benjamin and Cremonesi, Stefano",
    title = "{The Infrared Physics of Bad Theories}",
    eprint = "1707.03403",
    archivePrefix = "arXiv",
    primaryClass = "hep-th",
    reportNumber = "CERN-TH-2017-147",
    doi = "10.21468/SciPostPhys.3.3.024",
    journal = "SciPost Phys.",
    volume = "3",
    number = "3",
    pages = "024",
    year = "2017"
}

@article{Tong:2000ky,
    author = "Tong, David",
    title = "{Dynamics of N=2 supersymmetric Chern-Simons theories}",
    eprint = "hep-th/0005186",
    archivePrefix = "arXiv",
    reportNumber = "MIT-CTP-2985, KCL-TH-00-28",
    doi = "10.1088/1126-6708/2000/07/019",
    journal = "JHEP",
    volume = "07",
    pages = "019",
    year = "2000"
}

@article{Assel:2018exy,
    author = "Assel, Benjamin and Cremonesi, Stefano",
    title = "{The Infrared Fixed Points of 3d $\mathcal{N}=4$ $USp(2N)$ SQCD Theories}",
    eprint = "1802.04285",
    archivePrefix = "arXiv",
    primaryClass = "hep-th",
    reportNumber = "CERN-TH-2018-031",
    doi = "10.21468/SciPostPhys.5.2.015",
    journal = "SciPost Phys.",
    volume = "5",
    number = "2",
    pages = "015",
    year = "2018"
}

@article{badusp,
    author = "Comi, Riccardo and Garavaglia, Sebastiano and Giacomelli, Simone and Pasquetti, Sara",
    title = "{The partition function of good and bad symplectic SQCD}",
    journal = "In progress"
}

@article{cbgauge,
    author = "Comi, Riccardo and Garavaglia, Sebastiano and Singh, Palash",
    title = "{Gauging symmetries on the Coulomb branch}",
    journal = "In progress"
}

\end{document}